\pdfoutput=1

\documentclass[final,numrefs,sort&compress,noinfo]{nddiss2e}
\usepackage{comment}
\usepackage{epstopdf}
\usepackage{graphicx}
\usepackage{gensymb}
\usepackage{amsmath}
\usepackage{amsthm}
\usepackage{anyfontsize}
\usepackage{booktabs}

\DeclareMathOperator*{\argmax}{argmax}
\DeclareMathOperator*{\argmin}{argmin}
\newcommand*{\rom}[1]{\expandafter\@slowromancap\romannumeral #1@}

\begin{document}

\frontmatter 

\title{ENGINEERING AND ECONOMIC ANALYSIS FOR ELECTRIC VEHICLE CHARGING INFRASTRUCTURE --- PLACEMENT, PRICING, AND MARKET DESIGN}
\author{Chao Luo}
\work{Dissertation} 
\degaward{Doctor of Philosophy} 
\advisor{Yih-Fang Huang}
\department{Electrical Engineering}

\maketitle
%
%

\copyrightholder{Chao Luo} 
\copyrightyear{2017}           
\makecopyright

\begin{abstract}
The objective of this dissertation is to study the interplay between large-scale electric vehicle (EV) charging and the power system. In particular, we address three important issues pertaining to EV charging and integration into the power system: (1) charging station placement, (2) pricing policy and energy management strategy, and (3) electricity trading market and distribution network design to facilitate integrating EV and renewable energy source (RES) into the power system.

Regarding the charging station placement problem, we propose a multi-stage consumer behavior based placement strategy with incremental EV penetration rates and model the EV charging industry as an oligopoly where the entire market is dominated by a few charging service providers (oligopolists). A nested logit model is employed to characterize the charging preference of the EV owners. The optimal placement policy for each service provider is obtained by solving a Bayesian game. We also developed a simulation toolkit called ``The EV Virtual City" based on Repast. We observe that service providers prefer clustering instead of separation in the EV charging market.

As for the problem of pricing and energy management of EV charging stations, we provide guidelines for charging service providers to determine charging price and manage electricity reserve to balance the competing objectives of improving profitability, enhancing customer satisfaction, and reducing impact on the power system. In the presence of renewable energy integration and energy storage system, EV charging service providers must deal with a number of uncertainties, e.g.,  charging demand volatility, inherent intermittency of renewable energy generation, and wholesale electricity price fluctuation. We propose a new metric to assess the impact on power system without needing to solve complete power flow equations. Two algorithms --- stochastic dynamic programming (SDP) algorithm and greedy algorithm (benchmark algorithm) --- are applied to derive the pricing and electricity procurement strategy. We find that the charging service provider is able to reshape spatial-temporal charging demands to reduce the impact on power grid via pricing signals.

The last technical contribution of this dissertation is on the design of a novel electricity trading market and distribution network, which provides a platform to support seamless RES integration, grid to vehicle (G2V), vehicle to grid (V2G), vehicle to vehicle (V2V), and distributed generation (DG) and storage. We apply a sharing economy model to the electricity sector to stimulate different entities to exchange and monetize their underutilized electricity. We propose an online advertisement-based peer-to-peer (P2P) electricity trading mechanism. A fitness-score (FS)-based supply-demand matching algorithm is developed by considering consumer surplus, electricity network congestion, and economic dispatch. We compare the FS matching algorithm with the first-come-first-serve (FCFS) algorithm. The simulation results show that the FS matching algorithm outperforms the FCFS algorithm in terms of network congestion management, electricity delivery delay probability, energy efficiency, and consumer surplus.
\end{abstract}

\tableofcontents
\listoffigures
\listoftables



\begin{acknowledge}
I would like to thank my advisor Dr. Yih-Fang Huang, whose expertise, understanding, generous guidance and support made it possible for me to work on a research topic that is of great interest to me. I am grateful for his insightful and constructive suggestions on my research. After five-year intensive PhD study with him, I have gained not only the abilities of critical thinking and rigorous mathematical analysis, but also presentation and communication skills, which will be invaluable treasures in my life.

I would like to thank Dr. Vijay Gupta, Dr. Ken Sauer, Dr. Hai Lin, and Dr. Peter Bauer for being my PhD defense committee members. I really appreciate their time and efforts to review my dissertation and provide meaningful comments on how to improve it.

I would like to thank my parents and my sister for their continued love, help, support and encouragement. I thank them for always being there and going through the happiness and sorrow with me.

I would like to thank all my friends, colleagues, my roommate and many visiting scholars for their genuine friendship and support. I would always remember the crazy cheers and cadences in the football stadium, delicious cuisine, interesting parties, funny Halloweens, etc.

I would also like to thank University of Notre Dame for providing such a beautiful and divine campus for me. Wherever I go, and whatever happens in my life I know I always have a home at Notre Dame to renew my heart and refresh my spirit.

\end{acknowledge}

\begin{symbols}
\sym{L}{total number of candidate locations}
\sym{N}{total number of EVs}
\sym{$$\psi_{j,k}^n$$}{charging demand}
\sym{$$s_{j,k}$$}{placement indicator}
\sym{$$F_{j,k}$$}{charging station setup cost}
\sym{$$R_k$$}{total revenue}
\sym{$$c_{j,k}$$} {locational marginal price}
\sym{$$\Pi_k$$} { total profit}
\sym{$$U_k$$} { overall utility }
\sym{$$w$$} {coefficient of EV charging penalty}
\sym{$$\Upsilon_k$$} {average service probability }
\sym{$$\Xi_k$$} { average service coverage }
\sym{$$t_k$$} {average charging time}
\sym{$$p_k$$} {retail charging price}
\sym{$$i_n$$} {income of the $n$th EV owner }
\sym{$$d_{j,k}^n$$} {deviating distance}
\sym{$$z_{j,k}^n$$} {indicator of destination}
\sym{$$d_{th}$$} {distance threshold}
\sym{$$r_{j,k}$$} {indicator of restaurant}
\sym{$$g_{j,k}$$} {indicator of shopcenter}\
\sym{$$m_{j,k}$$} {indicator of supermarket }
\sym{$$\mathbf{P}_{\textrm{g}}^{\textrm{base}}$$} {active power vector without EV}
\sym{$$\mathbf{P}_{\textrm{g}}^{\textrm{EV}}$$} {active power vector with EV}
\sym{$$\mathbf{Q}_{\textrm{g}}^{\textrm{base}}$$} {reactive power vector without EV}
\sym{$$\mathbf{Q}_{\textrm{g}}^{\textrm{EV}}$$ }{reactive power vector with EV}
\sym{$$v_i$$} {voltage at bus $i$ }
\sym{$$\phi_{ik}$$} {voltage angle between bus $i$, $k$}
\sym{$$\alpha$$} {coefficient of $t_k$}
\sym{$$\beta$$} {coefficient of $p_k/i_n$}
\sym{$$\mu_k$$} {coefficient of $d_{j,k}$}
\sym{$$\eta_k$$} {coefficient of $z_{j,k}$}
\sym{$$\gamma_k$$} {coefficient of $r_{j,k}$}
\sym{$$\lambda_k$$} {coefficient of $g_{j,k}$}
\sym{$$\delta_k$$} {coefficient of $m_{j,k}$}
\sym{$$K$$}{Number of planning horizons}
\sym{$$s_j$$}{The $j$-th EV charging station}
\sym{$$p_{kj}$$}{Charging price of the $j$-th charging station at the $k$-th horizon}
\sym{$$c_k$$}{Real-time wholesale price at the $k$-th horizon}
\sym{$$E$$}{Electricity storage capacity}
\sym{$$I_k$$}{Remaining electricity at the beginning of $k$-th horizon}
\sym{$$W_k$$}{Profit of the $k$-th horizon}
\sym{$$W_{\textrm{min}}$$}{Threshold of minimum profit}
\sym{$$G_k$$}{Customer satisfaction at the $k$-th horizon}
\sym{$$F_k$$}{Impact at the $k$-th horizon}
\sym{$$o_k$$}{Electricity purchase at the $k$-th horizon}
\sym{$$u_k$$}{Renewable energy at the $k$-th horizon}
\sym{$$\eta_s$$}{Unit storage cost (in\$/MWh)}
\sym{$$\eta_c$$}{Charging efficiency}
\sym{$$\eta_d$$}{Discharging efficiency}
\sym{$$\phi_k$$}{Total charging demand at $k$-th horizon}
\sym{$$\gamma_{i,j}$$}{Price elasticity coefficient}
\sym{$$P_i$$}{Active power of $i$-th bus}
\sym{$$Q_i$$}{Reactive power of $i$-th bus}
\sym{$$G_{ik}$$}{Conductance of $ik$-th element in admittance matrix}
\sym{$$B_{ik}$$}{Susceptance of $ik$-th element in admittance matrix}
\sym{$$S_i^{\textrm{Ac}}$$}{Active power sensitivity of $i$-th bus}
\sym{$$S_i^{\textrm{Re}}$$}{Reactive power sensitivity of $i$-th bus}
\sym{$$Q$$}{Requested electricity quantity}
\sym{$$\theta$$}{Power factor}
\sym{$$\theta^{\textrm{min}}$$}{Minimum power factor}
\sym{$$\theta^{\textrm{max}}$$}{Maximum power factor}
\sym{$$T^{\textrm{min}}$$}{Minimum delivery starting time}
\sym{$$T^{\textrm{max}}$$}{Maximum delivery starting time}
\sym{$$p^{\textrm{bid}}$$}{Bid price}
\sym{$$p^{\textrm{offer}}$$}{Offer price}
\sym{$$R$$}{Delivery rate (MW)}
\sym{$$R^{\textrm{min}}$$}{Minimum delivery rate}
\sym{$$R^{\textrm{max}}$$}{Maximum delivery rate}
\sym{$$S(i,j)$$}{Fitness score}
\sym{$$D_{i,j}$$}{Predicted electricity delivery cost}
\sym{$$Z(i,j)$$}{Same zone indicator}
\sym{$$E(j)$$}{Renewable energy indicator}
\sym{$$\mathcal{K}^h$$}{Active supply-demand pair set}
\sym{$$\mathcal{J}^h_k$$}{Route set}
\sym{$$\beta_{k,j,l}^h$$}{Power flow on branch (in MW)}
\sym{$$Q^h_k$$}{Scheduled power delivery rate (in MW)}
\sym{$$r_l$$}{Resistance of distribution line}
\sym{$$F^{\textrm{base}}_l$$}{Base power flow on branch (in MW)}
\sym{$$F^{\textrm{now}}_l$$}{Aggregated power flow on branch (in MW)}
\sym{$$M(i,j)$$}{Payment of $i$-th DemandAd to $j$-th SupplyAd}
\sym{$$C_l$$}{Branch capacity}
\end{symbols}

\mainmatter

%
%

%
%

%
%
%
%
%
%
%
%
%
%

%
%

\chapter{INTRODUCTION}

\section{Overview}
This chapter provides a brief introduction to the EV ecosystem, which includes EV and EV charging infrastructure, EV market and manufacturers, the relationship between EV charging and power systems, as well as challenges that the EV industry may face.

\begin{figure}[!htbp]
  \centering
  \includegraphics[width=6in]{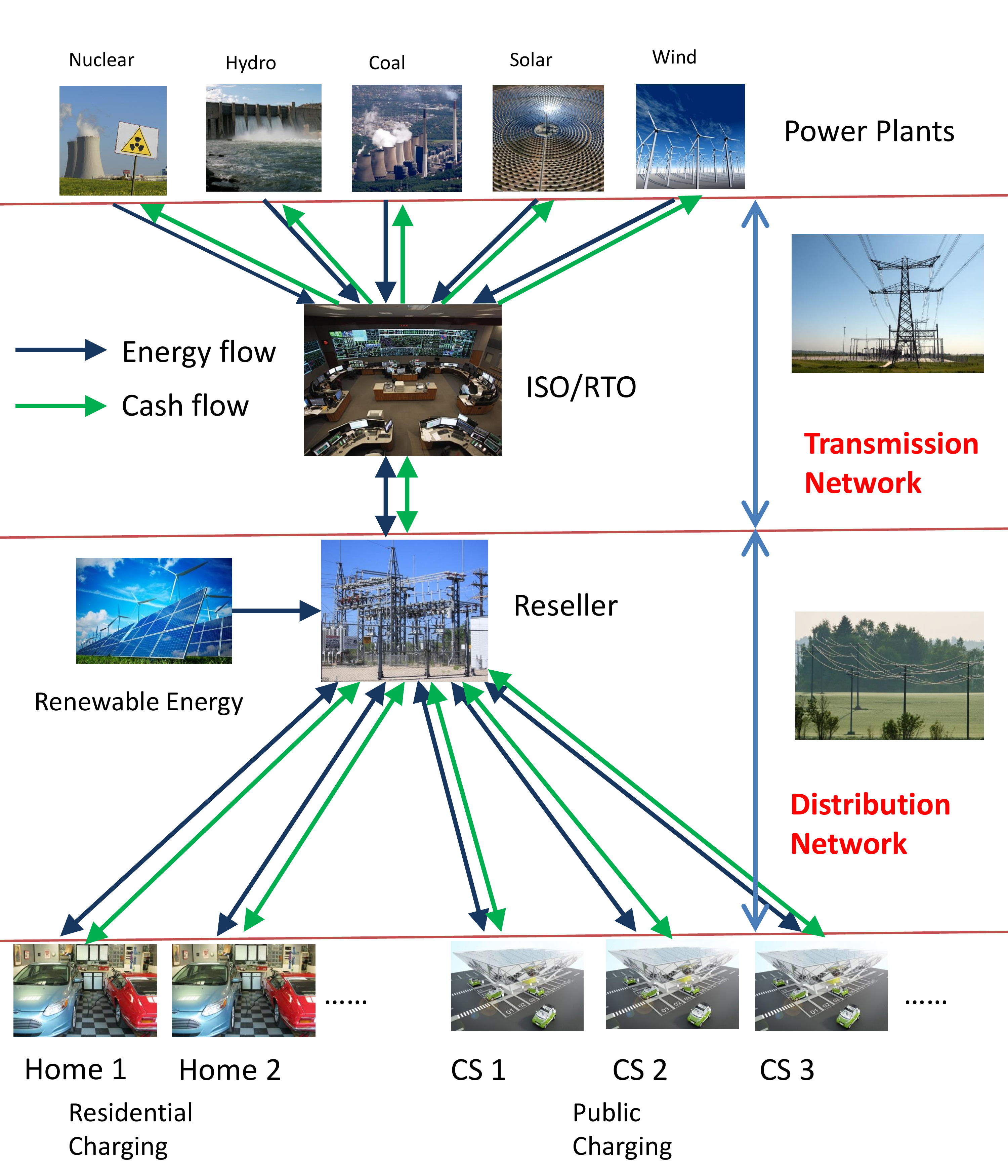}
  \caption{Energy Flow and Cash Flow in EV Ecosystem}\label{fig1_1}
\end{figure}

\section{EV Ecosystem}
 A complete EV ecosystem is illustrated in Fig. \ref{fig1_1}. Main entities like power plants, independent system operator (ISO) or regional transmission organization (RTO), EV charging infrastructure are included in this figure. The energy flow (blue line) and the cash flow (green line) show the interaction among different entities.

\subsection{Power Plant}
A power plant generates electricity by burning fossil fuels like coal or natural gas, or by using clean and renewable sources like nuclear energy, solar, or wind. A power plant can either sell the electricity to specific consumers through bilateral contracts (or futures), or sell the electricity in a wholesale market through auctions.

\subsection{ISO/RTO}
In the United States, ISO/RTO is an organization that coordinates, controls, and monitors the power grid. For instance, California ISO (CAISO), Southwest Power Pool (SPP), Midcontinent ISO (MISO) and PJM Interconnection are ISO/RTOs.

ISO/RTO determines the day-ahead clearing price and the ancillary service price through an auction between power plants and buyers. ISO/RTO also operates the electricity real-time market, where the electricity price is cleared in real-time and the power plants must deliver the committed electricity immediately.

\subsection{Reseller}
The resellers are buyers in the auction in the wholesale market. The load serving entity (LSE) and electricity marketers are typical resellers in the market.

\subsection{EV Charging Infrastructure}
EV owners can either install an EV charging station at home or go to a public charging station. So, EV charging infrastructure can be done at either private or public infrastructures.

\subsection{Electricity Transmission Network}
Electricity transmission is the long distance delivery of bulk electricity from power plants to substations, in which electricity is transferred using the transmission lines with high-voltage (i.e. 69 kV, 115 kV and 138 kV). The transmission lines are interconnected to form the transmission network.

\subsection{Electricity Distribution Network}
The electricity distribution network delivers electricity from the local substations to end users using a relatively low voltage (i.e. 120/240 V in North America, and 220, 230, or 240 V in Europe).

\section{Emerging EV Market}
As consumers are becoming more aware of the environmental impact of the greenhouse gas emissions as well as promotions and incentives from local governments, the demand for EVs has been growing over the past several years. From 2008 to June 2015, about 345,000 highway plug-in electric vehicles were sold in the United States, making it the largest plug-in EV market in the world \cite{cobb, cobb2}. The market share of plug-in EV of new registered cars in the US increased from 0.14\% to 0.37\% in 2012 and 0.62\% in 2013 and 0.72\% in 2014 \cite{evassociation}. According to Navigant Research \cite{navigant}, the global light duty EV market is expected to grow from 2.7 million vehicle sales in 2014 to 6.4 million in 2023.

\begin{figure}[htp]
  \centering
  \includegraphics[width=4in]{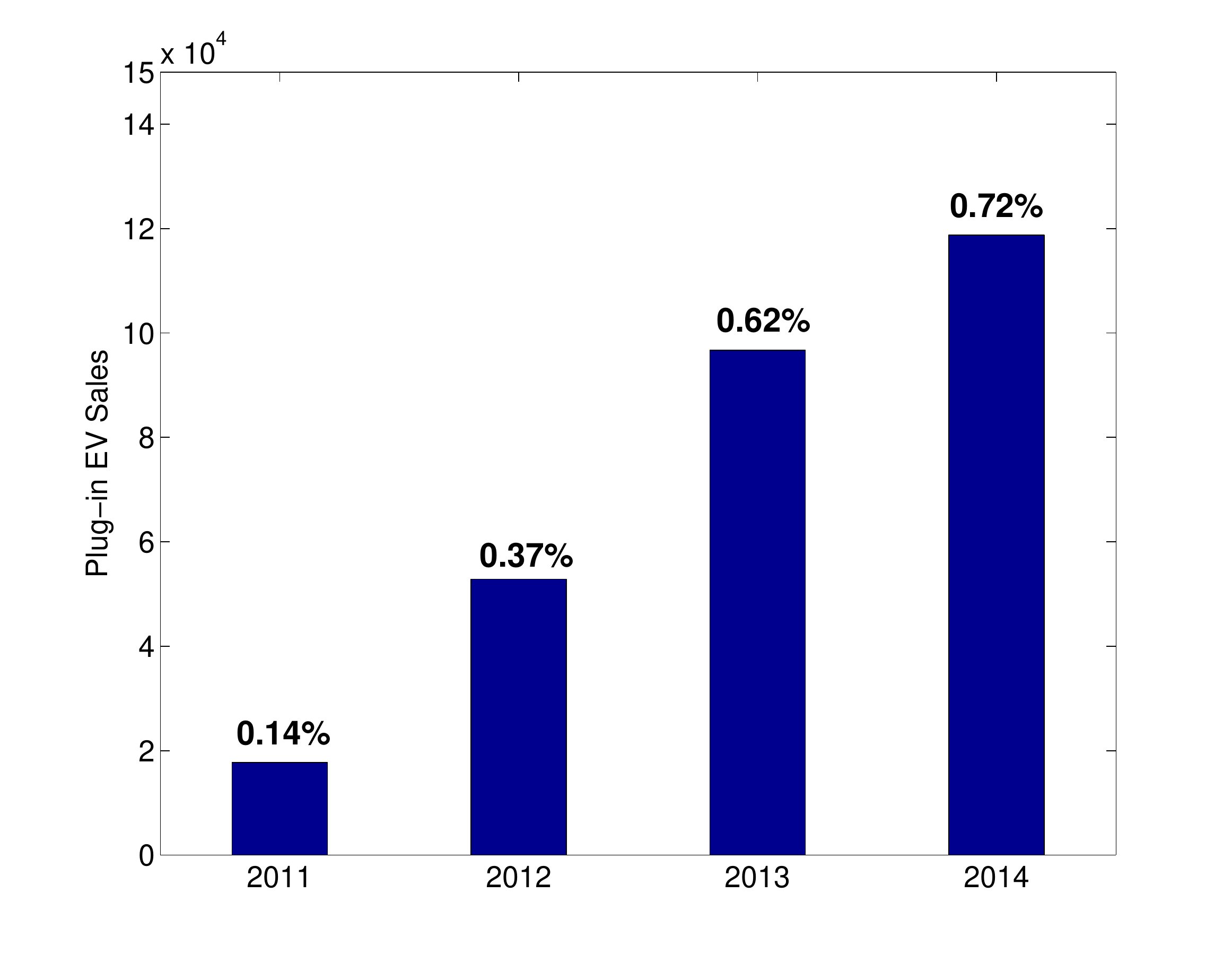}
  \caption{PHEV Sales}\label{fig1_2}
\end{figure}

In general, EVs can be classified into three main categories: Hybrid Electric Vehicle (HEV), Plug-in Electric Hybrid Vehicle (PHEV), and Battery Electric Vehicle (BEV) or All-Electric Vehicle (AEV).

\subsection{Hybrid Electric Vehicle (HEV)}
HEVs are primarily powered by the internal combustion engine (ICE) which uses gasoline. Each HEV also has an electric motor that uses the electricity stored in the battery. The battery gets recharged from the regenerative braking and the internal combustion engine. However, the battery cannot be plugged into the power system to get recharged. The electric powertrain helps HEVs achieve better fuel economy and lower emission than the conventional ICE vehicles. Many automakers have released their HEV-version cars like Honda Civic Hybrid, Ford Fusion Hybrid, BMW ActiveHybrid 3, etc.

\subsection{Plug-in Hybrid Electric Vehicle (PHEV)}
PHEVs are powered by both the internal combustion engine (ICE) and the electric motor. The ICE eliminates the ``range anxiety" since PHEVs can switch to the ICE when the battery is depleted. The battery can be plugged into the power grid to get recharged. Usually, PHEVs have a larger battery pack than the HEVs, which increases the ``all-electric range". Some representatives of PHEVs include Ford Fusion Energi, Chevrolet Volt, Cadillac ELR, etc.

\subsection{Battery Electric Vehicle/All Electric Vehicle (BEV/AEV)}
BEVs or AEVs are solely powered by an electric motor which uses the electricity from a battery. The battery gets recharged by plugging the vehicle into the power grid. BEVs do not rely on fossil fuels and they generate zero emissions. The representatives of BEVs are Fiat 500e, Nissan Leaf, Tesla Model S, etc.

Fig. \ref{fig1_2} shows the trend of EV market share based on the EV sales data from Electric Drive Transportation Association (EDTA). Note that the plug-in EVs (including PHEV and BEV) has a steady increase in market share from 2011 to 2014.

In this dissertation, we focus our attention on BEVs since there are more challenges and problems associated with this type of EV, as compared to the hybrids. In the subsequent discussions, we use EVs to refer BEVs.

\begin{figure}[htp]
  \centering
  \includegraphics[width=4in]{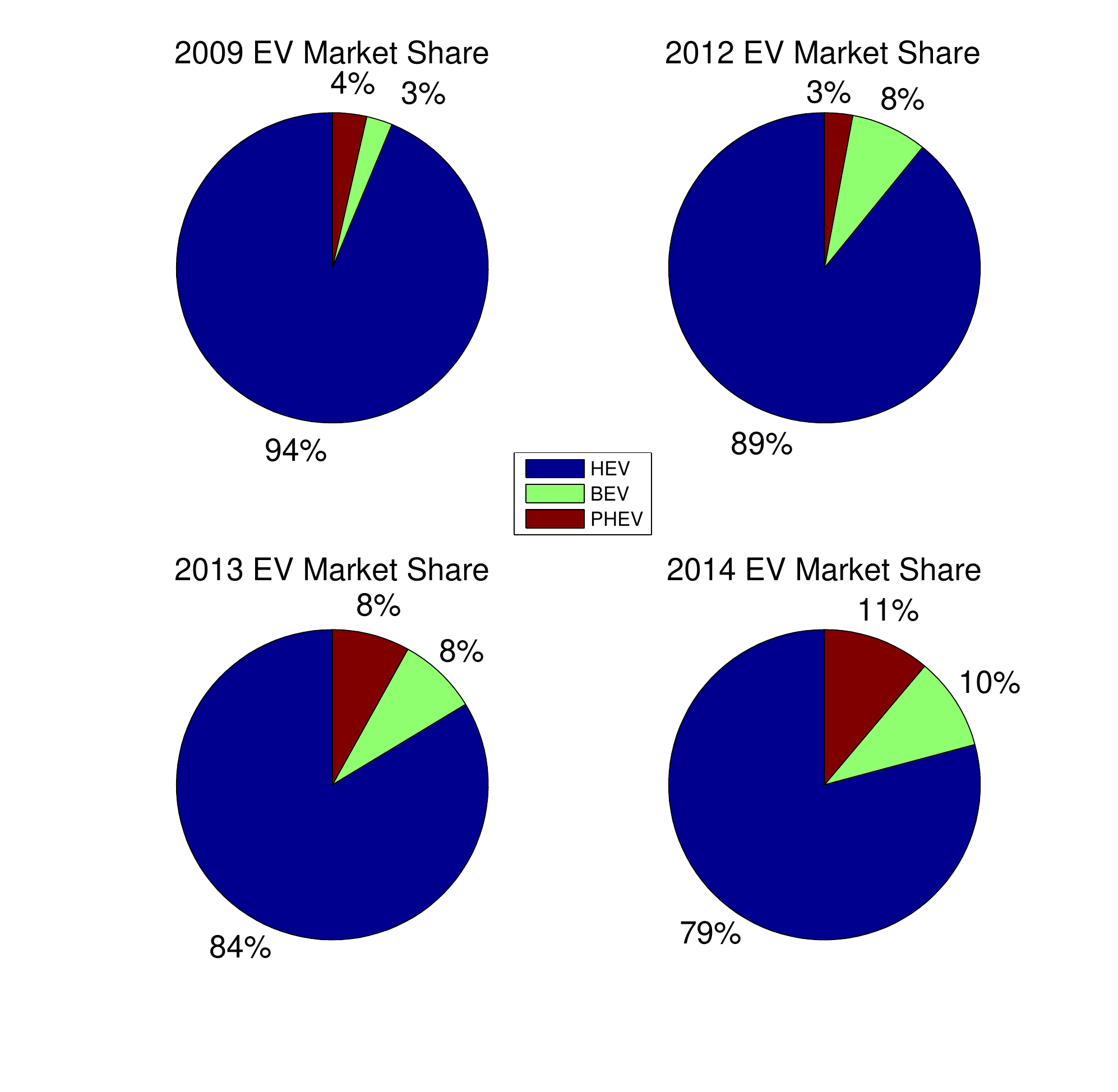}
  \caption{EV Market Share}\label{fig1_3}
\end{figure}

\section{Benefits of EVs}
In this section, we highlight some of the most prominent benefits of EV.

\subsection{Economic Growth}

Both individual EV owners and automobile manufacturers will benefit from a vibrant EV industry and the resulting growth in the entire supply chain. Many studies have shown that electricity is a cheaper fuel than gasoline to propel the vehicles \cite{EPRI, hadley, letendre2, parks, samaras, lilienthal}. In particular, \cite{EPRI2} showed that the equivalent cost of electricity as a gallon of gasoline is less than one dollar. In addition, Navigant Research predicted that the global demand for lithium ion batteries for the light duty fleet will increase from \$3.2 billion in 2013 to \$24.1 billion in 2023. The global revenue from the electric vehicle supply equipment (EVSE) is estimated to reach \$5.8 billion in annual revenue by 2022.

\subsection{Environmental Sustainability}

\begin{figure}[htp]
  \centering
  \includegraphics[width=4.0in]{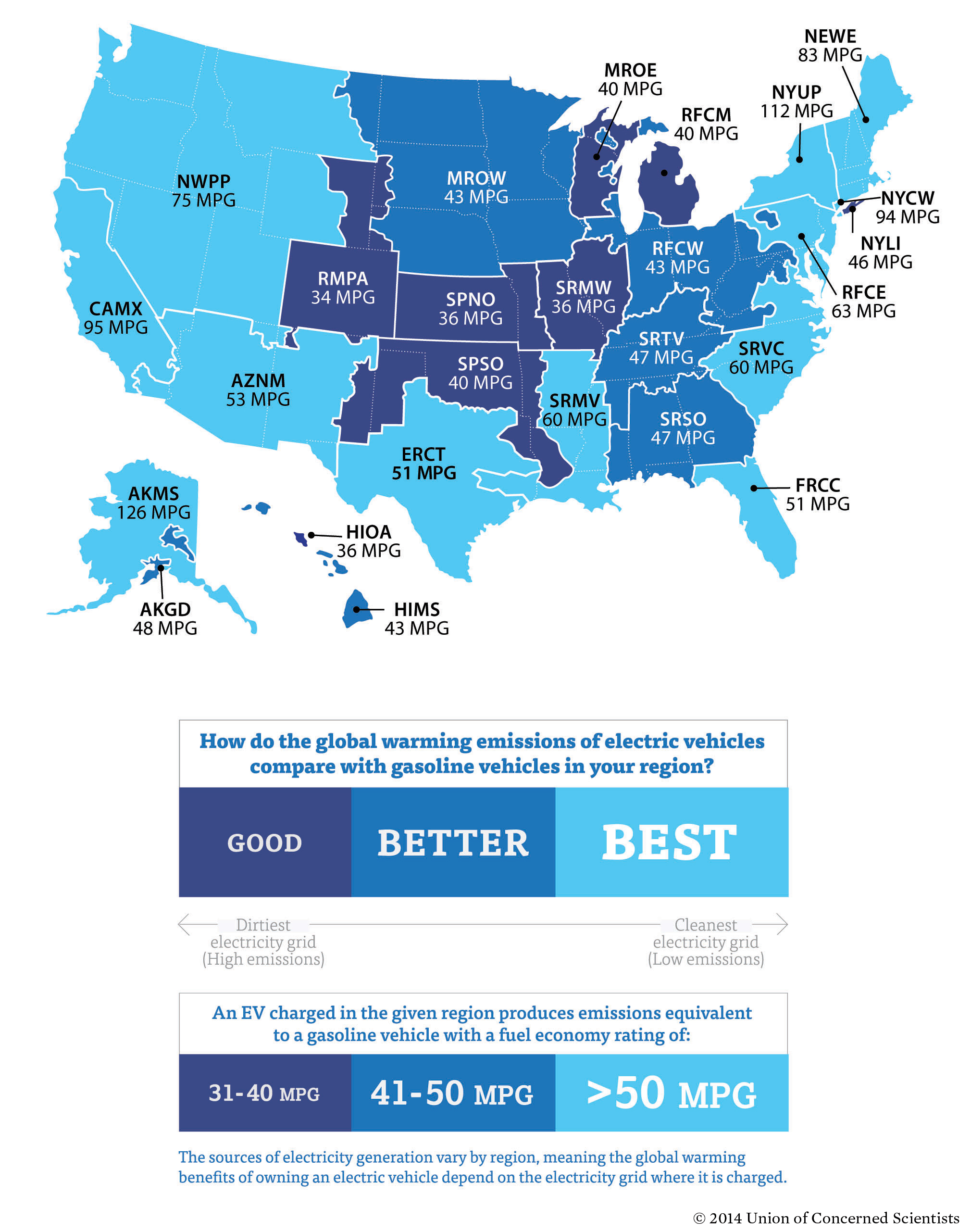}
  \caption{Comparison EV with ICE in Global Warming Emission}\label{EVvsICEemission}
\end{figure}

Massive adoption of EVs offers an opportunity to improve the air quality through increasing fuel efficiency, reducing or terminating greenhouse gas emissions \cite{EPRI, kintner, letendre, stephan, van, verzijlbergh}. Fig. \ref{EVvsICEemission} illustrates that driving on electricity offers a huge environmental benefit \cite{rnealer}. The electricity in the US is generated from a diverse portfolio of energy sources from natural gas to nuclear power, from hydroelectric to coal, from biomass to wind and solar. More importantly, the portion of renewable energy generation (e.g. wind, solar, and hydrodic) has increased over these years, which is a more desirable source to recharge EVs. A lot of efforts have been made to enhance the efficiency of electricity generation and distribution using smart grid technologies to make EV charging even more convenient.

\subsection{Energy Security}
 Relying on a single globally traded fuel, i.e. petroleum, is a potential threat to energy security. According to the US Energy Information Administration (EIA), soaring oil price has a negative impact on the US macroeconomic variables. The growing gasoline cost and the painful spikes in prices often have adverse impacts on family budget. EVs provide a viable solution to propel our transportation sector and alleviate dependence on oil. The diverse sources for electricity generation make electricity price remain relatively stable, which offers some protection for the EV owners against the volatile gasoline price.

\section{State of the Art of EV Technology}
The continuing evolution of EV technologies empowers EVs to compete with the ICE vehicles in the competitive automobile market. In particular, recent advances in powertrain, battery, propulsion motor, power converter, charging, and hybrid control technology make EVs a viable substitute for the conventional ICE vehicles. In this section, a brief summary to the state-of-the-art EV technologies will be given.

\subsection{Powertrain}
Generally, the EVs can be classified into three categories: HEV, PHEV, and BEV (all-electric vehicle). Both PHEVs and BEVs can be recharged by connecting the battery to the power system, so they are also called plug-in electric vehicles (PEVs). HEVs and PHEVs can be further divided into series HEVs and PHEVs, parallel HEVs and PHEVs, and series-parallel HEVs and PHEVs according to different configurations of the ICE and the electric motor (EM) \cite{chan, yong, momoh}. Since a BEV does not have ICE so the EM provides the entire power for the vehicle.

\underline{Series HEV and PHEV}: In series HEV and PHEV, the ICE mechanical output is first converted to electricity through a generator. After a power convertor, the generated electricity can be utilized to recharge the battery or to power the EM to propel the wheels. Another function of EM is to harvest the regenerative energy during braking. The decoupling between the engine and the driving wheel offers the advantage for HEV and PHEV design to place the generator set. However, series HEV and PHEV powertrain need three propulsion components: the ICE, the generator, and the EM, resulting in a lower energy efficiency.

 \begin{figure}[htp]
  \centering
  \includegraphics[width=4.0in]{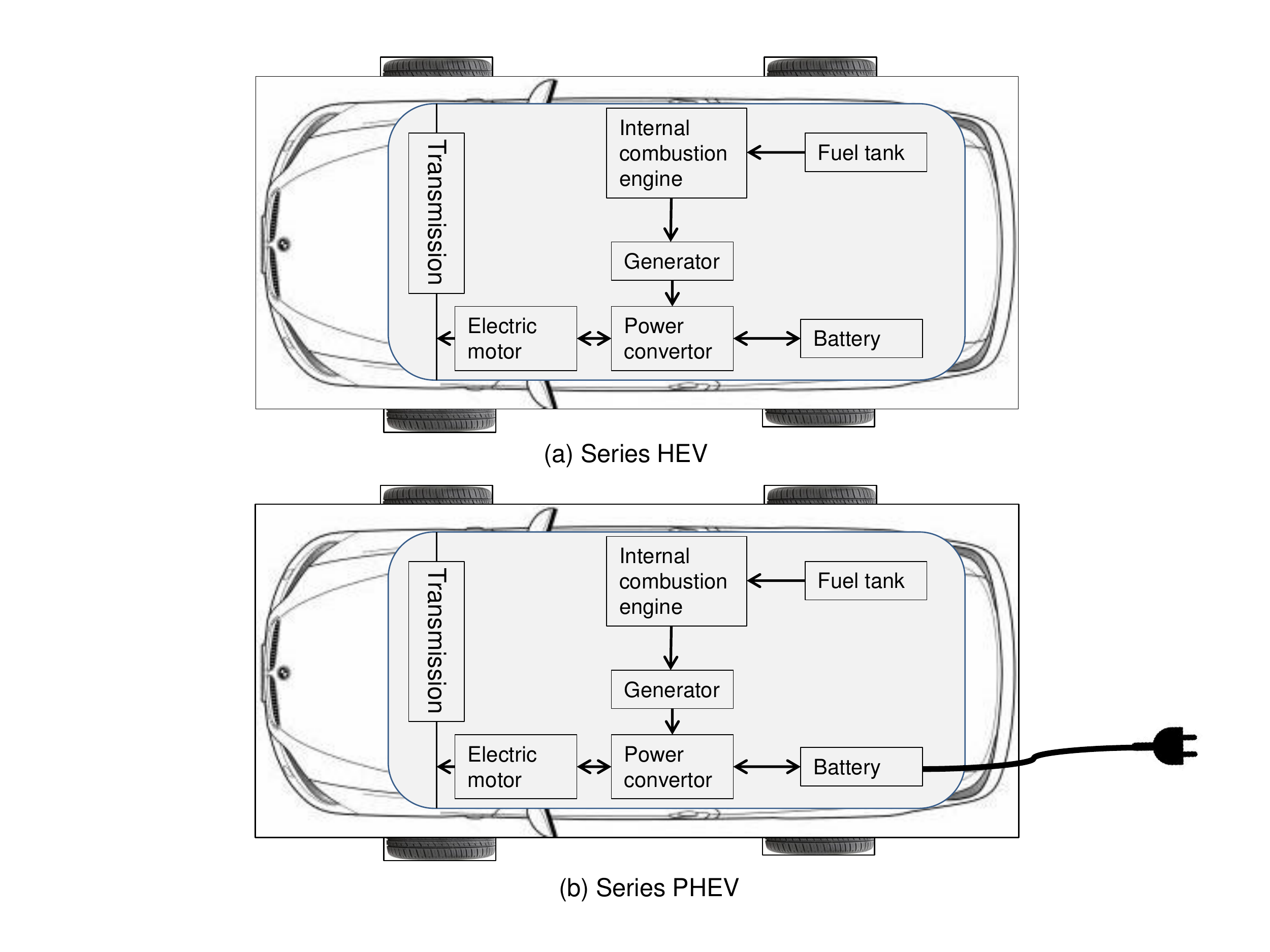}
  \caption{Powertrain: Series HEV and PHEV}\label{fig1_4}
\end{figure}

\underline{Parallel HEV and PHEV}: In contrast to series HEV and PHEV, parallel HEV and PHEV allow both ICE and EM to propel wheels in parallel. The ICE and EM are both coupled to the drive shaft of the wheels via two clutches. Similar to the series type, the EM can be used to harvest electricity during braking to recharge the battery or absorb the extra power produced by ICE. Compared with the series hybrid, the parallel HEV and PHEV need only two propulsion components---ICE and EM.  Since ICE and EM can propel the wheel simultaneously, the parallel type can have a smaller ICE and EM than the series type to achieve the same performance.

 \begin{figure}[htp]
  \centering
  \includegraphics[width=4.0in]{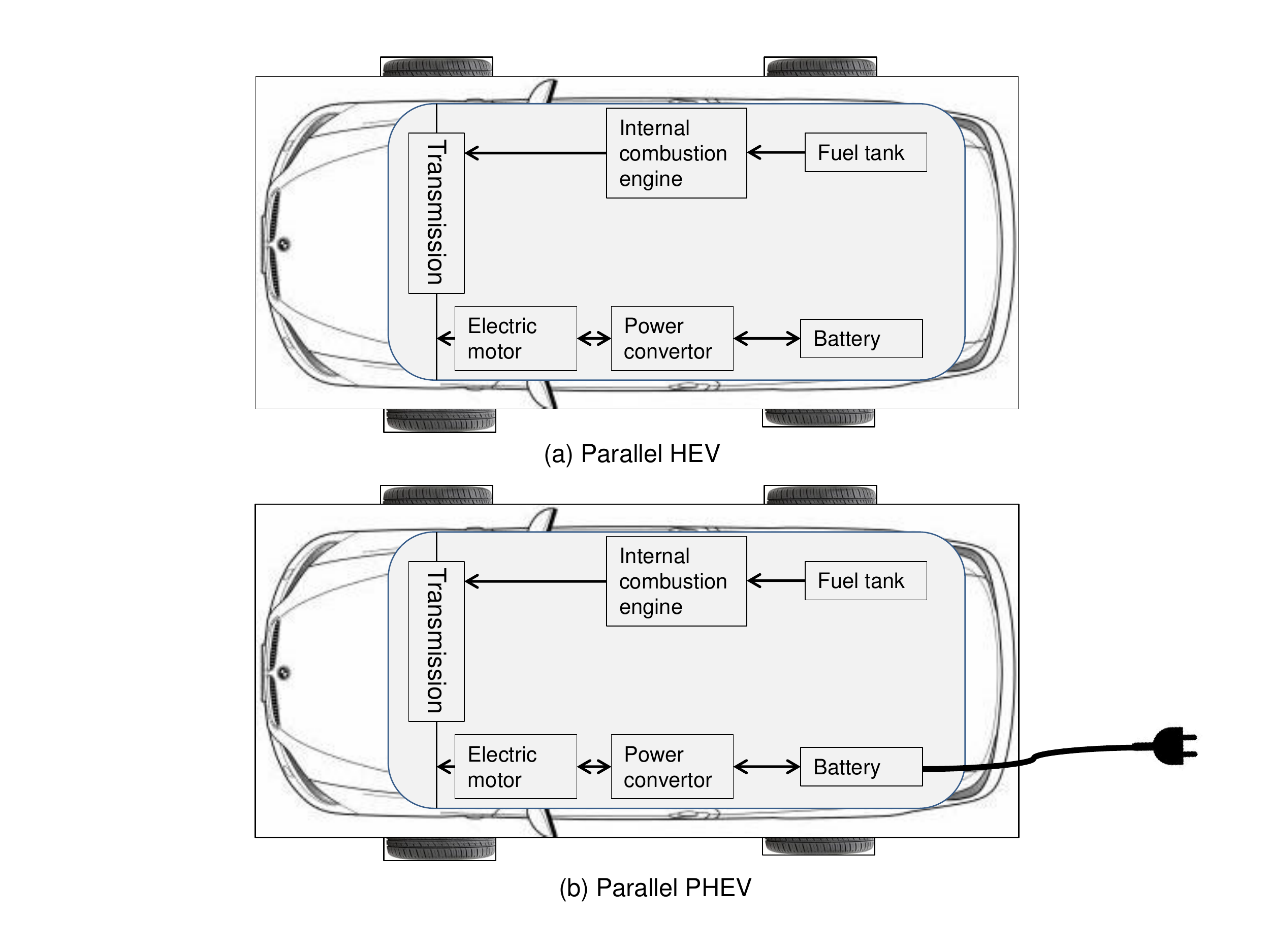}
  \caption{Powertrain:  Parallel HEV and PHEV}\label{fig1_5}
\end{figure}

\underline{Series-Parallel HEV and PHEV}: Series-parallel HEV and PHEV combine the advantages of both series hybrid type and parallel hybrid type. However, the series-parallel type is more complicated and expensive than either the series type or the parallel type. As the control and manufacturing technologies mature, many automobile manufacturers prefer to adopt this type of design.

 \begin{figure}[htp]
  \centering
  \includegraphics[width=4.0in]{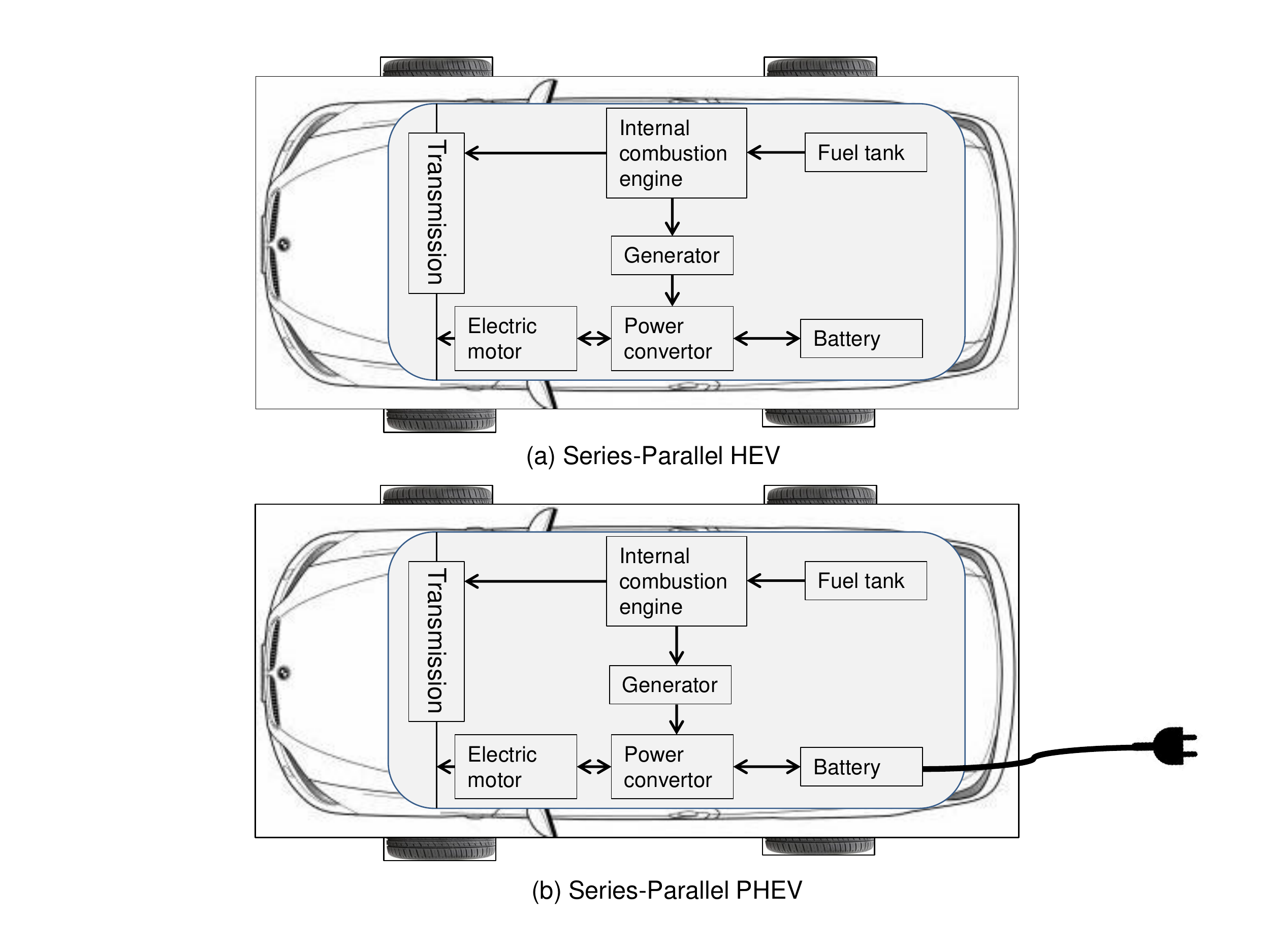}
  \caption{Powertrain:  Series-Parallel HEV and PHEV}\label{fig1_6}
\end{figure}

BEV: The BEVs do not have the ICE and battery is the sole energy source to propel the vehicle. The propulsion system is relatively simple compared to HEV and PHEV because there is no coupling between the ICE and EM. However, BEVs have the ``range anxiety" problem, since they are without the ICE as the backup energy source.

 \begin{figure}[htp]
  \centering
  \includegraphics[width=4.0in]{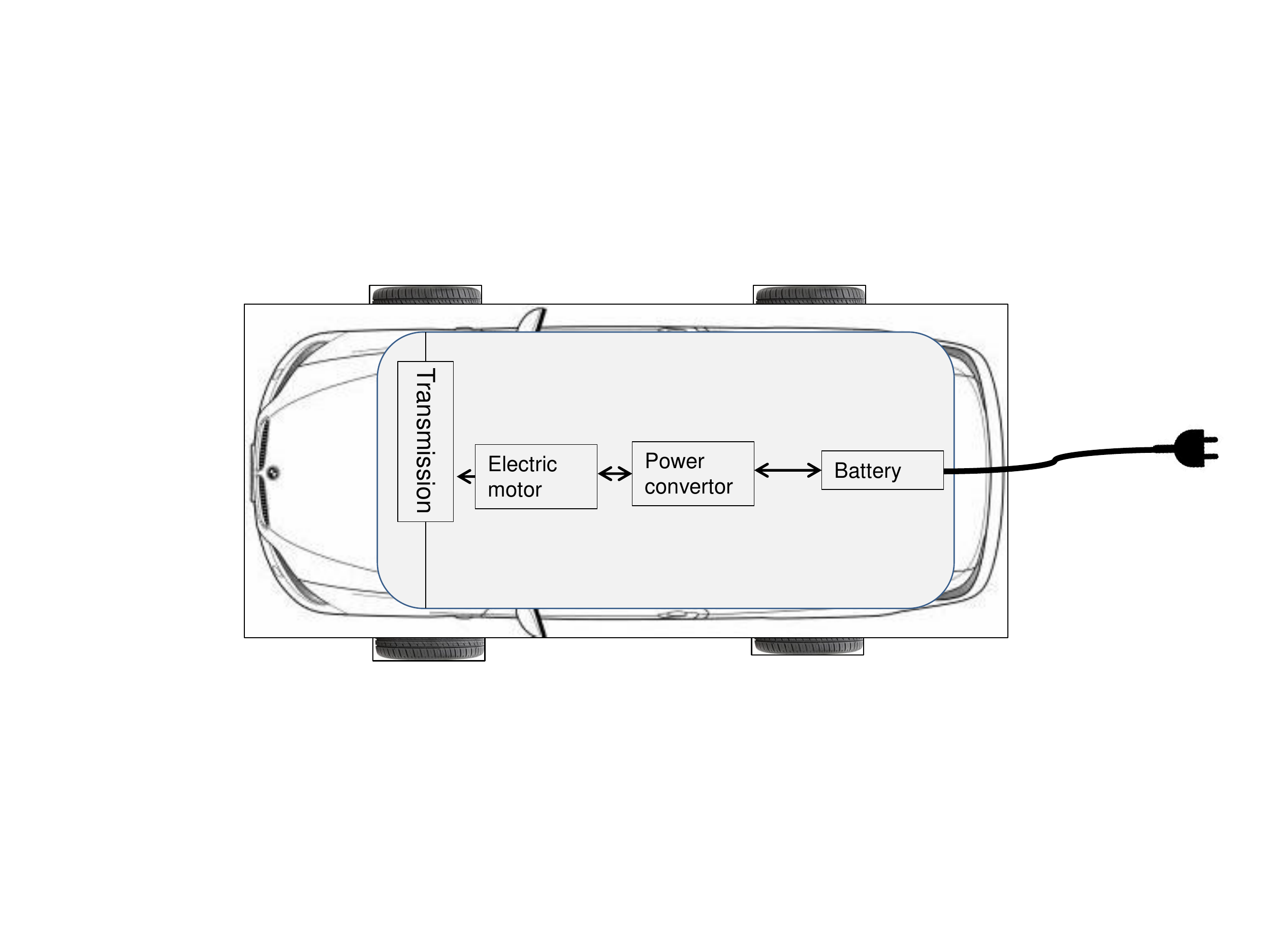}
  \caption{Powertrain:  BEV}\label{fig1_7}
\end{figure}

\subsection{Battery}
Battery is the only energy source for BEVs and one of two major energy sources for HEV and PHEV. It is always a challenging task for designers to find an appropriate battery for EVs with the desirable characteristics, such as high capacity, low cost, light weight, fast charging, safe operation, durability, long lifetime, and resistance to severe weather conditions. The limited battery capacity is one of the major concerns for potential EV consumers. The long charging time is another hinderance that discourages consumers to purchase EVs. Furthermore, the high cost of battery makes EVs more expensive than the ICE vehicles, which again places EVs in an unfavorable situation in the automobile market.

However, the EV battery research has made remarkable progresses in the past decades \cite{yong, divya, anderson}. Currently, various novel technologies are employed to refine the battery manufacturing process and many superior battery prototypes are under development in the labs \cite{battery}. In this subsection, we will briefly discuss the development of EV batteries and present several battery types that were used in the past or are used currently.

\underline{Lead-Acid Battery}: The first battery used in transportation is the lead-acid battery. The lead-acid battery technology is mature and inexpensive but it has many disadvantages, such as low capacity, heavy, and environmental pollution.

\underline{Nickel-Based Battery}: The lead-acid battery was soon replaced by the nickel-based battery, i.e., nickel-cadmium (NiCd) and nickel-metal-hydride (NiMH). The nickel-based battery has several advantages over the lead-acid battery, such as high cycle count, good load performance, simple storage and transportation, and good low-temperature performance. However, the nickel-based battery has some serious disadvantages like relatively low energy density, memory effect, toxic metal, and high self-discharge rate.

\underline{ZEBRA Battery}: ZEBRA battery utilizes the molten sodium aluminumchloride ($\textrm{NaAlCl}_4$ ) as the electrolyte. The negative electrode is the molten sodium and the positive electrode is the nickel in the discharged state and nickel chloride in the charged state. The operation temperature for ZEBRA battery is 245 $\textrm{C}^{\circ}$ (473 $\textrm{F}^{\circ}$). ZEBRA battery has the advantages of high energy density (5 times higher than lead acid battery), large cell, long life cycle, and low material cost. However, this battery type suffers from high internal resistance and high operating temperature, which requires delicate internal thermal management.

\underline{Lithium-Based Battery}: The advent of lithium-based battery gives new life to the EV industry. There are several battery types belonging to the lithium-based battery family---lithium-ion (Li-ion), lithium-ion polymer (LiPo), and lithium-iron phosphate ($\textrm{LiFePO}_4$). The lithium-based battery has the advantages of high energy density, low self-discharge rate, low maintenance, light, cheap, environmentally friendly, and fast charging. The limitations of this battery is that the internal resistance is relatively high. Thus high thermal runaway may lead to fire. It also requires sophisticated battery management circuit. The current EVs that are equipped with lithium-based battery include Nissan Leaf, Mitsubishi i-MiEV, Tesla Model S and Chevrolet Volt.

\underline{Emerging Battery Technologies}: Many other battery technologies are still in the experimental stage, e.g., metal phase, which gives superior performance. Those batteries are lithium-sulfur (Li-S), zinc-air (Zn-air) and lithium-air (Li-air). Those batteries offer excellent performance and will extend the all electric range and reduce the manufacturing cost of EVs.

\section{Challenges of Large-Scale EV Integration}
EVs exhibit many advantages over ICE vehicles, like higher fuel efficiency, lower operating cost, and zero greenhouse gas emissions. Nevertheless, many challenges remain as major obstacles for EV proliferation. In this section, we briefly discuss some of those challenges.

\subsection{Large-Scale EV Charging Scheduling}
Uncontrolled large-scale EV charging will undoubtedly place a heavy burden on the existing power system legacy. Researchers have done much to investigate the problem of optimal scheduling of EV charging by using historical data \cite{parks, denholm}, static time of usage based costs \cite{shao, hacker, karnama}, power load prediction \cite{rotering}, particle swarm optimization \cite{huston}, and non-cooperative games \cite{ma}.

\subsection{Limited Battery Capacity and Long Charging Time}
The onboard battery capacity is the bottleneck to increase the range for all-electric EVs. For instance, Nissan LEAF has a 24 kWh battery, which only provides a driving range of 84 miles. Chevrolet Spark EV has a 19 kWh battery with a driving range of 82 miles. Tesla Model S has a larger battery pack with a capacity of 70kWh, offering a 230 miles range. Moreover, it often takes up to several hours to fully charge the battery. Obviously, the long charging time is a severe drawback for EV compared to the ICE vehicles, which can be fully refuelled in only a few minutes.

\subsection{Insufficient EV Public Charging Infrastructure}
The scarcity of public charging stations is another major concern to potential EV consumers. Currently, most EV charging is done at the residence. The EV owners find it inconvenient to go to the commercial charging stations either because there are not enough charging stations available or those charging stations are not deployed in locations that can be accessed easily. In this dissertation, we propose an EV charging station placement strategy, which provides guidelines on determining how many charging stations we need and when and where to deploy them.

\subsection{Charging Station Operation and Management}
The effective management of EV charging infrastructure is crucial to EV ecosystem. Currently, there is no proper model available for charging station operation in terms of charging price setting, and energy management strategy. A proper model should take into account the profitability of charging stations, consumer satisfaction, and power system stability.

\subsection{Liberal Electricity Trading Market and Distribution Network}
The current wholesale electricity market is not conductive for the interplay between EVs and RESs. In addition, the current electricity distribution network cannot naturally support intermittent renewable energy generation. We need to redesign the electricity trading market and distribution network to facilitate large-scale EV and RES integration.

In essence, this dissertation addresses some of those challenges that include public charging infrastructure, charging station operation and management, and electricity trading market and distribution network.

\section{Dissertation Organization}
This dissertation is organized as follows: Chapter 2 discusses the interaction between EV charging and the power system. An optimal EV charging station placement strategy is studied in Chapter 3. In Chapter 4, a dynamic pricing and electricity management model is proposed for EV charging service providers. In Chapter 5, we present the design of a novel electricity trading market and distribution network for EVs and RESs. Conclusion and future work are presented in Chapter 6.

%
%

%
%
%
%
%
%
%
%
%
%

%
%

\chapter{INTERPLAY BETWEEN EV CHARGING AND THE POWER SYSTEM}

\section{Overview}
This chapter presents an overview on the EV charging system including existing charging standards, electric vehicle supply equipment (EVSE), and onboard charger. In addition, the mutual interaction between EV charging and the power system will be discussed and some relevant technologies like vehicle to grid (V2G) and grid-to-vehicle (G2V) will also be considered.

\section{EV Charging System}
An EV charging system includes a suite of software and hardware. Many countries have released their own EV charging standards. Accordingly, EV charging equipment manufacturers produce different charging connectors.

\subsection{EV Charging Standard}
Currently, there are several popular standards for EV charging. In North America, the Society of Automotive Engineers (SAE) has released the standard of electrical connectors for electric vehicles --- SAE J1772, with a formal title of ``SAE Surface Vehicle Recommended Practice J1772 (SAE J1772), SAE Electric Vehicle Conductive Charge Coupler". In Japan, CHAdeMO Association, formed by the Tokyo Electric Power Company, Nissan, Mitsubishi, Fuji Heavy Industries, and Toyota, has released the CHAdeMO standards for DC fast charging \cite{japanstandard}. In Europe, German Association of the Automotive Industry (VDA) has released the standard of VDE-AR-E 2623-2-2 \cite{europestandard}.

\underline{SAE J1772 Standard}: SAE J1772 \cite{toepfer} specifies three charging levels with different voltages and currents.

\begin{table}[htbp]
\center
\caption{CHARGE METHOD ELECTRICAL RATINGS (NORTH AMERICA)}\label{ev_charging_level}
 \begin{tabular}{cccc}
  \toprule
  Level& Voltage(V)&Current (A) & Charging Time (hour)\\
  \midrule
 Level 1 &120&12& 16-18\\
 Level 2 & 208 to 240&32& 3-8\\
 Level 3 & 600 max&400 max& $<$0.5 \\
  \bottomrule
 \end{tabular}
\end{table}

Level 1 Charging: Level 1 Charging is operated at 120 volts with single phase alternating current (AC). It may take 16-18 hours to charge an EV. Level 1 Charging usually takes place at home and all EVs come with the Level 1 Charging cord.

Level 2 Charging: Level 2 Charging is operated at the voltage ranging from 208 volts to 240 volts with single phase AC. It may take 3-8 hours to fully charge the battery pack. Level 2 Charging is apt for overnight or long-length charging and is the preferred charging method for both public and private facilities. They are typically found in shopping malls, parking lots, and commercial buildings. They can also be charged at residential homes.

Level 3 Charging : Level 3 Charging uses direct current (DC) with a maximum voltage of 600 volts. It takes less than 30 minutes to charge most of battery packs to 80\% full. The time to full charge is not too much more than what it takes to fill the gas tank at a gas station. However, the equipment can be quite expensive.

\underline{CHAdeMO Standard}: To regulate the DC fast charging, CHAdeMO Association introduced the CHAdeMO standard. According to CHAdeMO standard, DC fast charging should recharge the battery pack to at least 80\% in half an hour using the optimal DC charger of 50 kW. DC fast charging is performed through an external dedicated EV charging equipment, usually located in public areas (like parking lot, shopping mall, etc.) CHAdeMO standard has been the Japanese national standard since 2012.

\underline{VDE-AR-E 2623-2-2 Standard}: Originally proposed by Mennekes Elektrotechnik GmbH \& Co. KG in 2009 and later standardized by VDA in 2011, the VDE-AR-E 2623-2-2 EV charging connector has been widely used in Europe. VDA recommends that accessories and interfaces between the power supply and the EVs that permit `fuelling' at 20, 32, 63 amps (single and three-phase current) and at 70 amps (single-phase current only) with a maximum operating voltage of 500 volts at 50-60 Hz. This range will cover the entire range of power supply range worldwide.

\underline{Tesla Charging Standard}: Tesla mobile charging unit comes with adapters allowing for every type of power outlets, from ordinary 120 volt 12 amp (NEMA 5-20)  and 240 volt 50 amp (NEMA 14-50) to SAE J1772 connectors and CHAdeMO connectors. In addition, Tesla Motors is building up its own Supercharger network which provides DC fast charging service for Tesla vehicles.

\subsection{Electric Vehicle Supply Equipment}
The electric vehicle supply equipment (EVSE) is a charging meter which draws electricity from the power system and feeds it to the onboard battery pack through a charging coupler. An EVSE can be installed  either at home or in public. A public charging station typically has multiple EVSEs. Currently, there are several EVSE manufacturers, like ChargePoint, Blink, AeroVironment, etc.

\subsection{EV Charging Coupler and Plug Receptacle}
An EV charging coupler is like a gas pump nozzle which connects the EVSE and the EV plug receptacle. In the U.S., most EVSEs and EVs are equipped with a standard charging coupler and receptacle based on the SAE J1772 standard. Any vehicle with a standard plug receptacle is able to use any J1772-compliant AC Level 1 or AC Level 2 EVSE.

As for DC fast Charging,  CHAdeMO \cite{japanstandard} is a widely used standard among EVs manufactured by Japanese automakers, such as Nissan Leaf and Mitsubishi vehicles. The Chevy Spark and the BMW i3 come with the SAE J1772 combined charging system (CCS), which uses a single receptacle for AC Level 1, AC Level 2, and DC Level 3. Additionally, Tesla Motor operates its own supercharger network, which is based on their own connector and currently only charges Tesla vehicles.

\section{EV Charging, Power System, and Electricity Market}
This subsection discusses the interplay among EV charging, the power system, and the electricity market.

\subsection{Power System and Control}
Although power system varies in size and structural components, they all share the following basic characteristics:

\begin{itemize}
\item Consist of three phase AC system operating at a constant voltage. Generation and transmission facilities use three phase equipments. Industrial loads are usually three phase; single phase residential and commercial loads are distributed evenly across the phase to effectively balance the three phases.\\

\item The power generators are synchronized to produce electricity using various kinds of energy sources, such as fossil fuels, solar, wind, nuclear, etc.\\

\item Transfer electricity to consumers across a wide area. This requires a complex transmission and distribution system comprising of many subsystems operating at different voltages.\\

\end{itemize}

\begin{figure}[htbp]
\centerline{\includegraphics[width=6in]{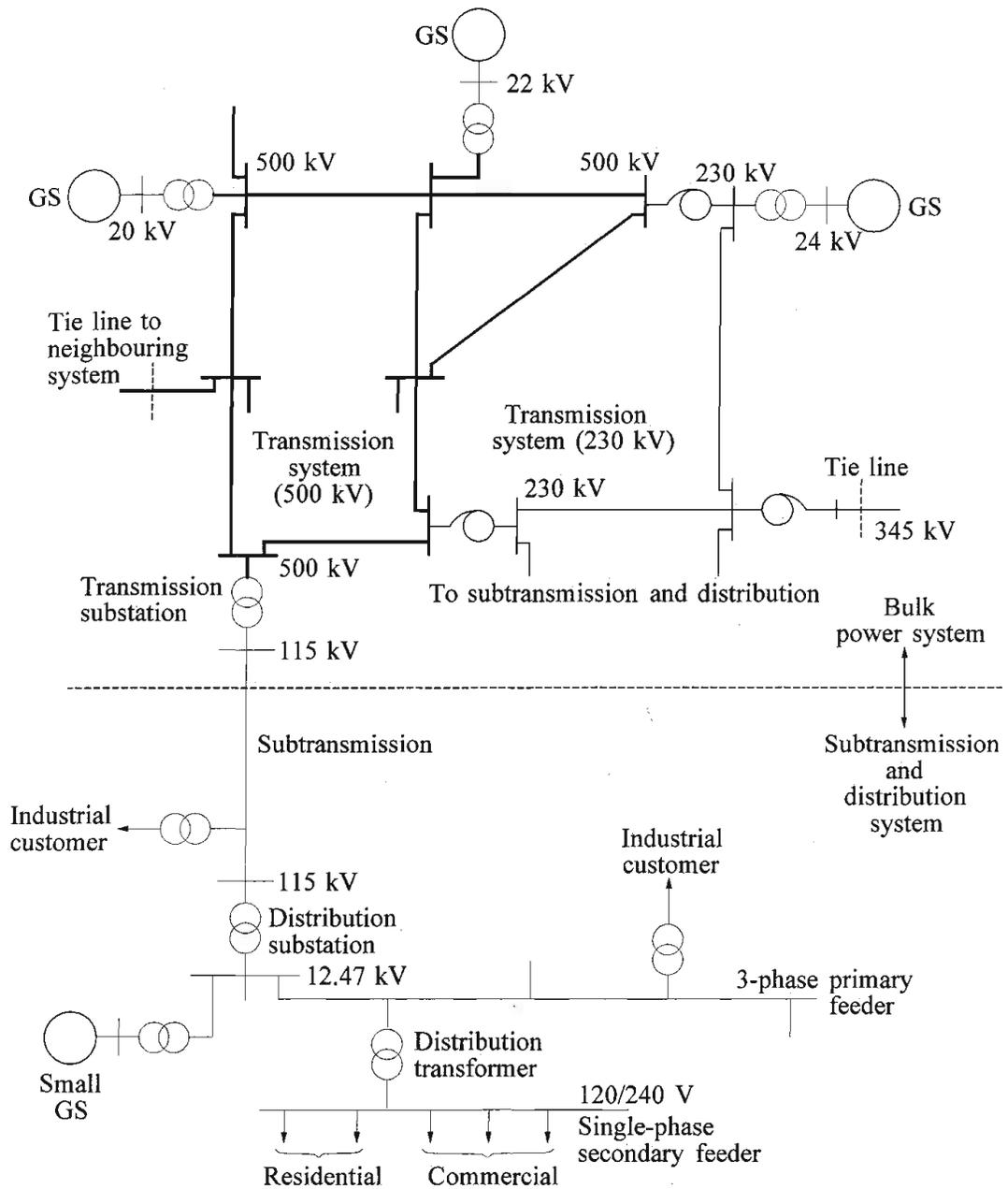}}
\center
\caption{Basic Elements of a Power System}
\label{powersystem}
\end{figure}

Fig. \ref{powersystem} shows the basic elements of a power system \cite{kundur}. Electricity is generated by the generation station (GS) and delivered to the end users through a complicated network of individual components, transmission lines, substations, feeders, etc.

In general, the electricity transmission network consists of three subsystems: transmission system, subtransmission system, and distribution system.

\underline{Transmission System}: It interconnects all major power plants and main load centers. It forms the backbone of the entire power system. It operates at the highest voltage levels (typically, above 230 kV). The voltage generated by the generators is usually in the range of 11 to 35 kV, which needs to be raised up to the transmission level voltage, say, 230 kV. When the electricity is delivered to the transmission substations, the voltage is stepped down to match the subtransmission levels (typically, 69 to 138 kV).

\underline{Subtransmission System}: It delivers the electricity in smaller quantities from transmission substations to distribution substations. Large industrial customers are typically supplied directly by the subtransmisison system. In modern electric power system, there is not a clear delimitation between a transmission system and a subtransmission system.

\underline{Distribution System}: It is responsible for delivering electricity to the end customers. Distribution voltages are typically in the rage of 4 to 34.5 kV. Small industrial customers are supplied by the primary feeders at this voltage level. The secondary distribution feeders supply residential or commercial customers at 120/240 V voltage level.

The most important task of an electric power system is to deliver electricity from the power generators to the end customers in a reliable and efficient manner. Power quality is one of the major issues which must be concerned with. Power quality can be described by a set of parameters, such as continuity of service, voltage variation, transient voltage or current, and harmonics. The power system has different levels of controllers that are intended for frequency control, voltage control, and reliability control. Basically, there are three groups of controllers in a power system: generating unit control, system generation control, and transmission control.

\underline{Generating Unit Control}: It mainly controls the input and output of a generator, which consists of the prime mover controller and the excitation controller. The prime mover controller is concerned with the speed regulation and control of energy supply system variables such as boiler pressure, temperature, and flow. The function of the excitation controller is to regulate the output voltage and reactive power.

\underline{System Generation Control}: It is responsible for coordinating the system generation and the system load and losses so that the desired frequency and power exchange with adjacent systems are maintained.

\underline{Transmission Control}: It has a myriad of power and voltage control devices, like static var compensators, synchronous condensers, capacitors and reactors, transformers, and HVDC transmission controls.

\subsection{ISO/RTO and Electricity Market}
In a deregulated electricity market, ISO/RTO is a nonprofit organization which is responsible for integrating a diverse mix of power resources into the power grid and coordinating generation and consumption. As a fair power grid operator, ISO/RTO has no financial interest in any market segment and different resources have an equal access to the transmission network. There are 9 ISO/RTOs in North America---California ISO (CAISO), Alberta Electric System Operator (AESO), Independent Electricity System Operator (IESO), Midcontinent ISO (MISO), Southwest Power Pool (SPP), Electric Reliability Council of Texas (ERCOT), ISO New England, New York ISO, and PJM Interconnection.

An ISO/RTO has a full model of the generation and transmission schedules to manage and avoid real-time congestions. An ISO/RTO also provides an integrated forward market (IFM) for trading and analyzing the electricity bids, transmission, capacity, and reserves needed to maintain grid stability. Additionally, the locational marginal price (LMP) is calculated by the ISO/RTO based on the cost of electricity generation and delivery.

Compared to regulated electricity markets, the presence of ISO/RTO has a few benefits:

\underline{Enhanced Reliability}: Since ISO/RTO is in charge of a large geographic area, and the market reliability can be improved through resource sharing, that allows excessive electricity in a local area to be transmitted to neighboring areas via an open market. The use of advanced technology and market-driven incentives improves the performance of power plants. Power plants tend to have lower outage rates in the unregulated market than in the monopoly market, because power plants have the motivations to keep the generators on line, especially during peak hours, to maximize their revenues.

\underline{Price Transparency}: The LMP mechanism creates a highly transparent system that calculates the prices based on the cost of electricity generation and delivery. In a monopoly market, customers and investors face the ``black box" pertaining to the information of prices and the locational cost of transmission, which inhibits investments in the grid.

\underline{Green Resources Integration}: ISO/RTO offers a fair platform for diverse electricity resources to compete with each other, bringing the cheapest electricity to customers. This non-discriminatory access to power grid opens doors for the low cost renewable energy.

\underline{Market Monitoring}: ISO/RTO plays a significant role in monitoring the electricity market. ISO/RTO calculates the LMP which truly reflects the supply and demand relationship. In addition, ISO/RTO can identify ineffective market rules and tariff provisions, identify potential anticompetitive market behaviors by participants and offer comprehensive market analysis to help in informed decisions.

\underline{Market Flexibility and Diversity}: The organized markets offer various electricity products and financial instruments which can be used to hedge price risks. Because average real-time energy prices are correlated to short-term forward bilateral prices, ISO and RTO markets foster forward contracting to stabilize prices. There are usually numerous sellers and buyers in the electricity wholesale market. Any entity can participate in the market if they satisfy the basic requirements.

\underline{Demand Response}: The ISO/RTO has access to power grid operation data, electricity usage data, generation and demand data. Facilitated by these data, the ISO/RTO can make informed decisions on pricing or set other rules to alter customer's usage behavior and shape the demand profile to optimize the grid.

As a core optimization algorithm used by ISO/RTO, Security Constrained Unit Commitment (SCUC) algorithm aims to determine the unit commitment (UC) and economic energy dispatch by taking into account the supply/demand bids, the ancillary service requirement, the transmission congestion and power balance. SCUC algorithm is used in the day-ahead market and the real-time market. SCUC algorithm employs Mixed Integer Programming (MIP) to effectively solve the optimization problem with various model requirements and constraints. The objective of SCUC algorithm is to minimize the overall cost of energy generation and ancillary services.

\begin{equation}
\begin{aligned}
\min\sum_{h=1}^T\sum_{i=1}^N&\biggl[SUC_i(1-U_{i,h-1})U_{i,h}+MLC_{i,h}U_{i,h}+
\int_{P_{\textrm{min},i}}^{P_{i,h}}C_{i,h}\left(P_{i,h}\right)dP \\
& +C_{i,h}^{\textrm{RU}}\cdot RU_{i,h}+C_{i,h}^{\textrm{RD}}\cdot RD_{i,h}+C_{i,h}^{\textrm{SP}}\cdot SP_{i,h}+
C_{i,h}^{\textrm{NS}}\cdot NS_{i,h}\biggl],
\end{aligned}
\end{equation}
where

$h$: Hour index

$T$: Total number of hours

$i$: Resource index

$N$: Total number of resources

$P_{i,h}$: Power output of resource $i$ in hour $h$

$RU_{i,h}$: Regulation up provided by resource $i$ in hour $h$

$RD_{i,h}$: Regulation down provided by resource $i$ in hour $h$

$SP_{i,h}$: Spinning reserve provided by resource $i$ in hour $h$

$NS_{i,h}$: Nonspinning reserve provided by resource $i$ in hour $h$

$C_{i,h}(P_{i,h})$: Cost (\$/hour) as a piece-wise linear function of output (MW) for resource $i$ in hour $h$

$C_{i,h}^{\textrm{RU}}$: Bid cost (\$/MW) of regulation up (MW) for resource $i$ in hour $h$

$C_{i,h}^{\textrm{RD}}$: Bid cost (\$/MW) of regulation down (MW) for resource $i$ in hour $h$

$C_{i,h}^{\textrm{SP}}$: Bid cost (\$/MW)of spinning reserve (MW) for resource $i$ in hour $h$

$C_{i,h}^{\textrm{NS}}$: Bid cost (\$/MW) of non-spinning reserve (MW) for resource $i$ in hour $h$

$SUC_i$: Start-Up Cost (\$/start) for resource $i$

$MLC_{i,h}$: Minimum Load Cost (\$/hour) for resource $i$ in hour $h$

$U_{i,h}$: Commitment status --- 0 if resource $i$ is off-line, and 1 if resource $i$ is online, in hour $h$

The constraints considered in SCUC encompass the power balance constraint, the ancillary service constraint, the transmission network constraint, and the inter-temporal constraint.

All ISO/RTOs have similar operation procedures to clear day-ahead market. In this section, we use CAISO as an example to demonstrate how the day-ahead wholesale market operates under the coordination of an ISO/RTO. Through three progressive stages (market power mitigation, integrated forward market, residual unit commitment), CAISO receives buy bids and sell offers, guarantees that the supply meets the demand, clears the prices and settles the transactions.

\underline{Market Power Mitigation (MPM)}: The ISO/RTO market aims to encourage competitive and efficient electricity consumption by ensuring that power generation offers are consistent with their marginal cost and the use of least costly centralized dispatch. The scheme of MPM are designed to ensure that the power generators are able to bid on their marginal costs, but not able to exercise market power. Market power is the ability of a provider to profitably raise the market price of a good or a service. End customers may suffer from electricity price rises if the power generators can exercise their market power to manipulate market prices.

Basically, there are two approaches to market power mitigation---``structural" approach and ``conduct and impact" approach \cite{mpm}. CAISO adopts the structural approach. For structural approach, generator offers are subject to mitigation if certain conditions are met. CAISO uses a formula to determine whether a given transmission constraint is structurally competitive or non-competitive. If the three largest resources available are jointly necessary to meet a given constraint relief demand, these three resources fail the test. Resources that fail the test in CAISO are mitigated to their reference level offers, which consist of a marginal cost estimate plus a ten percent adder. These rules are able to force the power generators to offer their bids close the short-term marginal cost to avoid being mitigated.

The internal monitor of CAISO will calculate the reference levels for incremental electricity offers, which are marginal cost plus a 10\% adder. Reference level calculation is based on many factors that include fuel prices and heat rates, a resource's lowest previous offers, lowest previous
prices at a resource's node, and opportunity cost.

\underline{Integrated Forward Market (IFM)}: In an IFM, ISO/RTO analyzes the energy and ancillary services market to calculate the transmission capacity needed (congestion management) and confirm the reserves to balance the supply and demand according to supply and demand bids. It guarantees that the generation plus imports equals to the load plus exports plus transmission losses. The ISO/RTO employs the LMP mechanism to calculate the market prices based on bids/offers submitted by buyers and sellers.

\underline{Locational Marginal Price (LMP)}: It is a mechanism to manage transmission congestion using market-based prices. LMP differs from place to place if transmission congestion occurs. Transmission congestion prevents the electricity of low-cost generators from satisfying all the loads and clearing the market. As a result, the low-cost generators have to ramp down to avoid transmission congestion. In essence, LMP is the marginal cost of supplying, at the least cost, the next increment of power demand at a specific location (node) in the power system, taking into account both the supply (generation or import) bids and the demand (load or export) offers and the physical characteristics of the transmission system and other operation constraints \cite{lmp}. Sometimes LMP is also called the ``node price".

The calculation of LMP is based on basic economic theory and power operation practice. The ISO/RTO determines the LMP at each node by maximizing the total social surplus under the transmission constraints and power losses. The total social surplus consists of the supplier surplus and the consumer surplus.

\begin{figure}[htbp]
\centerline{\includegraphics[width=3in, angle=-90]{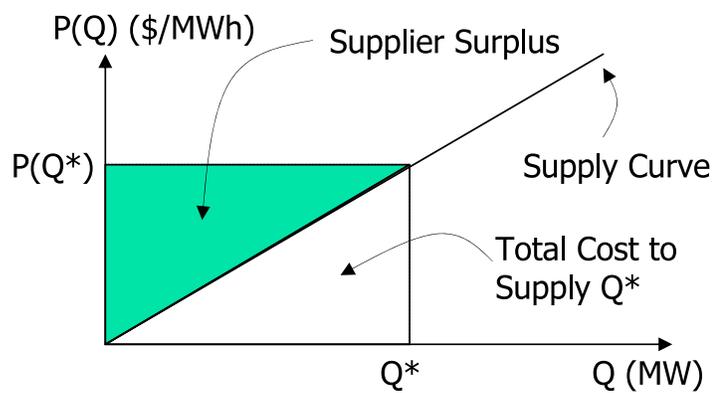}}
\center
\caption{Supplier Surplus}
\label{supplier_surplus}
\end{figure}

The supply curve in Fig. \ref{supplier_surplus} represents the marginal cost of supply. The y-axis $P(Q^*)$ corresponds to the minimum price-per-unit that the supplier hopes to be paid to produce the next increment $Q$ at the point of $Q^*$. The area under the curve up to $Q^*$ is the total cost ($\$/h$) of producing the quantity $Q^*$. In addition, the supply curve is monotone increasing. If the total quantity of $Q^*$ is priced at $P(Q^*)$, the green area shown in Fig. \ref{supplier_surplus} is the supplier surplus. The total revenue for selling $Q^*$ quantity is at least $P(Q^*)\times Q^*$ since the price can be greater than $P(Q^*)$. Suppose the total revenue is $P(Q^*)\times Q^*$, then the supplier surplus (area of upper triangle) equals to total revenue minus the total cost to supply (area of lower triangle).

\begin{figure}[htbp]
\centerline{\includegraphics[width=3in, angle=-90]{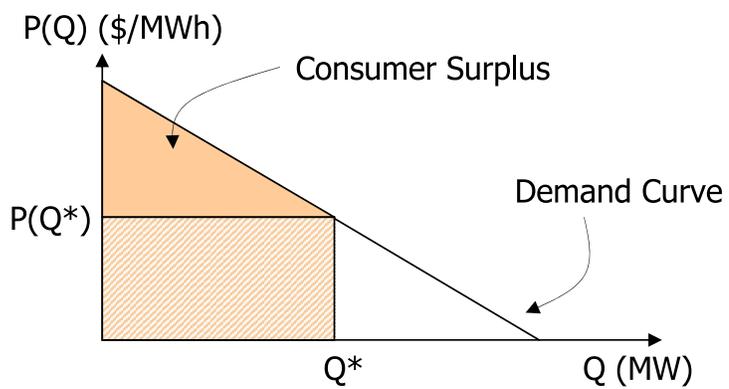}}
\center
\caption{Customer Surplus}
\label{customer_surplus}
\end{figure}

The demand curve in Fig. \ref{customer_surplus} represents the marginal benefit of demand. The y-aix $P(Q^*)$ corresponds to the price-benefit-per unit the consumer is willing to pay to consume the next increment of $Q$ at point $Q^*$. The area under the curve up to $Q^*$ is the total benefits ($\$/h$) to consume quantity $Q^*$. Also, the demand curve is monotone decreasing. If the total quantity of $Q^*$ is priced at $P(Q^*)$, the solid area shown is the customer surplus and represents the extra benefit the consumer acquires to
consume the quantity $Q^*$. The total payment for $Q^*$ is $P(Q^*)\times Q^*$ or less. Suppose the total payment is $P(Q^*)\times Q^*$, which is the area of the bottom square. The consumer surplus (area of upper triangle) = the total benefits (upper triangle + bottom square) minus total payment (the bottom square).

\begin{figure}[htbp]
\centerline{\includegraphics[width=3in, angle=-90]{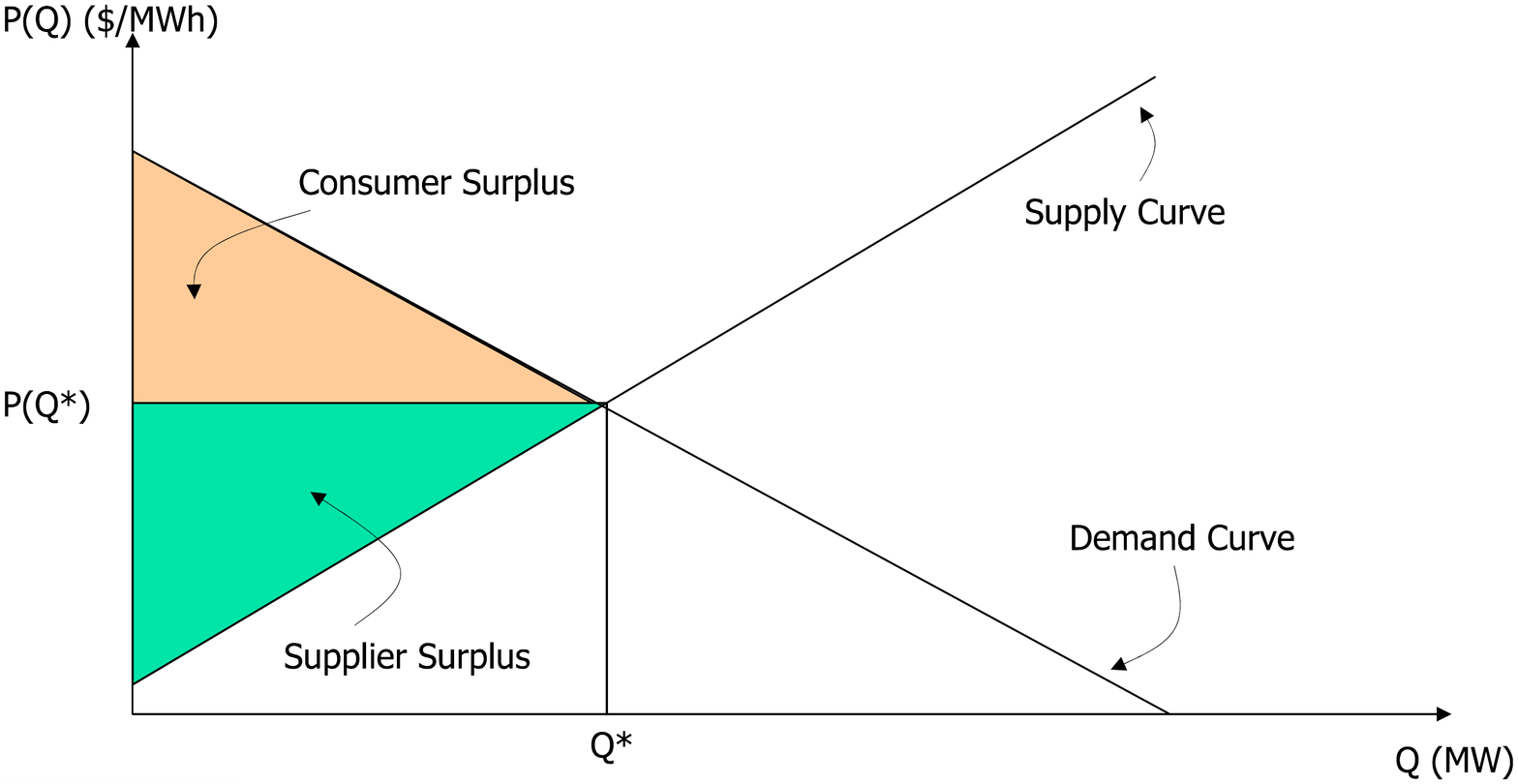}}
\center
\caption{Social Surplus}
\label{social_surplus}
\end{figure}

The total social surplus is the sum of the supplier surplus and the consumer surplus. As is shown in Fig. \ref{social_surplus}, the intersection of the demand curve and supply curve gives us the marginal clearing price, and the total social surplus is maximized under the condition that total supply equals total demand.

The ISO/RTO follows the same economic theory to calculate LMP, except that the transmission constraints, power grid operation practices, and transmission losses make the calculation more sophisticated. In this case, the incremental generator supply bids become the supply curve, and the decremental demand bids become the demand curve. LMP calculation is much more complex than the simple supply-demand curve relationship due to all transmission and other operation constraints. The transmission constraints and losses prevent the lowest-cost electricity from being delivered to all nodes, leading to LMP varying from location to location.

\underline{Residual Unit Commitment (RUC)}: In the integrated forward market, the ISO/RTO clears the prices and volumes by maximizing the total social surplus. If the clearing volume does not meet the forecast demand, the ISO/RTO needs to procure additional capacity from other markets to be available in real-time using the residual unit commitment process. This process ensures that there are enough capacity online to meet the forecast demand. All generators committed to residual unit commitment must be available online and submit an energy bit in the real-time market. The Resource Adequacy (RA) program developed by CAISO guarantees that load serving entities have at least 115\% of the peak-hour demand available as capacity.

The real-time market opens when the results of the day-ahead market are published, and closes 75 minutes prior the trading hour. The real-time market consists of several processes: (1) the market power mitigation process; (2) the hour-ahead scheduling process; (3) real-time unit commitment (RTUC); (4) the fifteen-minute market; (5) the real-time dispatch (RTD).

\underline{Market Power Mitigation (MPM)}: Similar to MPM in day-ahead market, the real-time market employs MPM to create a valid bid-pool for real-time market optimization. MPM runs every 15 minutes and the determination of MPM is based on the non-competitive congestion test.

\underline{Hour-Ahead Scheduling Process (HASP)}: HASP is a special run of the real-time unit commitment process. HASP produces (1) advisory schedules for internal pricing nodes and intertie resources that do not have hourly block schedules/bids; (2) final schedules for intertie resources with hourly block bids for energy and ancillary services. HSAP is run once every hour.

\underline{Real-Time Unit Commitment Process (RTUC)}: RTUC is a continuous process running at an interval of 15 minutes. This process (1) produces bind and advisory fifteen-minute market (FMM) awards; (2) issues start-up instructions for fast start and short start resources; (3) issues shut-down instructions for resources that are not used in the grid; (4) produces transition decisions for multi-stage generation resources.

\underline{Fifteen Minute Market (FMM)}: FMM runs every 15 minutes. This process is responsible for (1) determining financially binding FMM schedules and corresponding LMPs for all Pricing Nodes, including all Scheduling Points; (2) determining financially and operationally binding Ancillary Services Awards and corresponding ASMPs procure required additional Ancillary Services and calculating ASMP; (3) determining LAP LMPs.

\underline{Real-Time Dispatch (RTD)}: RTD uses a Security Constrained Economic Dispatch (SCED) algorithm every 5
minutes throughout the trading hour to determine the optimal dispatch instructions to balance supply and demand. RTD can operate in three modes: RTED (real-time economic dispatch), RTCD (real-time contingency dispatch) and RTMD (real-time manual dispatch).

\subsection{Interplay between EV Charging and Power System}
In this section, we consider the interaction between EV charging and the power system. In particular, we discuss the impact of EV charging on the power system. Additionally, we will investigate how EVs can be used as mobile energy storage to inject electricity into power grid via the emerging V2G technology.

\underline{Grid-to-Vehicle (G2V): Challenge and Opportunity of Power Grid}: The principal task of the power grid is to deliver economical and reliable electricity to end customers. Nevertheless, many studies have shown that simultaneous large-scale EV charging can disrupt the normal operation of the existing power system with respect to frequency variation, voltage imbalance, and severe power loss \cite{lopes, kinter, scott}. In regards to power supply, \cite{hadley, letendre, parks, kintner, stephan, lemoine} have shown that the existing power generators can support up to 30\% to 40\% PEV penetration rate without increasing generation capacity if the EV charging is optimally scheduled. Additionally, large-scale EV charging can also overload the distribution network \cite{ipakchi, clement, fernandez, kelly, schneider, shireen, taylor}. \cite{moses, verz} have demonstrated that high EV charging demand causes temperature increase and the AD-DC conversion of EV can lead to harmonic distortion, which will shorten the life span of upstream components like transformers and cables.

Generally, voltage and frequency are considered as the major variables to assess the power quality. The frequency of a generator is calculated using the following formula.

\begin{equation}
f=\frac{PN}{60},
\end{equation}
where $P$ is the number of stator pole pairs in the rotor, $N$ is the rotational speed of the rotor in rpm (revolutions per minute).

To maintain a constant frequency, the consumption and generation of active power should always be balanced. For instance, if the consumption exceeds the generation at any time, then the extra power is supplied by the rotational inertia of the generator by decreasing the speed, which results in the downward drift of frequency \cite{renac}.

Literature abounds in addressing EV charging scheduling. In \cite{chen, caramanis, gan, sojoudi, sortomme}, the authors have proposed different frameworks to coordinate EV charging to ensure stable and economical operation of the power grid. To protect the distribution network from overloading, various strategies have been proposed, e.g., demand response \cite{callaway}, time of use meters, and resource scheduling algorithms \cite{ipakchi, galus, dyke}.

In addition, the EV charging service provider can benefit from the burgeoning EV charging market. As the EV charging market rises, the EV charging stations are expected to gradually take over the market from the conventional gas stations. The charging service provider can make a good deal of profit by providing charging service to EVs. According to Navigant Research, global revenue from EVSE charging services will grow from \$81.1 million in 2014 to \$2.9 billion by 2023.

\underline{Vehicle-to-Grid (V2G): Challenge and Opportunity of Power Grid}: In addition to drawing electricity from the power system, EVs can also inject electricity into the system through a bidirectional power flow channel. Various studies \cite{tomic, letendre, brooks, yannick, kempton, brooks2, gao, uchukwu} have shown that the battery packs on EVs can be utilized as mobile energy storage, which inject electricity into the power system to provide ancillary services, like frequency regulation, spinning reserve, voltage control, reactive power compensation, etc. EV owners receive payments from the power aggregators by selling electricity to the power system, therefore, reducing the overall operation cost of EVs \cite{kempton2, lipman, kempton3, petersona, tanma}. Studies \cite{kemptonwind, hnguyen, nsendama, lopes, kempton4} have explored how to employ EVs to mitigate the power fluctuation arising from intermittent renewable energy generation. \cite{gparsons} have conducted a survey on customers' willingness to pay for V2G, and their findings suggest that the V2G concept is more inclined to be accepted if the power aggregators offer either pay-as-you-go service or advanced cash payment. The V2G communication protocols were developed and tested in report \cite{kgowri}. Additionally, the V2G concept has been implemented and validated using hardware and software simulator in \cite{kempton4, svandael, kempton5, yma, msingh}. Centralized and decentralized V2G mechanisms have been implemented at University of Delaware \cite{svandael}. A testbed of V2G for frequency regulation and energy storage in PJM system has been developed in \cite{kempton5}.

\begin{figure}[htp]
  \centering
  \includegraphics[width=6.0in]{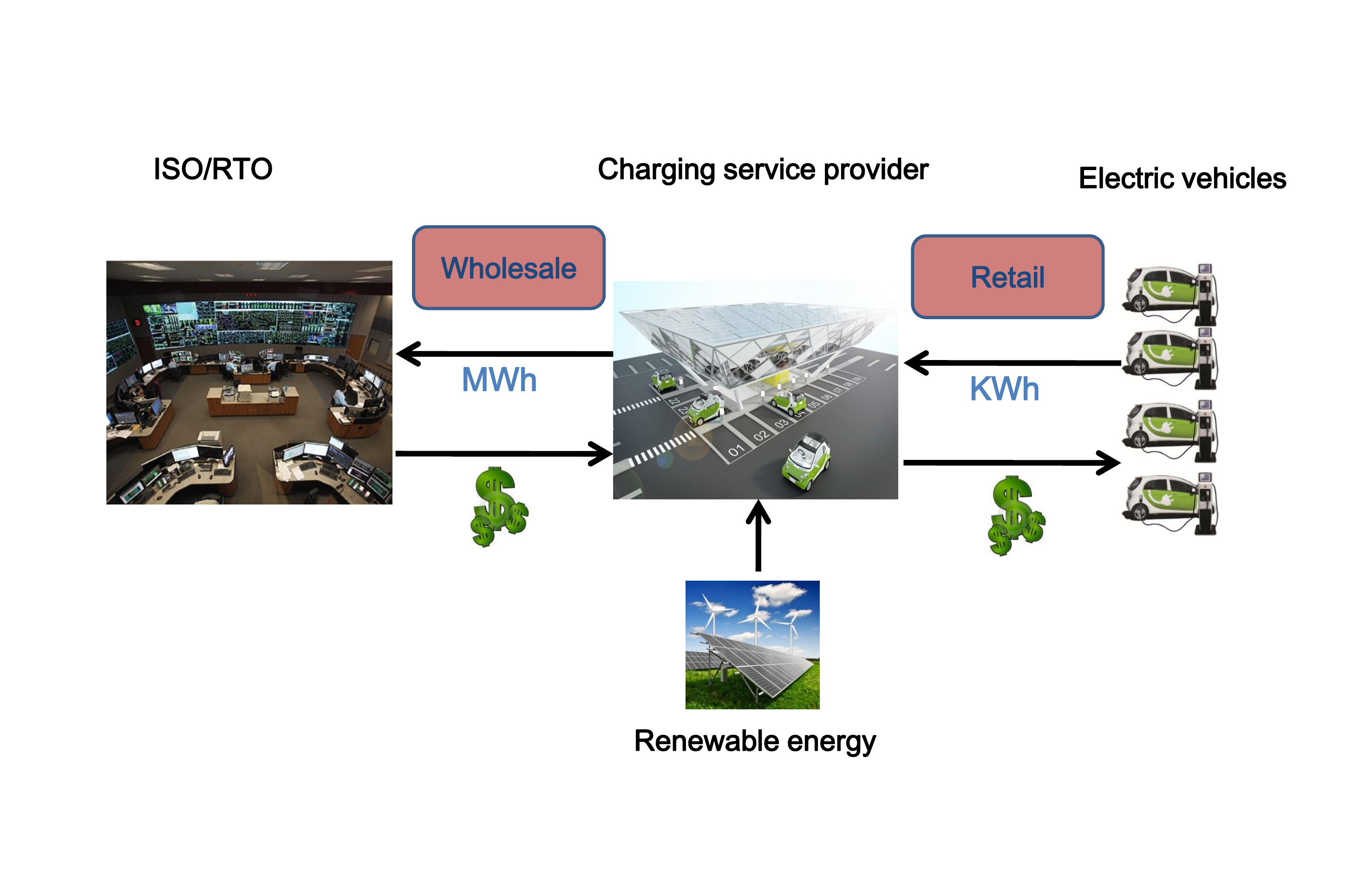}
  \caption{Vehicle to Grid Mechanism}\label{v2g}
\end{figure}

Although the electricity from EV battery packs is more costly than that from power plants (e.g., \$ 0.30/kWh versus \$ 0.05/kWh), it is still a compelling alternative to provide ancillary services because the price of power in ancillary consists of two parts---capacity price and energy price. If an EV is committed to provide ancillary services, the payment includes a capital price for being available to quickly respond to power grid signals, and an energy price for the actual energy output. Basically, the V2G exhibits several advantages over the power plants in providing ancillary services: (1) the capital cost of generation or storage equipment for V2G is low, (2) V2G can respond quickly, and (3) V2G has the ability to operate in the V2G mode without serious maintenance penalties.

While V2G has the aforementioned advantages, it has two main limitations, namely, battery degradation and DC/AC power conversion loss. Note that frequent charging and discharging will reduce the battery's life span. Therefore, battery degradation cost is an important factor that cannot be neglected in V2G. \cite{tomic} presents a formula to assess battery degradation cost.

\begin{equation}
c_{\textrm{d}}=\frac{c_{\textrm{bat}}}{L_{\textrm{ET}}}=\frac{E_{\textrm{s}}c_{\textrm{b}}+c_1t_1}{L_{\textrm{C}}E_{\textrm{s}}\textrm{DoD}},
\end{equation}
where $c_{\textrm{bat}}$ is the battery replacement cost in US dollars (capital and labor costs), $L_{\textrm{ET}}$ is the the battery lifetime energy throughput for a particular cycling regime in kWh, $E_{\textrm{s}}$ is the battery capacity in kWh, $c_{\textrm{b}}$ is the cost of battery replacement in in US\$/kWh, $c_1$ is the cost of labor in US\$/h, $t_1$ is the labor time for battery replacement, $L_{\textrm{C}}$ is the battery lifetime in cycles, and DoD is the maximum discharge rate of battery, usually usually 80\% for NiMH and 100\% for Li-Ion batteries.

Additionally, there are several challenges to implement V2G, such as the coordination between EVs and power grid, pricing, hardware, power quality assurance, etc. In summary, while V2G may seem like a viable alternative energy source for power generation, its realization may still be quite some time away.

%
%
%
%
%
%
%
%
%
%

%
%
\chapter{PLACEMENT OF EV CHARGING STATIONS --- BALANCING THE BENEFITS AMONG MULTIPLE ENTITIES}

\section{Overview}
In a study of placement of electric vehicle (EV) charging stations, this section proposes a multi-stage placement policy with incremental EV penetration rates. A nested logit model will be employed to analyze the charging preference of the individual consumer (EV owner), and predict the aggregated charging demand at the charging stations. The EV charging industry is assumed to be an oligopoly where the entire market is dominated by several charging service providers (oligopolists). We plan to use the Bayesian game to model the strategic interactions among the service providers and derive the optimal placement policy at the beginning of each planning stage. To derive the optimal placement policy, we consider both the transportation network graph and electric power network graph. In addition, we also propose to develop a simulation
software---EV Virtual City 1.0 using Java to investigate the interactions among the consumers (EV owner), the transportation network graph, the electric power network graph, and the charging stations.

\section{Motivation and Related Work}
The continued technological innovations in battery and electric drivetrain have made electric vehicles (EVs) a viable solution for a sustainable transportation system. Currently, most EV charging is done either at residences, or for free at some public charging infrastructure provided by municipalities, office buildings, etc. As the EV industry continues to grow, many more commercial charging stations need to be strategically added and placed. Development of effective management and regulation of EV charging infrastructure needs to consider the benefits of multiple constituencies--consumers, charging station owners, power grid operators, local government, etc. In this section, we propose a placement policy aiming at striking a balance among the profits of charging station owners, consumer satisfaction, and power grid's reliability.

Our work is motivated by the desire of those service providers to make a forward-looking decision on charging station placement to obtain a good return on the investment. We plan to use the most up-to-date information (i.e., travel pattern, traffic flow, road network, power grid, etc.) to make the best-effort decisions on charging station placement, hoping that service providers will have a good chance to profit over the next few years. In this model, we do not consider some factors like uncertainties in fuel prices, climate change, population migration etc., which are random and unpredictable in a long period of time. Instead, we believe some revenue management techniques (i.e., realtime pricing) may be applied to deal with the potential effects of those factors in our future work.

We assume that the service providers aim to strike a balance between the competing goals of maximizing the profits and minimizing the disturbance to electric power network due to large-scale EV charging. Accordingly, we construct a utility function that incorporates both the total profits and the impacts of EV charging on the electric power grid. Each charging service provider attempts to maximize their own expected utility function while satisfying the Quality-of-Service (QoS) constraints through choosing the optimal locations of charging station. Note that consumers usually have different preferences (tastes) over certain products and services. The nested logit model will be used to analyze and predict the charging preference of EV owners. As the EV penetration rate increases, the existing charging stations may no longer satisfy the QoS constraints
and a new stage shall be initiated to place more charging stations. At the beginning of each stage, the service providers predict the charging demand of each charging station candidate using the nested logit model. The optimal placement strategy is obtained through a Bayesian game.

There is a growing literature addressing the issues relevant to EV charging station placement. \cite{ge, mehar, frade, yi} formulated charging station placement as an optimization problem. However, they did not take into account the overall consumer satisfaction and the impacts of EV charging on electric power network in their works. Besides, their optimization models were formulated from the perspective of a central urban planner rather than that of service providers in a free competitive market. In \cite{sweda}, the authors presented a strategy to deploy charging stations by analyzing the patterns of residential EV ownership and driving activities. In their work, they deploy the new charging stations either randomly with no weight or only based on the weights of population. They did not consider the mobility of EVs and the overall consumer experience. Bernardo et al. \cite{bernardo} proposed a discrete choice model (DCM) based framework to study the optimal locations for fast charging stations. They treat each charging station as a player in a noncooperative game. However, the underlying assumption in their work is that each player has complete information about other players, which is too restrictive and infeasible in a practical competitive market. In this section, we propose a Bayesian game framework that does not require the complete information of other players.

The main contributions of our work are summarized as follows:

\begin{itemize}
\item A multi-stage charging station placement strategy with incremental EV penetration rates is first formulated, which takes into account the interactions among EVs, road network, and the electric power grid.
\item  A nested logit model is then employed to characterize the overall consumer satisfaction and predict the aggregated charging demand, which provides insights into the preferences and decision-making processes of EV owners.
\item An oligopolistic market model of EV charging service providers is studied and a Bayesian game framework is applied to analyze the strategic interactions among service providers.
\item A simulation software has been developed to analyze the interplay among EV owners, road network, power grid, urban infrastructure and charging stations.
\end{itemize}

\section{Preliminaries on Oligopoly and Consumer Behavior Analytics}
In this section, we will introduce basic knowledge on oligopoly and consumer behavior analytics.

\subsection{Oligopoly Market}
An oligopoly \cite{stigler, bulow} is a market form which is dominated by a small number of sellers (oligopolists). Oligopoly can reduce competition and lead to higher prices for consumers through different forms of collusion. In contrast to oligopoly, monopoly is a market form which is controlled by a single seller. There are many oligopolistic market examples, such as the Pepsi and Coke in the soft drink industry and the four wireless providers in US---Verizon, AT\&T, T-Mobile, and Sprint. The retail gasoline market is another example of an oligopoly because a small number of firms control a large majority of the gasoline market.

For this type of market, each oligopolist is aware of the actions of the other oligopolists, and the decisions of one oligopolist can significantly affect or be affected by others. According to game theory, strategic planning by oligopolists needs to take into account the possible responses of the other market participants.

\underline{Characteristics of Oligopoly}: Oligopoly has several characteristics:
\begin{itemize}
\item Profit maximizer: An oligopolist maximizes its profits.
\item Ability to set price: Oligopolies all have the market power to set the prices.
\item Entry and exit: The cost to enter the market can be high. The most important barriers are government regulation, patents, key technologies, and strategic actions from incumbent firms to impede the entry of nascent firms.
\item Product differentiation: The product in an oligopoly can be homogeneous (i.e. steel, gasoline) or differentiated (i.e. wireless services, automobile).
\item Knowledge about each other: Oligopolists have perfect knowledge about their own production and demand. However, usually the information about other oligopolists are incomplete.
\item Strategic interaction: An oligopoly consists of a few giant companies. Each oligopolist's action can influence the price of the entire market. Therefore, an oligopolist will take into account the possible actions of the other oligopolists when determine its own actions.
\end{itemize}

\underline{Outcomes of Oligopoly}: The competition among the oligopolists can result in different outcomes depending on how they interact with other. In some markets, the firms may employ very restrictive transaction practices (collusion, market sharing etc.) to raise prices and restrict production in much the same way as a monopoly. An example of such oligopoly market structure is the Organization of the Petroleum Exporting Countries (OPEC), which can greatly impact on the international price of oil. In some markets, there may exist a recognized leader who sets the product price and the other market participants just follow the price. This type of market is called price leadership. In other markets, the competition among the oligopolists can be extremely fierce. They compete with each other in prices and productions. This may ultimately lead to perfect competition.

\underline{Mathematical Models of Oligopoly}: Various models have been employed to analyze oligopoly, among which Cournot-Nash model and Betrand model are the most famous. There are several classic papers comparing and discussing the two models \cite{kreps, vives, singh, vives2}.

Cournot model describes a market where the firms compete on the quantities instead of prices and each firm makes an output decision under the assumption of the other firm's output is fixed. The firms will adjust their output quantities in response to the other firms' actions. It may take a series of actions and reactions before the firms finally reach the Cournot-Nash Equilibrium, in which situation no firm will change their output quantities given that they are aware the reactions of the other firms to any changes.

For example, we assume there are two firms in a market which produce homogeneous product. Suppose the output of firm 1 is $q_1$ and the output of firm 2 is $q_2$. Since the two firms produce homogeneous product, the price should be the same when they reach the Nash Equilibrium. We suppose the equilibrium price is a function of $q_1$ and $q_2$, i.e. $f(q_1+q_2)$. The cost of the two firms are denoted using $C_1(q_1)$ and $C_2(q_2)$, respectively. The profit of each firm is the total revenue minus the cost.

\begin{equation}
\Pi_j=f(q_1+q_2)q_j-C_j(q_j).\;\;\;(j=1,2)
\end{equation}

The best response is to maximize $\Pi_j$ by finding the optimal $q_j$ given that $q_i$ is fixed ($i\neq j$). Take the first derivative of $\Pi_j$ with respect to $q_j$ and set it to zero.

\begin{equation}
\frac{\partial\Pi_j}{\partial q_j}=q_j\frac{\partial f(q_1+q_2)}{\partial q_j}+f(q_1+q_2)-\frac{\partial C_j(q_j)}{\partial q_j}=0.
\end{equation}
The value of $q_j$ satisfying the equation above is the best responses. The Nash Equilibria are the pairs of $q_1$ and $q_2$ are satisfying the set of equations simultaneously.

In contrast to Cournot model, Bertrand model describes an market where the firms compete with each other in terms of product prices rather than product quantities. We assume there are at least two firms dominating the market, and the firms will not cooperate with each other. They try to maximize their profits by setting the optimal prices taking into account the possible pricing policies of the other firms.

Different from Cournot model, we need to consider two cases---homogeneous product case and heterogeneous product case in Bertrand model, which lead to very interesting results.

Homogeneous Product Case:  We assume firm 1 and firm 2 produce exactly same product. For simplicity, we assume the two firms have identical constant marginal cost function. Let $MC$ represent the marginal cost of producing one unit of the product. The firms set prices at $p_1$ and $p_2$ with outputs of $q_1$ and $q_2$, respectively. Since the product is exactly the same, any firm with a lower price will grab all the consumers. Therefore, the prices of the two firms will be the same when they reach equilibrium, i.e. $p_1=p_2=p$, where $p$ is the single market price. Furthermore, we assume the market demand is expressed by a function $q=q(p)$. Under the equilibrium, firm 1 and firm 2 will split the market demand equally---each firm sells the same amount of products, $q_1=q_2=q/2$.

Heterogeneous Product Case: We assume the products of firm 1 and firm 2 are similar, but not exactly the same. In this case, the prices do not have to be the same in equilibrium since a lower price does not mean the firm will lose the market. In addition, we assume the demands of the two firms depend on the two prices: $q_1=q_1(p_1,p_2)$ and $q_2=q_2(p_1,p_2)$. For the demand function of firm $i$ $q_i$, the partial derivative with respect to $p_i$ should be negative, indicating that the price rise will result in a decrease in demand and the partial derivative with respect to $p_j$ should be positive, indicating that the price rise of the other firm will result in an increase in its own demand. The profits of the two firms are given by

\begin{equation}
\Pi_1 = p_1q_1(p_1,p_2)-C_1(q_1(p_1,p_2)),
\end{equation}
and

\begin{equation}
\Pi_2 = p_2q_2(p_1,p_2)-C_2(q_2(p_1,p_2)).
\end{equation}

We take the first derivative of $\Pi_1$ and $\Pi_2$ with respect to $p_1$ and $p_2$, respectively. The best responses of $p_1$ and $p_2$ are obtained by letting the derivatives equal to zero and solve the set of equations. This is also called the first-order conditions (FOC).

\begin{equation}
\begin{aligned}
&\frac{\partial \Pi_1}{\partial p_1}=q_1(p_1,p_2)+p_1\frac{\partial q_1(p_1,p_2)}{\partial p_1}-
\frac{\partial C_1(q_1)}{\partial q_1}\frac{\partial q_1(p_1,p_2)}{\partial p_1}=0,\\
&\frac{\partial \Pi_2}{\partial p_2}=q_2(p_1,p_2)+p_2\frac{\partial q_2(p_1,p_2)}{\partial p_2}-
\frac{\partial C_2(q_2)}{\partial q_2}\frac{\partial q_2(p_1,p_2)}{\partial p_2}=0.
\end{aligned}
\end{equation}

We can rewrite it into a more compact way as follows,

\begin{equation}
\begin{aligned}
&\frac{\partial \Pi_1}{\partial p_1}=q_1+p_1\frac{\partial q_1}{\partial p_1}-MC_1\frac{\partial q_1}{\partial p_1}=0,\\
&\frac{\partial \Pi_2}{\partial p_2}=q_2+p_2\frac{\partial q_2}{\partial p_2}-MC_2\frac{\partial q_2}{\partial p_2}=0.
\end{aligned}
\end{equation}
where $MC_1$ and $MC_2$ are the marginal cost of firm 1 and firm 2, respectively.

\subsection{Consumer Behavior Analytics}
In consumer behavior analytics, discrete choice model (DCM) is employed to characterize, explain, and predict how a consumer will choose a certain product in the presence of a finite set of alternatives, for instance which car to buy, where to go to college, and which mode of transportation (bus, taxi, rail) to use to work \cite{train, morey, berry, benakiva}. DCM is a powerful tool that we can use to calculate the probability that a certain consumer will choose a certain product among a finite set of alternatives. The basic idea of DCM is that a consumer tries to maximize the total utility when making a decision in the presence of multiple choice alternatives.

We assume that there are some factors that determine a consumer's choice. Some of them are observed by the researchers, but some are not. The observed factors are labelled as $x$ and the unobserved factors are denoted as $\epsilon$ (we can call it the random term). How these factors influence the choice of a consumer is expressed using a function $y=h(x,\epsilon)$. The function characterizes the behavioral process of decision making of a consumer. Since $\epsilon$ is not observed, the choice of a consumer is not deterministic and cannot be exactly predicted by researchers. We assume that $\epsilon$ is a random variable with a probability density function (PDF) $f(\epsilon)$. The probability that a consumer chooses a particular product among a finite set of alternatives is just the probability that the unobserved factor such that the behavioral process results in the that outcome, i.e. $P(y|x)=\textrm{Prob}(\epsilon s.t. h(x,\epsilon)=y)$. The probability can be calculated by integration as follows,

\begin{equation}
\begin{aligned}
P(y|x)&=\textrm{Prob}\left(I[h(x,\epsilon)=y]=1\right)\\
&=\int I[h(x,\epsilon)=y]f(\epsilon)d\epsilon.
\end{aligned}
\end{equation}
where $I[h(x,\epsilon)=y]$ is an indicator function which takes value of 1 when the statement in the brackets is true and value of 0 otherwise.

\underline{Nested Logit Model}: DCM consists of many models, which encompasses the simple logit model, the nested logit model, the probit model, and the mixed logit model. The logit model is the most simple model with the assumption that the unobserved factors have iid extreme value distribution. Other complex models are developed based on logit model. In contrast to logit, nested logit model divides the set of alternatives into multiple small nests wherein alternatives are correlated and the alternatives across different nests are independent. Probit is derived under the assumption that the unobserved factors have a joint normal distribution, while the mixed logit allows the unobserved factors to have any distribution. In this subsection, we will primarily introduce the nested logit model, which will be utilized to characterize the EV charging demand in the following sections.

The nested logit model is appropriate for the situation where the set of alternatives faced by a consumer can be partitioned into several subsets, or nests. The nested logit model differs from the logit model in that it allows correlations among the alternatives.

Suppose that a consumer is facing a set of alternatives. Let the alternatives be partitioned into $K$ non-overlapping nests $B_1,B_2,\cdots,B_K$. The utility that consumer $n$ obtains from alternative $j$ in nest $B_k$ is denoted as $U_{nj}=V_{nj}+\epsilon_{nj}$, where $V_{nj}$ is
observed utility and $\epsilon_{nj}$ is the unobservable utility. The vector of unobservable utilities $[\epsilon_{n1},\epsilon_{n2},\cdots,\epsilon_{nJ}]$ has the generalized extreme value (GEV) distribution with the cumulative distribution function (CDF) as follows,

\begin{equation}
F(\epsilon_n)=\exp\left(-\sum_{k=1}^K\left(\sum_{j\in B_k}e^{-\epsilon_{nj}/\sigma_k}\right)^{\sigma_k}\right),
\end{equation}
where $\sigma_k$ is a measure of the degree of independence in the unobservable utility within nest $B_k$. For any alternatives $j$ and $i$ in nest $B_k$, the unobservable utility $\epsilon_{nj}$ and $\epsilon_{ni}$ are correlated. For any alternatives in different nests, the unobservable utilities are uncorrelated: $Cov(\epsilon_{nj},\epsilon_{nm})=0$, for $j\in B_k$ and $m\in B_l$ and $k\neq l$.

The probability that consumer $n$ will choose alternative $j$ is obtained by taking the expectation over the unobservable utilities as defined in Eq. (\ref{prob}).

\begin{equation}\label{prob}
\begin{aligned}
P_{nj}&=\mathbf{Prob}\left(V_{nj}+\epsilon_{nj} > V_{ni} + \epsilon_{ni}, \forall i \neq j \right)\\
&=\int_{-\infty}^{+\infty}F_{nj}(V_{nj}-V_{n1}+\epsilon_{nj},V_{nj}-V_{n2}+\epsilon_{nj},\cdots, V_{nj}-V_{nJ}+\epsilon_{nj})d\epsilon_{nj}
\end{aligned}
\end{equation}
where $F_{nj}$ is obtained by taking the derivative of $F(\epsilon_n)$ with respective to $\epsilon_{nj}$: $F_{nj}\frac{\partial F(\epsilon_n)}{\partial \epsilon_{nj}}$. Finally, the integration gives us the probability in the following form.

\begin{equation}\label{prob2}
P_{nj}=\frac{e^{V_{nj}/\sigma_k}\left(\sum_{i\in B_k}e^{V_{ni}/\sigma_k}\right)^{\sigma_k-1}}
{\sum_{l=1}^K\left(\sum_{i \in B_l}e^{V_{ni}/\sigma_l}\right)^{\sigma_l}}.
\end{equation}
See Appendix for more detailed derivation of Eq. (\ref{prob2}).

\section{Problem Formulation}
I postulate the problem of EV charging with an oligopolistic market structure that has multiple charging service providers (oligopolists). The service providers aim to maximize their expected utility while satisfying the QoS constraints by selecting optimal station placements.

Particularly, we consider the case of three service providers that offer three EV charging services \cite{toepfer}, namely, Level 1, Level 2, and Level 3 (see Table \ref{ev_charging_level} for details). Level 1 and Level 2 are AC charging. Level 3 charging is DC Fast charging. Let $\mathcal{O}=\{1,2,3\}$ denote the set of charging service providers. Moreover, we assume that service provider 1 offers Level 1 charging, service provider 2 offers Level 2 charging, and service provider 3 offers Level 3 charging. The three charging levels represent three charging services, which have different charging voltages and currents, charging times, and charging experiences. In economics, they are imperfect substitutes to each other. In our model, we are interested in investigating how the different charging services compete with each other in choosing locations and prices. Each service provider can run multiple charging stations. At each planning stage, service providers select some charging stations from a given set of candidates, denoted as $\mathcal{I}=\{1,2,3\cdots,L\}$. The set of EVs is denoted as $\mathcal{E}=\{1,2,3,\cdots,N\}$.

\subsection{Profit of EV Charging}
We assume that service providers run the charging stations like "chain stores", so charging stations affiliated with the same service provider have the same retail charging price. The charging stations purchase the electricity from the wholesale market at the locational marginal price (LMP). In a deregulated electricity market (like PJM, NYISO, NEISO, MISO, ERCOT, California ISO in USA, the New Zealand and Singapore markets), LMP is computed at every node (bus) by the market coordinator. LMP primarily consists of three components: system energy price, transmission congestion cost, and cost of marginal losses \cite{lmp}. Let $p_k$ represent the retail charging price of provider $k\;(k=1,2,3)$, and $p_{-k}$ be the retail charging prices of the other two service providers except $k$. Let $c_{j,k}$ be the LMP of the $j$th charging station candidate of service provider $k$, and $\psi_{j,k}$ be the predicted charging demand at the $j$th charging station candidate of service provider $k$. The vector $S_k=[s_{1,k},s_{2,k},\cdots, s_{L,k}]^{\textrm{T}}$ represents the placement policy of service provider $k$, where $s_{j,k}\in\{0,1\}$ is an indicator with $s_{j,k}=1$ implying that service provider $k$ will place the $j$th charging station. Let $S_{-k}$ represent the placement policies of the other two service providers, $\theta_{j,k}$ be the placement cost of the $j$th charging station. The total profit of service provider $k$ is

\begin{equation}
\Pi_k=p_k\Psi_k^{\textrm{T}}S_k-\textrm{diag}[C_k]\Psi_k^{\textrm{T}}S_k
-\Theta_k^{\textrm{T}}S_k,
\end{equation}
and the total revenue of service provider $k$ is,
\begin{equation}
R_k=p_k\Psi_k^{\textrm{T}}S_k-\textrm{diag}[C_k]\Psi_k^{\textrm{T}}S_k,
\end{equation}
where $\Psi_k=[\psi_{1,k},\psi_{2,k},\psi_{3,k},\cdots,\psi_{L,k}]^{\textrm{T}}$, $C_k=[c_{1,k},c_{2,k},\cdots,c_{L,k}]^{\textrm{T}}$ and\\ $\Theta_k=$ $[\theta_{1,k},\theta_{2,k},\cdots,\theta_{L,k}]^{\textrm{T}}$. The notation $\textrm{diag}[.]$ is an operator to create a diagonal matrix using the underlying vector, and $[.]^{\textrm{T}}$ is the transpose operation. In the equation above, the total sale is $p_k\Psi_k^{\textrm{T}}S_k$, the cost of purchasing electricity is $\textrm{diag}[C_k]\Psi_k^{\textrm{T}}S_k$, and the placement cost is $\Theta_k^{\textrm{T}}S_k$.

\subsection{Impact of EV Charging on Power Grid}
It is conceivable that the simultaneous large-scale EV charging can disrupt the normal operation of the power grid in terms of frequency variation, voltage imbalance, voltage variation, power loss. In a conventional power grid, the generators will cooperatively control the output of real power and reactive power to maintain system stability, perform frequency regulation and voltage regulation. The charging service providers must optimally place the charging stations to mitigate the ``disturbance" to the power grid. Accordingly, the overall utility function of service provider can be defined as:

\begin{equation}\label{omega}
U_k=\Pi_k-wB_k,
\end{equation}
where $\Pi_k$ is the total profits from EV charging. $B_k$ characterizes the penalty arising from large-scale EV charging. The variable $w$ is a weighting coefficient, reflecting the tolerance to the penalty. In the following section, we will further discuss how to develop a proper metric to evaluate the penalty $B_k$. We should note that the weighting factor $w$ in Eq. (\ref{omega}) offers a mechanism for the charging service providers to strike a balance between their own profit and the ``stress" their charging add to the power system.  If $w=0$, the impact of EV charging on the grid is not considered at all, and any non-zero value of $w$ implies some impact on the grid---the larger $w$ is, the larger the impact it is.  Generally, if the focus is on the charging provider's profit, a small $w$ is used.  In practice, the value of $w$ needs to be determined with the help of heuristic and empirical data.

\subsection{Quality-of-Service Metrics}
We use two quality-of-service (QoS) metrics for the service provider: (1) average service delay probability $\Upsilon_k$, (2) average service coverage $\Xi_k,\;(k=1,2,3)$.

\begin{equation}
\Upsilon_k=\frac{1}{N}\sum_{i=1}^N\upsilon_{i,k},
\end{equation}

\begin{equation}
\Xi_k=\frac{1}{N}\sum_{i=1}^N\xi_{i,k},
\end{equation}
where $\upsilon_{i,k}$ is the average service delay probability for the $i$th EV owner getting the EV charged at service provider $k$. For the $i$th EV owner, $\upsilon_{i,k}$ is defined as the ratio of the number of delayed charging to the total number of charging attempts; $\xi_{i,k}$ is the average number of accessible Level $k$ charging stations along the route from origin to destination. Notice that $\upsilon_{i,k}$ and $\xi_{i,k}$ are two random variables depending on the travel patterns of all EVs, the urban road network and the charging stations. It is difficult to use a simple formula to compute them. Instead, we employ Mento Carlo method to estimate those two values.

\subsection{Multi-Stage Charging Station Planning Scheme}
At each planning stage, the service providers aim at solving the following fundamental problem to obtain the optimal placement policy subjected to the QoS constraints.

\begin{equation}\label{obj}
\begin{aligned}
&[S_k^T|S_1^{T-1},S_2^{T-1},S_3^{T-1}]=\\
&\argmax_{\substack{{s_{1,k},\cdots,s_{L,k}}\\{s_{j,k}\in\{0,1\}}}}\left\{\mathbb{E}_{S_{-k}}[U_k]|S_1^{T-1},
S_2^{T-1},S_3^{T-1}\right\},
\end{aligned}
\end{equation}
subject to
\begin{equation}
\Upsilon_k\leq\Upsilon^0,
\end{equation}
\begin{equation}
\Xi_k\leq\Xi^0,
\end{equation}
where $\mathbb{E}_{S_{-k}}[.]$ denotes the expectation over $S_{-k}$, and $S_k^T$ is the placement policy at stage $T$. The variables $\Upsilon^0$ and $\Xi^0$ are the predetermined QoS constraints.

\subsection{Emerging Problems}
To solve the optimization problem defined as Eq. (\ref{obj}), we are confronted with three principal questions: (1) How to predict the aggregated charging demand $\psi_{j,k}$ at each charging station candidate? (2) How to find an appropriate metric to characterize the impacts of EV charging on the power grid? (3) How to derive the optimal placement policy in an easier way? For the first question, we propose to employ the nested logit model to estimate the charging demand using a nested logit model. As for the second question, we will discuss the how EV charging may impact the power grid, and propose a metric to assess the impacts of EV charging. For the last question, notice that the optimization problem formulated by Eq. (\ref{obj}) is intractable since the optimal placement decision for every service provider also depends on the decisions taken by other service providers. To this end, we employ a Bayesian game model to characterize the strategic interaction and price competition among the service providers.

\section{EV Charging Demand Estimation}
In our model, the aggregated charging demand at a charging station candidate is defined as the sum of the product of the probability that EV owners choose that charging station and the electricity required to charge the EVs. The charging behaviors of EV owners may be influenced by many factors that include the charging price, travel cost, amenities at or near the charging station, the travel purpose, EV owner's income, etc. We plan to the nested logit model to characterize the attractiveness of a charging station.

The nested logit model is widely used in the analysis and prediction of a consumer's choice from a finite set of choice alternatives. The main idea of nested logit model is that a consumer is a utility maximizer. The consumer will choose the very product which brings him/her the maximum utility.

In our problem, the utility that the $n$th EV owner can obtain from choosing charging station $j\;(j=1,2,\cdots,L)$ of service provider $k\;(k=1,2,3)$ is denoted as $U_{j,k}^n=\overline{U}_{j,k}^n+\epsilon_{j,k}^n$, where $\overline{U}_{j,k}^n$ is the observable utility and $\epsilon_{j,k}^n$ is the unobservable utility. The vector of unobservable utility $\epsilon^n=[\epsilon_{1,1}^n,\cdots,\epsilon_{L,1}^n,\epsilon_{1,2}^n,\cdots,\epsilon_{L,2}^n, \epsilon_{1,3}^n,\cdots,\epsilon_{L,3}^n]^{\textrm{T}}$ is assumed to have a generalized extreme value (GEV) distribution with cumulative distribution function (CDF).

\begin{equation}\label{cdf}
F(\epsilon^n)=\exp\left(-\sum_{k=1}^3\left(\sum_{l=1}^Le^{-\epsilon_{l,k}^n/\sigma_k}\right)^{\sigma_k}\right),
\end{equation}
where $\sigma_k$ is a measure of the degree of independence in the unobservable utility among the charging stations owned by service provider $k$. For nested logit model, $\epsilon_{j,k}$ is correlated within each charging level, and uncorrelated across different charging levels.

For nested logit model, we can decompose the observable utility $\overline{U}_{j,k}^n$ into two components---the utility of choosing service provider $k$ and the utility of choosing a charging station $j$. In addition, we assume home charging is the ``outside good" in this market \cite{berry, salop}. Thus, $\overline{U}_{j,k}^n$ for EV owner $n$ can be expressed as

\begin{equation}
\overline{U}_{j,k}^n=\overline{W}_{k}^n+\overline{V}_{j,k}^n,
\end{equation}
where $\overline{W}_{k}^n$ is the observable utility of choosing service provider $k$ (choosing nest $k$), and $\overline{V}_{j,k}^n$ is the observable utility of choosing charging station $j$ given that service provider $k$ has been chosen; $\overline{W}_{k}^n$ and $\overline{V}_{j,k}$ are linear weighted combinations of attributes of the charging stations and the EV owner.

Note that the retail charging price and the charging time are the two important factors differentiating the three charging services. In addition, we assume the income of EV owners will also play a role in choosing charging services. In contrast to our previous work \cite{luo}, we use a different formula to calculate $\overline{W}_k^n$ here.

\begin{equation}
\overline {W}_k^n =\alpha \frac{1}{t_k}+\beta \frac{p_k}{i_n},
\end{equation}

where $t_k$, $p_k$ and $i_n$ represent, respectively, the averaged charging time, the retail charging price, and the income of the $n$th EV owner; $ \alpha, \beta$ are the corresponding weighting coefficients. This model is similar to Ben-Akiva and Lerman's utility model in their study of public transportation mode \cite{benakiva}. The value of $\alpha$ is positive because shorter charging time implies a better charging service experience, therefore, leading to higher utility. The value of $\beta$ is negative because a higher retail charging price always results in less utility. However, the retail charging price is divided by the income, which reflects that the retail charging price for the EV owners becomes less important as their income increases. As an ``outside good", the utility of home charging is normalized, i.e. $\overline{W}_0^n=0$.

Furthermore, we define $\overline{V}_{j,k}^n$ as follows,
\begin{equation}
\begin{aligned}
\overline{V}_{j,k}^n=&
\mu_kd_{j,k}^n+\eta_kz_{j,k}^n+\gamma_kr_{j,k}+\lambda_kg_{j,k}+\delta_km_{j,k},
\end{aligned}
\end{equation}
where $z_{j,k}^n$ is the destination indicator. If the $j$th charging station  is near the EV owner's travel destination (within a threshold distance $d_{th}$), $z_{j,k}^n=1$, otherwise, $z_{j,k}^n=0$. $d_{j,k}^n$ is the deviating distance due to EV charging. We use Dijkstra's shortest path algorithm \cite{cormen} to calculate a travel route for each EV owner from his/her origin to destination. If an EV owner needs to go to charging station $j$ halfway, we define the deviating distance The variable $d_{j,k}^n$ as the route length of this new route minus the route length of the original route. Additionally, each candidate charging station has a vector of characteristics $[r_{j,k}, g_{j,k}, m_{j,k}]^{\mathrm{T}}$, which characterizes the attractiveness of this charging station in terms of those amenities. For instance, if there exists a restaurant near location $j$, $r_{j,k} = 1$, otherwise $r_{j,k} = 0$. Similarly, $g_{j,k}$ and $m_{j,k}$ are the indicators for shopping center and supermarket, respectively. The corresponding weighting coefficients are $\mu_k,\eta_k,\gamma_k,\lambda_k,\delta_k$.

The EV owner's choice is not deterministic due to the random unobservable utility. However, we can derive the probability that an EV owner will choose a certain charging station by taking the expectation over the unobservable utilities. Similar to Eq. (\ref{prob2}), the probability that the $n$th EV owner will choose the $j$th charging station of service provider $k$ is

\begin{equation}
\Phi_{j,k}^n=\frac{e^{\overline{U}_{j,k}^n/\sigma_k}\left(\sum_{l=1}^Le^{\overline{U}_{l,k}^n/\sigma_k}\right)^{\sigma_k-1}}
{\sum_{t=1}^3\left(\sum_{l=1}^Le^{\overline{U}_{l,t}^n/\sigma_t}\right)^{\sigma_t}}.
\end{equation}

All coefficients in the nested logit model can be estimated and calibrated from preference survey data. The nested logit model enables us to compute the probability that an EV owner will go to a certain charging station, even though, an EV owner's decision may not always comply with the calculated probabilities. An individual EV owner may go to a fixed charging station at his/her discretion. However, employing the nested logit model provides a statistically meaningful prediction for the charging demand based on ensemble averages.

Once the EV owners' choice probability is computed, we can predict the charging demand of a charging station.  Let $q_n\;(n=1,2,\cdots,N)$ denote the total electricity (measured in kWh) that the $n$th EV owner purchases from the charging station, and $q_n$ is a random variable uniformly distributed in the range $[Q_a, Q_b]$, where $Q_a$ and $Q_b$ are, respectively, the lower and upper limit of charging demand for all EVs. The total predicted charging demand of charging station $j$ of service provider $k$ is
\begin{equation}
\psi_{j,k}=\sum_{n=1}^Nq_n\Phi_{j,k}^n.
\end{equation}

\section{Impact of EV Integration on Power Grid}
The main function of the power grid is to deliver electricity to users reliably and economically. However, large-scale EV integration can potentially disrupt the normal operation of power grid in terms of system stability, severe power loss, frequency variation, voltage imbalance, etc. Generally, the variations in voltage and frequency of electricity are considered as the major factors to characterize the power quality.

Assume that the power system has $M$ generators and $D$ buses (substations), and the power flow study approach \cite{grainger} is applied to solve the voltage, real power, and reactive power in the power system. Consider the node power equations, which can be written as real and reactive power for each bus.

\begin{equation}
0=-P_i+\sum_{k=1}^N|v_i||v_k|(G_{ik}\cos\phi_{ik}+B_{ik}\sin\phi_{ik}),
\end{equation}
\begin{equation}
0=-Q_i+\sum_{k=1}^N|v_i||v_k|(G_{ik}\sin\phi_{ik}-B_{ik}\cos\phi_{ik}),
\end{equation}
where $P_i$ and $Q_i$ are, respectively, the injected real power and reactive power at bus $i$. The variable $G_{ik}$ is the real part of the element in the bus admittance matrix corresponding to the $i$th row and $k$th column, and $B_{ik}$ is the imaginary part of the element. $\phi_{ik}$ is the voltage angle between the $i$th bus and the $k$th bus. $|v_i|$ and $|v_k|$ are the voltage magnitudes at bus $i$ and bus $k$, respectively.

Generally, the power plants hope that the load is predictable and stable (or at least slow-varying). If the load fluctuates too much, the power plants have to ramp up and ramp down frequently, resulting in low efficiency and high cost from committing spinning reserve \cite{rebours}. Therefore, the fluctuation of the active power and reactive power at the generators with and without EV charging can be used as a metric to evaluate this ``stress". In particular, we use the 2-norm deviation of generating power (real power and reactive power) of all generators in the power system to calculate the impacts of EV charging.

\begin{equation}
B=||\mathbf{P}_{\textrm{g}}^{\textrm{base}}-\mathbf{P}_{\textrm{g}}^{\textrm{EV}}||^2_2+
||\mathbf{Q}_{\textrm{g}}^{\textrm{base}}-\mathbf{Q}_{\textrm{g}}^{\textrm{EV}}||^2_2,
\end{equation}
where $\mathbf{P}_{\textrm{g}}^{\textrm{base}}=[P_1^{\textrm{base}},P_2^{\textrm{base}},\cdots,P_M^{\textrm{base}}]$ is a vector representing the active power generated by the $M$ generators under the base power load scenario (without EV charging), and $\mathbf{P}_{\textrm{g}}^{\textrm{EV}}=[P_1^{\textrm{EV}},P_2^{\textrm{EV}},\cdots,P_M^{\textrm{EV}}]$ is the vector of active power with EV charging (i.e., base power load superposed by EV charging load). Similarly, $\mathbf{Q}_{\textrm{g}}^{\textrm{base}}=[Q_1^{\textrm{base}},Q_2^{\textrm{base}},\cdots,Q_M^{\textrm{base}}]$ is the vector of reactive power of the base load, and $\mathbf{Q}_{\textrm{g}}^{\textrm{EV}}=[Q_1^{\textrm{EV}},Q_2^{\textrm{EV}},\cdots,Q_M^{\textrm{EV}}]$ is the vector of reactive power with EV charging. For a specific power system, $\mathbf{P}_{\textrm{g}}^{\textrm{base}},\mathbf{P}_{\textrm{g}}^{\textrm{EV}},\mathbf{Q}_{\textrm{g}}^{\textrm{base}}$, and $\mathbf{Q}_{\textrm{g}}^{\textrm{EV}}$ can be calculated through solving global power flow.

\section{Spatial Competition and Optimal Placement through a Bayesian Game}
It is pivotal for firms to choose the right location and product to compete with rivals in the same industry. Business locations will affect business competition, and conversely intensive competition will affect how firms choose the appropriate locations. One question arises naturally is whether or not firms from the same industry like to cluster their stores. There are some classical literature on spatial competition, e.g. Hotelling's location model \cite{hotelling} and Salop's circle model \cite{salop}. Firms have incentives for both clustering and separation. On one hand, firms prefer clustering so that they can learn from each other on how to improve manufacturing and research productivity \cite{glaeser,shaver}, and learn demand from each other to reduce the cost of searching for the optimal location. Firms also cluster for the labor pool and supplies. In addition, firms can benefit from the spinoffs that are located near parent firms. On the other hand, the fear of intensive price competition due to clustering may motivate the firms to separate from each other.

The EV charging service providers face the similar dilemma. Therefore, we need to investigate how the service providers will interact with each other in choosing their charging station locations and setting the retail charging prices. In practice, the exact placement costs and utility functions of the competing service providers may not be known to the service provider \emph{a priori}. We thus formulate the problem as a Bayesian game \cite{harsanyi} among the service providers at each planning stage.

The main components of a Bayesian game include a set of players $\mathcal{I}$, a strategy space $S_k$, a type space $\Theta_k$, a payoff function $u_k$,  and a joint probability of the types $f(\Theta_1,\Theta_2,\Theta_3)$. We will construct the utility function of each service provider. The Bayesian Nash Equilibirum (BNE) of charging station placement strategy can be derived by the first order of conditions (FOC).

Before proceeding to analyze the Bayesian game, we need the following assumptions.

$\mathbf{Assumption\;1}$: $f(S_{-k})$ is binomially distributed with parameter 0.5, \emph{i.e.} $S_{-k}\backsim \textrm{Binomial}(2L, 0.5)$.

 $\mathbf{Remark\;1}$: The distribution of $S_{-k}$ reflects player $k$'s conjecture on how other players will act during the game. Each player can form their conjectures about other players according to their beliefs about the competitors. For instance, a player may be risk neutral, risk aversion or risk seeking. For simplicity, we assume that $S_{-k}$ has a binomial distribution with parameter 0.5. However, the theoretical analysis can be applied to any other distribution of $S_{-k}$.

$\mathbf{Assumption\;2}$: All service providers in the market are Bertrand competitors.

$\mathbf{Remark\;2}$: Bertrand competitors are players that do not cooperate with each other. Their goal is to maximize their own utility. They will not form any type of ``coalition" to manipulate the market.

For each player, the Bayesian Nash Equilibirum (BNE) of placement policy can be derived from Eq. (\ref{obj}). To solve Equation (\ref{obj}), we need to know the retail charging prices of all the service providers. In a Bertrand competition, the retail prices for every combination of the charging station placement policies are determined by the first order of conditions (FOC):
\begin{equation}\label{eq1_3}
\frac{\partial\Pi_1}{\partial p_1}=\sum_{n=1}^N\sum_{j=1}^{L}q_ns_{j,1}\left[\Phi_{j,1}^n+(p_1-c_{j,1})\frac{\partial\Phi_{j,1}^n}{\partial p_1}\right]=0
\end{equation}
\begin{equation}\label{eq2_3}
\frac{\partial\Pi_2}{\partial p_2}=\sum_{n=1}^N\sum_{j=1}^{L}q_ns_{j,2}\left[\Phi_{j,2}^n+(p_2-c_{j,2})\frac{\partial\Phi_{j,2}^n}{\partial p_2}\right]=0
\end{equation}
\begin{equation}\label{eq3_3}
\frac{\partial\Pi_3}{\partial p_3}=\sum_{n=1}^N\sum_{j=1}^{L}q_ns_{j,3}\left[\Phi_{j,3}^n+(p_3-c_{j,3})\frac{\partial\Phi_{j,3}^n}{\partial p_3}\right]=0
\end{equation}
where $c_{j,1},c_{j,2},$ and $c_{j,3}$ represent the LMP at each charging station candidate.

$\mathbf{Remark\;3}$: For simulation simplicity, we assume the charging stations affiliated to the same service provider have the same retail charging prices ($p_1, p_2, \textrm{ and } p_3$). However, our analysis can be easily generalized to the case where each charging station sets its own retail price. Note that those retail prices obtained from Eq. (\ref{eq1_3})-(\ref{eq3_3}) may not be the real-time prices used in practice. They are only the equilibrium prices in this market under the assumption of Bertrand competition. They can be interpreted as the averaged charging prices of the service providers over a long period of time. In practice, the service providers take turns to set the retail price in response to the prices of the competitors. Additionally, if some of the other factors change (i.e. consumer's preference, crude oil price soaring, etc.), the existing equilibrium breaks and a new equilibrium must be computed using the same procedure.

$\mathbf{Theorem\;1}$ [Strategy Decision Condition]: Service provider $k$ will choose placement policy $l(l=1,2,3,\cdots,2^L)$ if the type space $\Theta_k$ falls into the hypervolume specified by
\begin{equation}
\begin{aligned}
&\mathcal{H}(l)=\\
&\{\Theta_k\in \mathbb{R}_+^L:\Theta_k^{\textrm{T}}\left(S_{k,j}-S_{k,l}\right)-(\mathbb{E}R_{k,j}-\mathbb{E}R_{k,l})\\
&\;\;\;\;\;\;\;\;\;\;\;\;\;\;\;\;+w(B_{k,j}-B_{k,l})>0;\forall j \neq l\},
\end{aligned}
\end{equation}
where $S_{k,j}$ and $S_{k,l}$ denote the placement strategy $j$ and $l$, respectively. $\mathbb{E}R_{k,j}$ and $\mathbb{E}R_{k,l}$ denote the expected total revenue with deployment strategy $j$ and $l$, respectively.

\begin{proof}
Each service provider has $L$ location candidates, so there are $2^L$ different placement policies. The type space can be seen as an $L$-dimensional space, and $\Theta_k=[\theta_{1,k},\theta_{2,k},\cdots,\theta_{L,k}]^{\textrm{T}}$ represents a point in this space.

By Eq. (\ref{obj}), strategy $l$ is optimal if
\begin{equation}
\begin{aligned}
&\mathbb{E}[R_{k,l}]-\Theta_{k}^{\textrm{T}}S_{k,l}-wB_{k,l}>\mathbb{E}[R_{k,j}]-\\
&\;\;\;\;\;\;\;\;\;\;\Theta_{k}^{\textrm{T}}S_{k,j}-wB_{k,j};(j=1,2,\cdots,2^L, j \neq l).
\end{aligned}
\end{equation}
Rearranging the terms, we get
\begin{equation}
\begin{aligned}
&\Theta_k^{\textrm{T}}\left(S_{k,j}-S_{k,l}\right)-(\mathbb{E}R_{k,j}-\mathbb{E}R_{k,l})+\\
&\;\;\;\;\;\;\;w(B_{k,j}-B_{k,l})>0;(j=1,2,\cdots,2^L, j \neq l),
\end{aligned}
\end{equation}
where each inequality represents a hyperplane and the intersection of all the inequalities defines a hypervolume in the type space.
\end{proof}

\section{Simulation Platform Development and Case Study}
\begin{figure}[htbp]
\centerline{\includegraphics[width=4in]{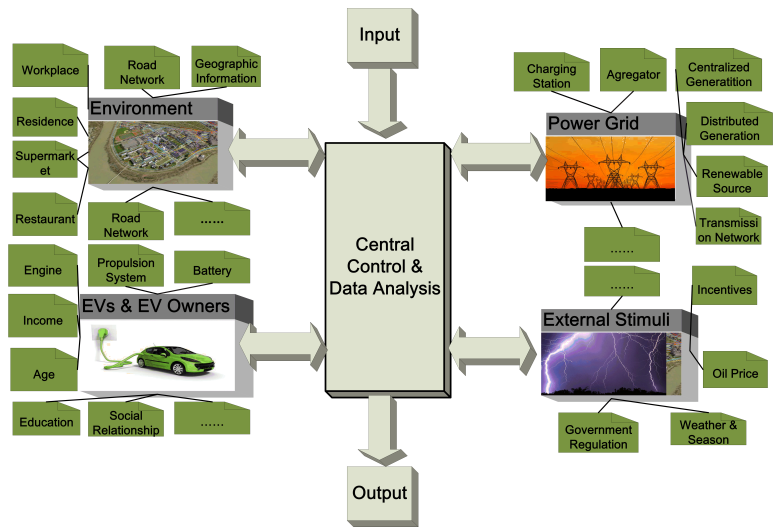}}
\center
\caption{The Architecture of The EV Virtual City 1.0}
\label{architecture}
\end{figure}

We propose to develop a general-purpose simulation software---The EV Virtual City 1.0 using Repast \cite{repast}. Our simulation software is designed to construct a virtual digital city by integrating a variety of data and information, such as geographic information, demographic information, spatial infrastructure data, urban road network graph, electric power network graph, travel pattern, diurnal variation in traffic flow, seasonal fluctuation of driving activities, social interaction, etc. The platform is flexible that one can include or exclude many modules to satisfy different simulation needs. See Fig. \ref{architecture} for the architecture of the simulation software.

\begin{figure}[htbp]
\centerline{\includegraphics[width=4in]{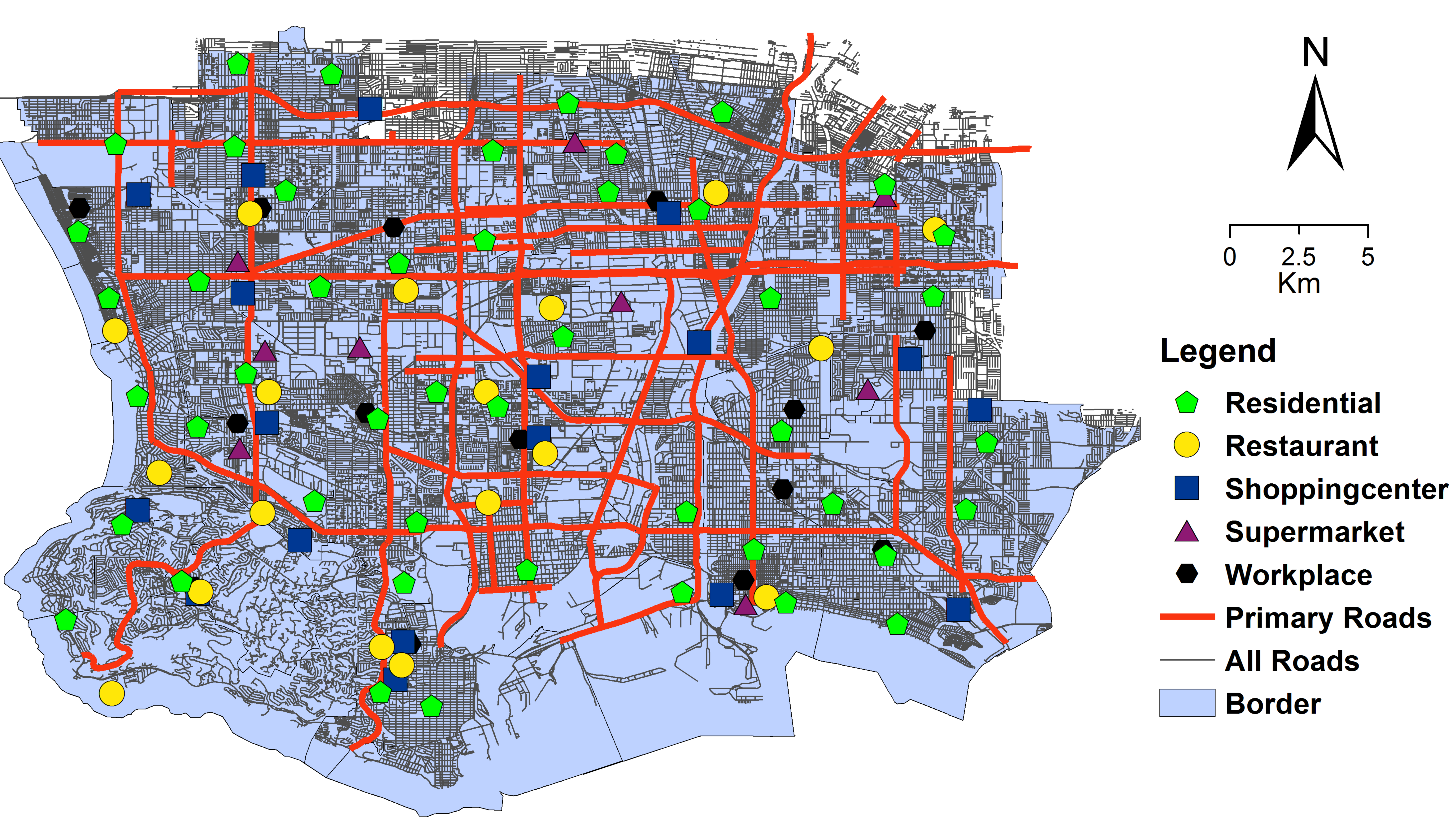}}
\center
\caption{Roads and Buildings of San Pedro District}
\label{road}
\end{figure}

We will conduct a case study using the data of San Pedro District of Los Angeles. We import the shapefiles of zip code tabulation area (ZCTA) and road network data from the U.S. Census Bureau into our simulation software. In addition, we have calculated the centroids of locations of residence, restaurants, supermarkets, shopping centers and workplaces using Google Maps, see Fig. \ref{road}.

\begin{figure}[htbp]
\centerline{\includegraphics[width=4in]{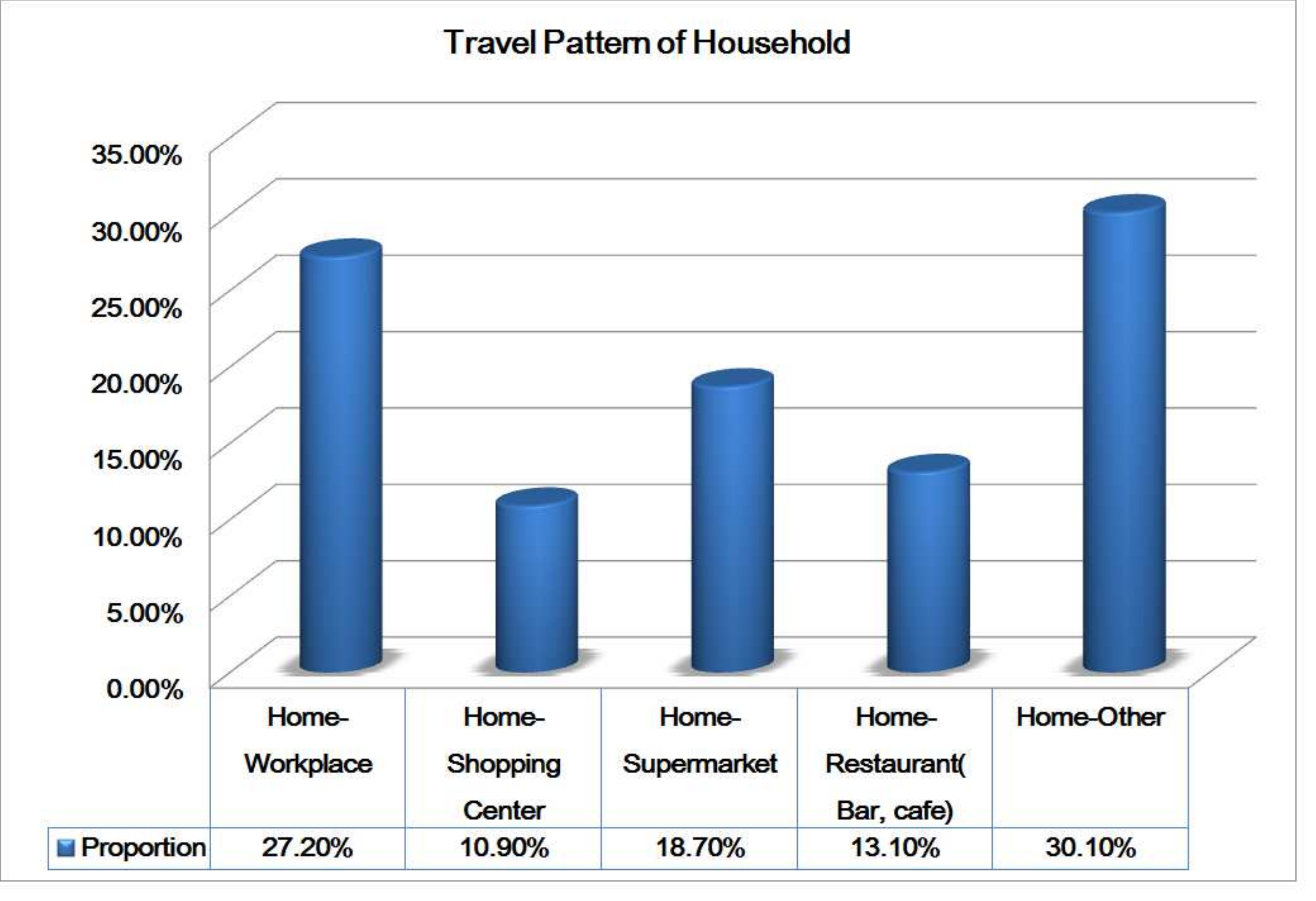}}
\center
\caption{The Statistics of Travel Patterns}
\label{statistic}
\end{figure}

Since travel pattern also plays a significant role in analysing the charging behavior of EV owners, it is necessary to have a thorough study on the statistics of travel pattern. From the 2009 National Household Travel Survey (2009 NHTS) \cite{nhts}, we obtained the travel pattern statistics. See Fig. \ref{statistic}.

\begin{figure}[htbp]
\centerline{\includegraphics[width=4in]{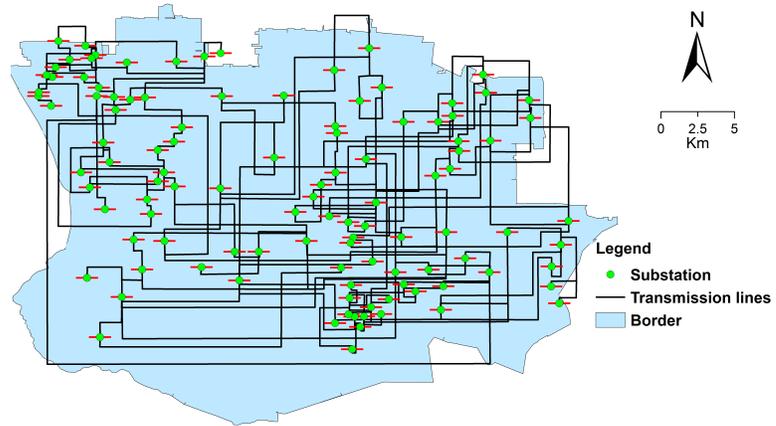}}
\center
\caption{IEEE 118-Bus Power System Test Case}
\label{118bus}
\end{figure}

From the California Energy Commission website, we obtained the maps of transmission line and substations of San Pedro District. This area has 107 substations in total. Thus we use the IEEE 118-bus power system test case in our simulation. For each charging station placement policy, we used MATPOWER \cite{zimmerman} to calculate the LMP of each bus and the output power of each generator with and without EV charging. See Fig. \ref{118bus}.

\begin{figure}[htbp]
\centerline{\includegraphics[width=4.2in]{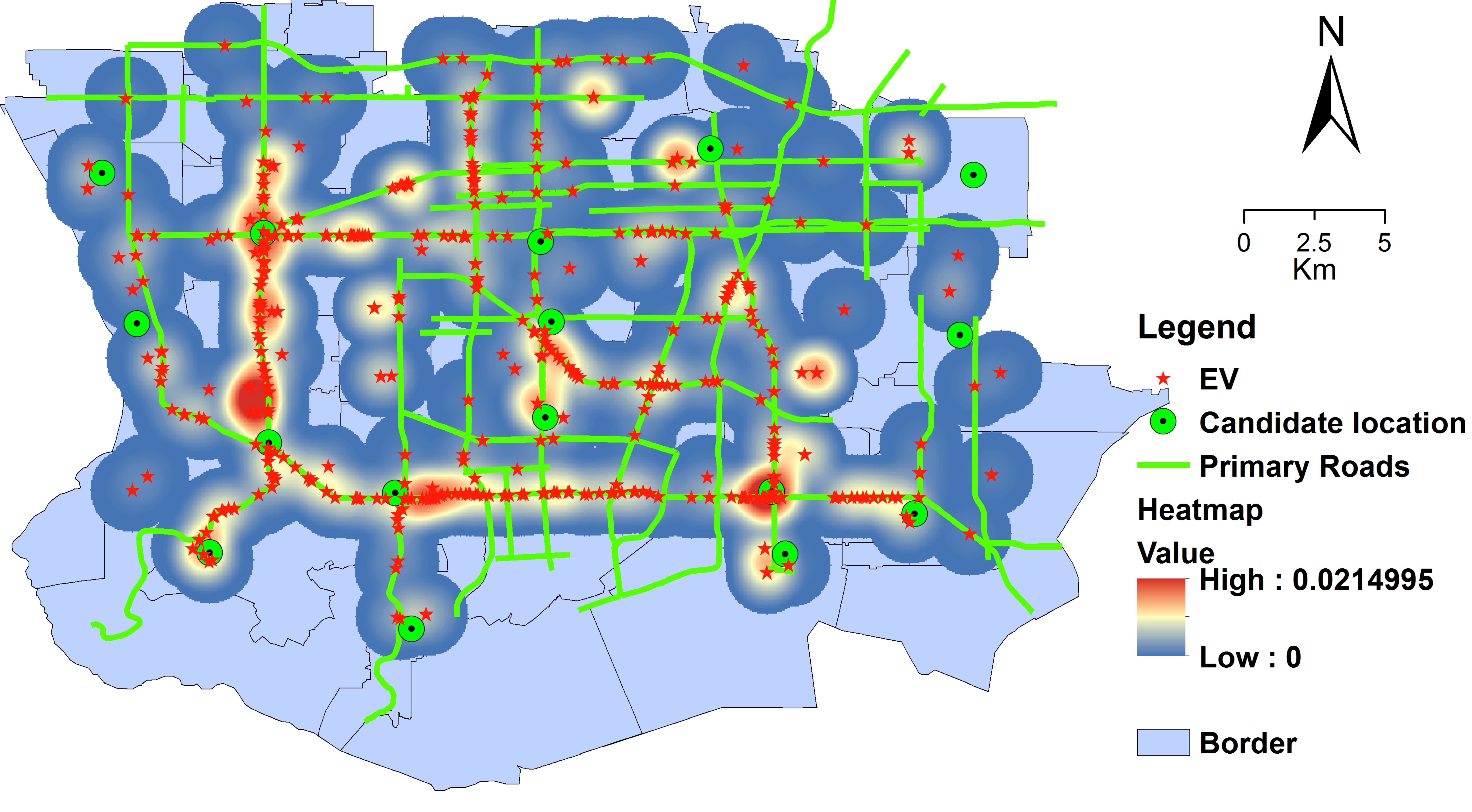}}
\center
\caption{A Snapshot of EVs Movement}
\label{movement}
\end{figure}

\begin{figure}[htbp]
\centerline{\includegraphics[width=4.2in]{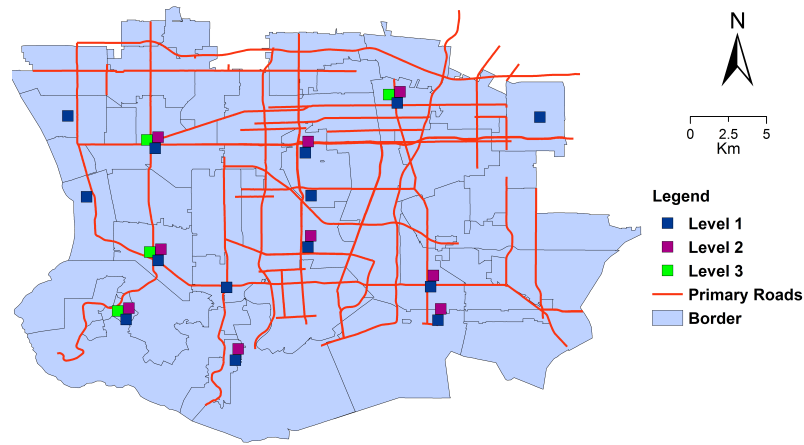}}
\center
\caption{Stage 1 Placement}
\label{stage1}
\end{figure}

\begin{figure}[htbp]
\centerline{\includegraphics[width=4.2in]{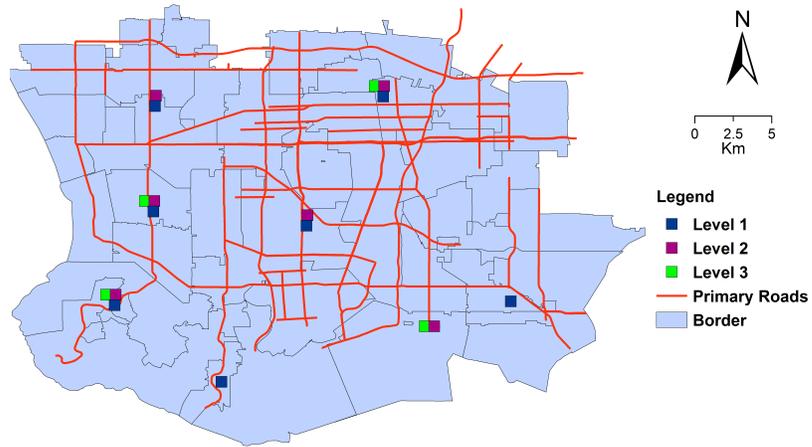}}
\center
\caption{Stage 2 Placement}
\label{stage2}
\end{figure}

\begin{figure}[htbp]
\centerline{\includegraphics[width=4.2in]{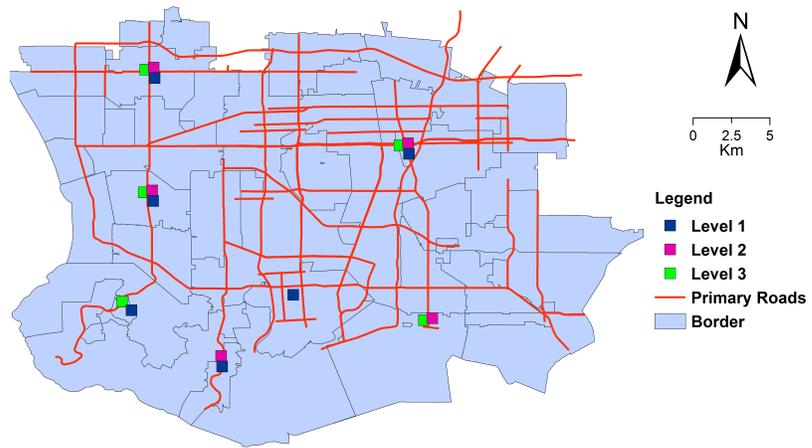}}
\center
\caption{Stage 3 Placement}
\label{stage3}
\end{figure}

\begin{figure}[htbp]
\centerline{\includegraphics[width=4.2in]{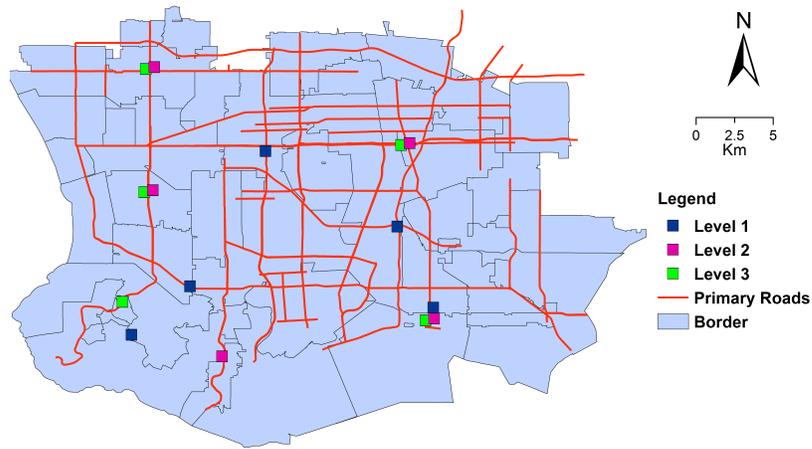}}
\center
\caption{Stage 4 Placement}
\label{stage4}
\end{figure}

\begin{figure}[htbp]
\centerline{\includegraphics[width=4.2in]{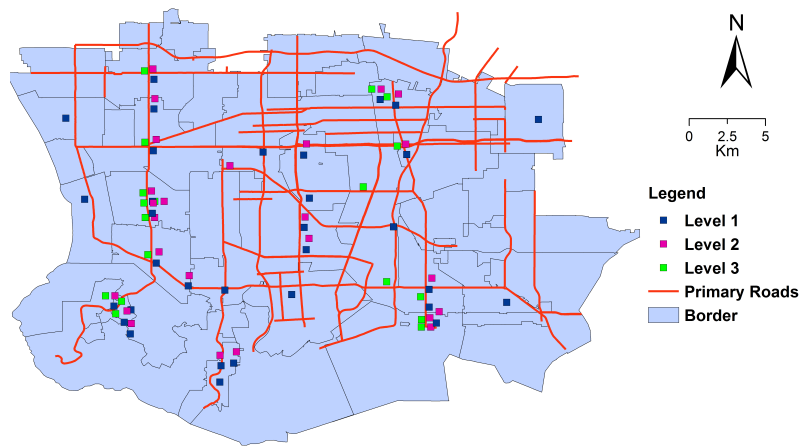}}
\center
\caption{Superimposing All Stages}
\label{overview}
\end{figure}

\begin{figure}[htbp]
\centerline{\includegraphics[width=4in]{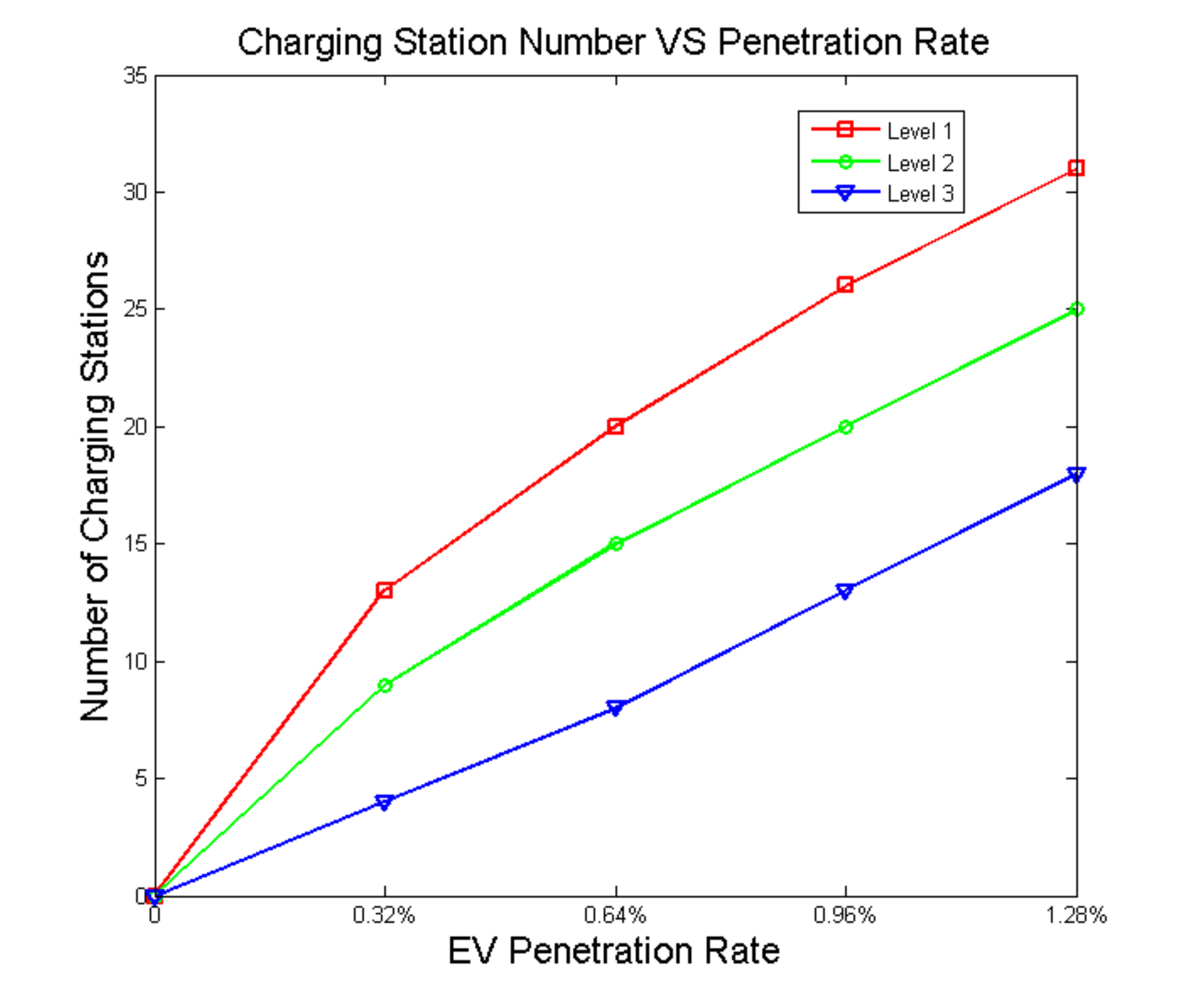}}
\center
\caption{Charging Station verses EV Penetration Rate}
\label{line}
\end{figure}

A snapshot of the moving EVs is shown in Fig. \ref{movement}. Each red star represents an EV owner. The traffic flow heatmap of EV owners is also plotted in this figure. Figs. \ref{stage1} to \ref{stage4} correspond to the charging station placement for stages 1 to 4, respectively. Fig. \ref{overview} is an overview of charging station placement by superposing Figs. \ref{stage1} to \ref{stage4}. Fig. \ref{line} shows how the number of charging station increases as the EV penetration rate increases.

\section{Conclusion}
From the simulation results, we can make the following observations:

1. The optimal charging station deployment is consistent with the EV traffic flow heatmap. This suggests that our model can adequately capture the mobility of EVs and provide EV owners with convenient charging services.

2. As for the number of charging stations, Level 1 charging station is predominant over Level 2 and Level 3. Level 3 has the least number of charging stations. Notice that it takes a much longer time to finish charging for Level 1, so Level 1 service provider must place more charging stations to meet the average delay probability constraint. The difference in quantity also reveals that the service providers have different marketing strategies. Service provider 1 tries to place the charging stations evenly across the entire area, while service provider 3 is more likely to place the charging stations at some ``hot" locations.

3. The number of charging stations grows almost linearly with the number of EVs except for the initial stage. At the initial stage, Level 1 and Level 2 service providers tend to place more charging stations than the next stages. This is because service providers must place more charging stations to meet the average service coverage constraints. As the number of charging stations increases, however, the service coverage constraint is less of a concern for the service providers.

4. Service providers prefer clustering instead of spatial separation. The three service providers have segmented the EV charging market by providing three different products (different charging level services) in terms of voltage, current, charging speed and charging price. Due to product differentiation, they significantly soften the price competition so that they do not need to spatially separate from each other to further relax competition. This observation supports the opinions in \cite{airmen}-\cite{jbaum} that firms do not have to maximize differentiation in every characteristic of the product. In general, differentiation in one dominant characteristic is sufficient to soften price competition.

%
%

%
%
%
%
%
%
%
%
%
%

%
%
\chapter{STOCHASTIC DYNAMIC PRICING FOR EV CHARGING STATIONS WITH RENEWABLES INTEGRATION AND ENERGY STORAGE}

\section{Overview}
This chapter studies the problem of stochastic dynamic pricing and energy management policy for electric vehicle (EV) charging service providers. In the presence of renewable energy integration and energy storage system, EV charging service providers must deal with multiple uncertainties --- charging demand volatility, inherent intermittency of renewable energy generation, and wholesale electricity price fluctuation. The motivation behind our work is to offer guidelines for charging service providers to determine proper charging prices and manage electricity to balance the competing objectives of improving profitability, enhancing customer satisfaction, and reducing impact on power grid in spite of these uncertainties. We propose a new metric to assess the impact on power grid without solving complete power flow equations. To protect service providers from severe financial losses, a safeguard of profit is incorporated in the model. Two algorithms --- stochastic dynamic programming (SDP) algorithm and greedy algorithm (benchmark algorithm) --- are applied to derive the pricing and electricity procurement policy. A Pareto front of the multi-objective optimization is derived. Simulation results show that using SDP algorithm can achieve up to 7\% profit gain over using greedy algorithm. Additionally, we observe that the charging service provider is able to reshape spatial-temporal charging demands to reduce the impact on power grid via pricing signals.

\section{Motivation and Related Work}
Electric vehicles (EVs) exhibit many advantages over fossil fuel driven vehicles in terms of operation and maintenance cost, energy efficiency, and gas emission \cite{asimpson, mhajian, rsioshansi}. However,  the fear of limited driving distance (range anxiety) is hanging over EV drivers' heads like the Sword of Damocles. To alleviate this range anxiety, the capacity of on-board battery should be increased and more EV charging stations should be deployed. Intensive research work has been carried out to study how to strategically deploy charging stations \cite{cluo1, cluo2, yiz2, alam, yli}. Currently, EV charging service is primarily provided for free as one of the employee benefits in some organizations or as a perk to those owners of some specific EV models (e.g. Tesla). There is a lack of viable and profitable pricing and energy management model for public charging stations. Our goal is to offer guidelines for charging service providers to make informed and insightful decisions on pricing and electricity procurement by jointly optimizing multiple objectives under uncertainties.

There is a growing literature aimed at providing guidelines for economic operation of EV charging stations. In \cite{yguo1, yguo2}, the authors studied a dynamic pricing scheme to improve the revenue of an EV parking deck. However, their model did not take into account customer satisfaction and the impact on power grid due to EV charging. In \cite{jfoster, shan, esortomme, sbeer}, several algorithms have been proposed for a power aggregator to manage EV charging loads and submit bids to electricity market to provide regulation service (RS). Game theory based approaches have been used to model the interplay among multiple EVs or between EVs and power grid in \cite{yhan, aovalle, hnguyen, rcouillet}.  Yan \emph{et al.} presented a multi-tier real time pricing algorithm for EV charging stations to encourage customers to shift their charging schedule from peak period to off-peak period \cite{qyan}. Nevertheless, they did not consider that some customers may strategically change their charging schedule in response to pricing signals. In \cite{dban}, Ban \emph{et al.} employed multi queues to model the arrivals and departures of EVs among multiple charging stations. Pricing signals were used to guide EVs to different charging stations to satisfy the predefined quality of service (QoS); but the interactions between EV charging and power grid was not analysed in their model.  A distributed network cooperative method was proposed to minimize the charging cost of EVs while guaranteeing that the aggregated load satisfies safety limits \cite{nrahbariasr}. Their model, however, did not incorporate renewable energy generation and consider charging demand volatility.

In our model, we take a comprehensive view of these interweaving issues pertaining to EV charging pricing and energy management. Specifically, we formulate our problem to simultaneously optimize multiple objectives --- improving the profit, enhancing the customer satisfaction, and reducing the impact on power grid in the light of renewable energy generation and energy storage. Our model takes into account multiple uncertainties including charging demand volatility, inherent intermittency of renewable energy generation, and real time wholesale electricity price fluctuation. For each type of uncertainty, an appropriate model is proposed and incorporated in the overall optimization framework. Finally, a stochastic dynamic programming (SDP) algorithm is employed to derive the charging prices and the electricity procurement from the power grid for each planning horizon. Besides, SDP algorithm has been used for water reservoir operation in \cite{cozelkan, akerr}.  In terms of the electricity retail market, a game theory based dynamic pricing scheme is studied in \cite{ljia} which also takes into account renewable integration and local storage.

The main contributions of our work are as follows:
\begin{itemize}
\item We proposed a multi-objective optimization framework to solve the problem, and the solutions provide us insights into how to make a tradeoff among multiple objectives of the profitability, the customer satisfaction, and impact on power grid, and offer guidance to set charging prices to balance the charging demand across the power system.
\item We used Newton's method to derive a fast-computing metric to assess the impact of EV charging on power grid, which frees us from solving the complete nonlinear power flow equations. This metric also can be used to analyze other electric load's impact on power grid.
\item We derived the active power and reactive power sensitivities for the load buses in a power system which can serve as a guideline for EV charging station placement to alleviate the charging stress on the power grid.
\item In terms of market risk, we introduced a safeguard of profit for EV charging service providers, which raises a warning when the profit is likely to reach a dangerous threshold. This mechanism is beneficial for the charging service provider to safely manage its capital and avoid severe financial losses.
\end{itemize}

\section{Preliminary on Dynamic Programming}
 DP is an effective tool to solve complex optimization problems by partitioning it into multiple simpler subproblems \cite{bertsekas, nemhauser}. DP is used to solve the problems that have optimal substructures. Having formulated an optimization problem with a large number of variables and constraints, we need to find an efficient approach to solve the problem. DP takes a sequential or multistage decision process containing independent variables and converts it into a series of simpler subproblems, each involving only a few variables. After solving the subproblems, we can then obtain the solution to the original problem by combining the results of the subproblems.

\subsection{Basic Model}
The basic model for DP has two assumptions: (1) an underlying discrete-time dynamic system, and (2) a cost function that is additive over time. For a deterministic system, the system dynamics, under the influence of decisions at the discrete instants of time, is expressed as follows

\begin{equation}
x_{k+1}=f_k(x_k,u_k),\;\;\;k=0,1,2,\cdots,N-1,
\end{equation}
where
\begin{itemize}
\item $k$: index of the discrete time instant,
\item $x_k$: the state of the system at time $k$,
\item $u_k$: the decision variable at time $k$,
\item $N$: the total number of horizons,
\end{itemize}
and $f_k$ is the system state evolution function.

The cost incurred at time $k$ is denoted by $g_k(x_k,u_k)$, and the total cost is obtained by summing the cost of all horizons

\begin{equation}
g_N(x_N)+\sum_{k=0}^{N-1}g_k(x_k,u_k),
\end{equation}
where $g_N(x_N)$ is a terminal cost incurred at the end of this process. The optimization is taken over the control variables $u_0,u_1,u_2,\cdots,u_{N-1}$.

\begin{equation}
J_0(x_0)=\max_{u_0,u_1,\cdots,u_{N-1}}\left[g_N(x_N)+\sum_{k=0}^{N-1}g_k(x_k,u_k)\right].
\end{equation}
Apply the state evolution function $x_{k+1}=f_k(x_k,u_k)$, we can rewrite the optimization problem as

\begin{equation}
J_0(x_0)=\max_{u_0,u_1,\cdots,u_{N-1}}\left[g_N(x_{N-1},u_{N-1})+\sum_{k=0}^{N-1}g_k(x_0,u_0,u_1, \cdots ,u_k)\right].
\end{equation}
Since horizon 0 does not depend on $u_1,\cdots,u_{N-1}$, we can decompose the optimization problem in the following way

\begin{equation}
\begin{aligned}
J_0(x_0)&=\max_{u_0}\left[g_0(x_0,u_0)+\max_{u_1,\cdots,u_{N-1}}\left(\sum_{k=1}^{N-1}g_k(x_1,u_1,u_2,\cdots,u_k)+g_N(x_{N-1},u_{N-1})\right)\right]\\
&=\max_{u_0}\left[g_0(x_0,u_0)+J_1(x_1)\right]\\
&=\max_{u_0}\left[g_0(x_0,u_0)+J_1(f_0(x_0,u_0))\right].
\end{aligned}
\end{equation}

\subsection{Principle of Optimality}
\noindent $\mathbf{Principle\;of\;Optimality}$: Let $\pi^*=\{u_0^*,u_1^*,\cdots,u_{N-1}^*\}$ be an optimal policy for the basic problem. Consider the subproblem where we are at $x_i$ at horizon $i$ and wish to minimize the ``cost-to-go" from horizon $i$ to horizon $N-1$

\begin{equation}
g_N(x_N)+\sum_{k=i}^{N-1}g_k(x_k,u_k).
\end{equation}
Then the truncated policy $\{u_i^*,u_{i+1}^*,\cdots,u_{N-1}^*\}$ is optimal for the subproblem.

A proof of the principle of optimality (by contradiction) simply states that if the remaining decisions were not optimal then the whole policy could not be optimal. Specifically, if the truncated policy $\{u_i^*,u_{i+1}^*,\cdots,u_{N-1}^*\}$ was not optimal as stated, we would be able to further reduce the cost by switching to an optimal policy for the subproblem once we reach $x_i$.

The principle of optimality suggests that the optimal policy can be constructed in a piecemeal manner. First, we construct an optimal policy for the ``tail subproblem" involving the last horizon, then we extend the optimal policy to the ``tail subproblem" involving the last two horizons, and continue in this manner until the optimal policy for the entire problem is constructed.

\section{Problem Formulation}
\begin{figure}[htbp]
\centering
\includegraphics[width=1\textwidth]{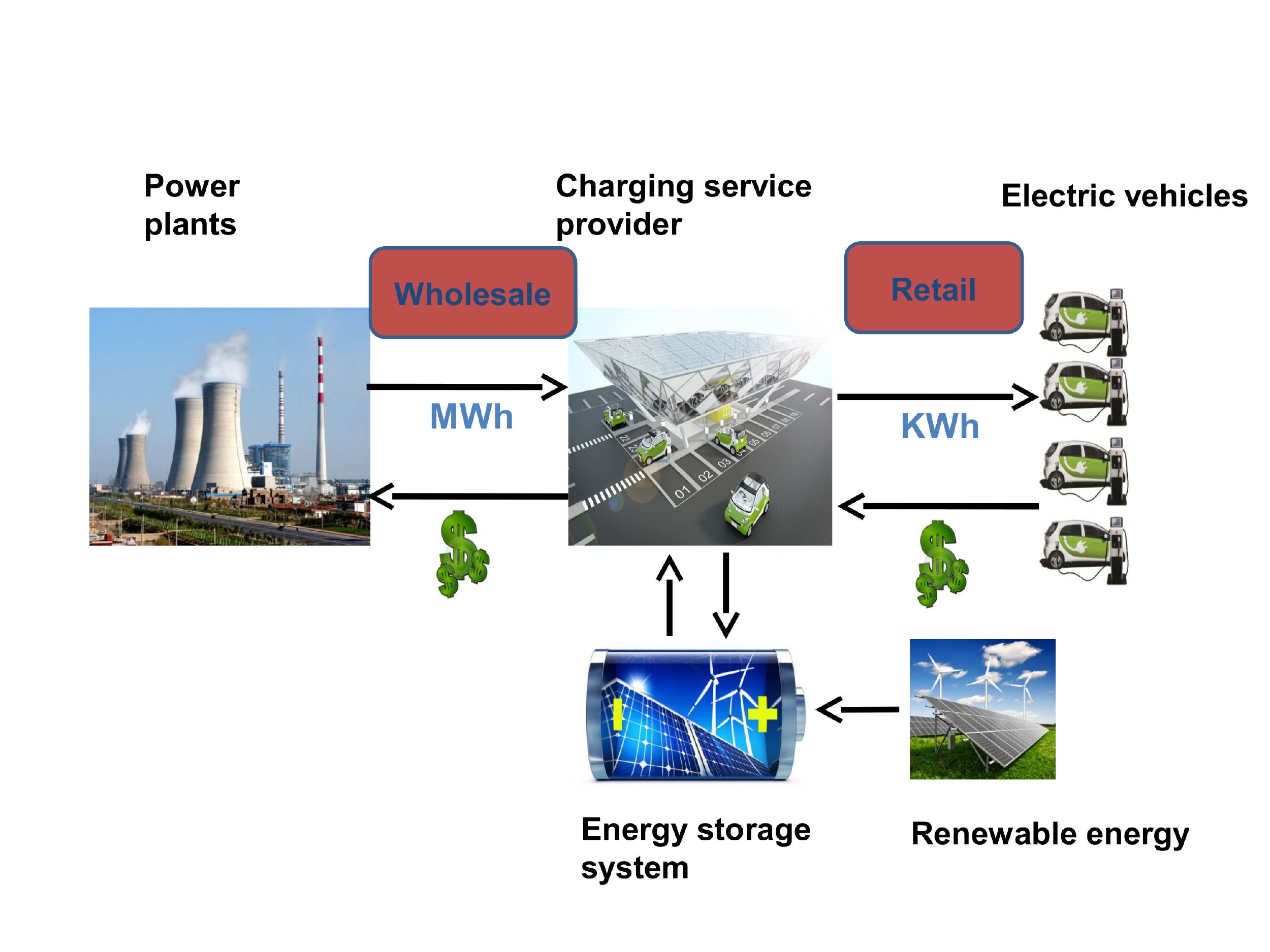}
\caption{The EV Charging Market}
\label{ev_market}
\end{figure}

In this study, we assume that an EV charging service provider operates a set of charging stations within a large region. As a mediator between the wholesale market and end customers (EVs), the charging service provider procures electricity from the wholesale market and resells it to EVs. We also assume that the service provider is able to harvest renewable energy (i.e. solar or wind power) and save it in an energy storage system. An overview of the EV charging service provider's model is illustrated in Fig. \ref{ev_market}.

\subsection{Profit of Charging Service Provider}
In the United States, the Independent System Operator (ISO) or the Regional Transmission Organization (RTO) collects supply offers from power plants and demand bids from load serving entities (LSEs) or market participants, calculates the day-ahead wholesale prices and real time spot prices, coordinates and monitors the economic dispatch of electricity across a vast region \cite{rhuisman, ghamoud, mventosa}. We assume that the charging service provider is an LSE, who purchases electricity from the wholesale real time market and resells it to EVs. Let $\mathcal{S}=\{s_1,s_2,\cdots,s_L\}$ denote the charging stations operated by the service provider. A day is divided into $K$ planning horizons. At the start of each horizon, the service provider will publish new charging prices during this horizon. Price differentiation is allowed across charging stations.  Let $\mathcal{P}=\{p_{k1},p_{k2},\cdots,p_{kL}\},k=1,2,\cdots,K$ denote the charging prices in the $k$-th horizon, and $o_k$ denote electricity procurement from the wholesale real time market. We use wholesale real time electricity market prices in our theoretical analysis. Let $\mathcal{C}=\{c_1,c_2,\cdots,c_K\}$ represent wholesale real time electricity prices. In addition, we assume that the service provider has an energy storage system with capacity $E$ MWh. Let $I_k$ denote the electricity in the storage at the beginning of the $k$-th horizon, and $u_k$ be the renewable energy generation during the $k$-th horizon. The profit made in the $k$-th horizon is given by

\begin{equation}
\begin{aligned}
W_k&=\sum_{j=1}^Lp_{kj}d_{kj}-c_ko_k-\\
&\eta_s(I_k+\eta_cu_k+\eta_co_k-\frac{1}{\eta_d}\sum_{j=1}^Ld_{kj} + w_k),
\end{aligned}
\end{equation}
where $d_{kj}$ corresponds to the charging demand (electricity consumption) at the $j$-th station in the $k$-th horizon, $\sum_{j=1}^Lp_{kj}d_{kj}$ is the total revenue, $c_ko_k$ is the cost of electricity procurement, and $\eta_s$(\$/MWh) is the unit storage cost, which includes capital cost and maintenance cost. Besides, $\eta_c$ ($0<\eta_c<1$) and $\eta_d$ ($0<\eta_d<1$) are charging efficiency and discharging efficiency, respectively. And $w_k$ is the process noise of the energy storage system, which has a Gaussian distribution with zero mean and variance $\sigma_w^2$.

\subsection{Customer Satisfaction}
\begin{figure}[htbp]
\centering
\includegraphics[width=1\textwidth]{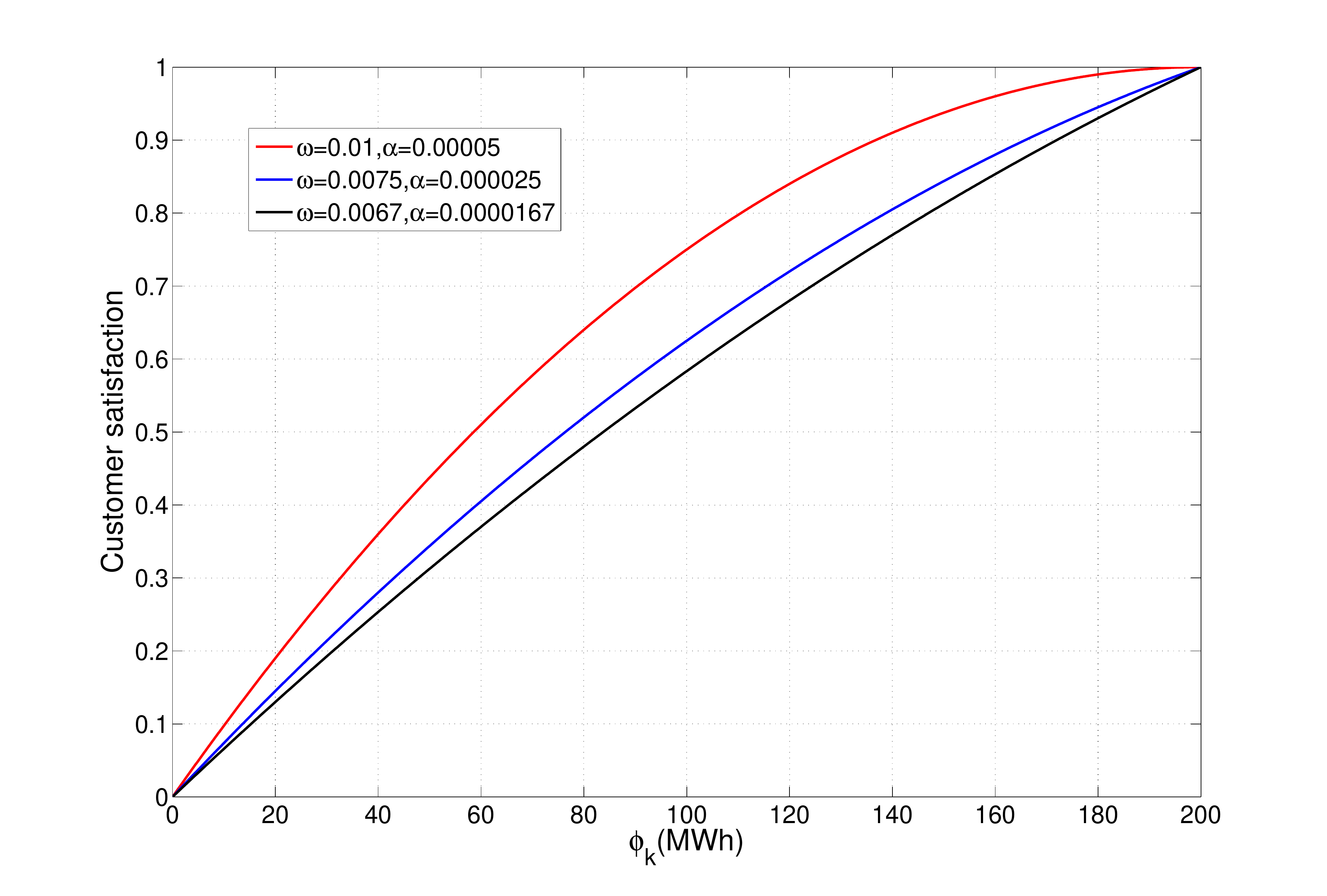}
\caption{Sample Customer Satisfaction Functions ($E=200$)}
\label{satisfaction}
\end{figure}

Customer satisfaction helps to build up customer loyalty, which can reduce the efforts to allocate market budgets to acquire new customers. Poor customer satisfaction will discourage people to purchase EVs, affecting the development of entire EV industry. Customer satisfaction is one of the objectives in our multi-objective optimization framework. Several customer satisfaction evaluation methods have been investigated in \cite{pyang, mfahrioglu, rfaranda}. In this paper, we consider the market-level customer satisfaction instead of the individual-level satisfaction. We use a quadratic function to formulate the overall customer satisfaction of all EVs in a horizon, namely,

\begin{equation}
\label{satisfactioneq}
G_k=-\frac{\alpha}{2}\phi_k^2+\omega\phi_k,\;0\leq \phi_k \leq E
\end{equation}
where $E$ is the electricity storage capacity, $\omega$ and $\alpha$ are shape parameters, $\phi_k$ is the aggregated charging demand (electricity consumption) of all EVs in the $k$-th horizon which is defined as,

\begin{equation}
\phi_k=\sum_{j=1}^Ld_{kj}.
\end{equation}
Eq. (\ref{satisfactioneq}) with different shape parameters is plotted in Fig. \ref{satisfaction}. In plotting Fig. \ref{satisfaction}, we choose the shape parameters $\alpha$ and $\omega$ such that the concave function $G_k$ has a minimum of 0, which indicates that EV drivers have the least satisfaction, and a maximum of 1, which indicates that they have the most satisfaction. Note that Eq. (\ref{satisfactioneq}) is a non-decreasing function with a non-increasing first order derivative. This implies that customer satisfaction will always grow as the total charging demand $\phi_k$ increases, but the growth rate will decrease and customer satisfaction tends to get saturated as the total charging demand approaches the storage capacity $E$. This is a standard assumption following the law of diminishing marginal utility (Gossen's First Law) in economics \cite{gheinrich}.

\subsection{Impact on Power Grid}
Large-scale EV charging presents a substantial load to power networks \cite{hturker, kkumar}. Many studies have shown that uncoordinated EV charging can affect the normal operation of power grid in terms of severe power loss, voltage variation, frequency deviation, and harmonics problems \cite{jlopes, kschneider, wsu, pkundur, renac}. Usually, grid frequency can be well maintained either by the power generator side using automatic gain control (AGC) \cite{pkundur} or by the load side using certain demand response techniques \cite{czhao, mmoghadam}. In our study here, we only consider the impact of voltage variation (magnitude and phase). In addition, we assume that a higher-level entity like an aggregator or ISO/RTO can take care of network transmission constraint issues within the power system under its supervision, so the EV charging service provider does not need to worry about transmission constraint problem. Let $F_k$ denote the impact of EV charging on power grid at the $k$-th horizon.

\begin{equation}\label{impact}
F_k=f(d_{k1},d_{k2},\cdots,d_{kL}),
\end{equation}
where $d_{kj}$ is the charging demand at the $j$-th charging station in the $k$-th horizon, and $f(.)$ is a function to be discussed in Section \rom{4}. Function $f(.)$ should reflect the basic assumption that the impact on power grid increases when the charging demands increase.

\subsection{Multi-Objective Optimization Framework}
A multi-objective optimization problem arises naturally from the fact that the charging service provider needs to balance multiple competing objectives --- maximizing profit, maximizing customer satisfaction, and minimizing the impact on power grid. For the $k$-th horizon, we formulate the multi-objective optimization as follows,

\begin{equation}\label{wholeproblem}
\begin{aligned}
&\max_{\mathbf{X}_k}\left\{\mathbb{E}(W_k),\mathbb{E}(G_k),\mathbb{E}(-F_k)\right\}\\
&\;\;\;\;\textrm{s.t.  } \mathbf{X}_k\in U(\mathbf{X}_k),
\end{aligned}
\end{equation}
where $\mathbf{X}_k=[p_{k1},p_{k2},\cdots,p_{kL},o_k]^{\textrm{T}}$ is the vector of decision variables, and $\mathbb{E}(.)$ represents the expectation operation.

Clearly, there are several approaches to solve multi-objective optimization problems: weighted sum approach, adaptive weighted sum approach, $\epsilon$-constraint approach, a priori approach and a posteriori approach \cite{clhwang, kmiettinen, ikim}, among others. The weighted sum approach is not suitable for obtaining the whole Pareto front if the main objective function is not convex. In this paper, we use an adaptive weighted sum approach discussed in \cite{ikim} to derive the Pareto front of Eq. (\ref{wholeproblem}). The main idea of the adaptive weighted sum approach is that firstly we use the ordinary weighted sum to obtain the basic shape of Pareto front, and then refine it by recursively reducing mesh size within the Pareto front. First, we rewrite the problem as follows,

\begin{equation}
\begin{aligned}
&\max_{\mathbf{X}_k}\mathbb{E}(\Pi_k)=\mathbb{E}\left\{\lambda_1\frac{W_k}{W_k^{\textrm{max}}}+\lambda_2\frac{G_k}{G_k^{\textrm{max}}}
-\lambda_3\frac{F_k}{F_k^{\textrm{max}}}\right\}\\
&\;\;\;\;\textrm{s.t.  } \mathbf{X}_k\in U(\mathbf{X}_k),
\end{aligned}
\end{equation}
where $\lambda_1,\lambda_2,$ and $\lambda_3$ are nonnegative coefficients, satisfying the constraint of $\lambda_1+\lambda_2+\lambda_3=1$.  Different weight vectors ($\lambda_1, \lambda_2, \lambda_3$) generates different convex Pareto optima. The non-convex part of this Pareto front can be found in the refinement phase. Additionally, $W_k^{\textrm{max}}$, $G_k^{\textrm{max}}$, and $F_k^{\textrm{max}}$ are the maximum values of each objective function in the $k$-th horizon.

Our ultimate goal is to maximize the aggregated utility across multiple horizons.

\begin{equation}\label{eq1}
\begin{aligned}
&(\mathbf{X}_1^*,\cdots,\mathbf{X}_K^*)=\argmax_{\mathbf{X}_1,\cdots,\mathbf{X}_K}
\left\{\sum_{k=1}^K\mathbb{E}(\Pi_k)\right\},\\
&\;\;\;\;\;\;\textrm{s.t.  } \mathbf{X}_k\in U(\mathbf{X}_k), k=1,2,\cdots,K.
\end{aligned}
\end{equation}

To solve this multi-horizon and multi-objective optimization problem, we face several challenges: (1) How do we accurately estimate the charging demand $d_{kj}$ at each charging station? (2) How do we develop an appropriate metric to assess the impact on power grid defined in Eq. (\ref{impact})? (3) How should we incorporate a safeguard of profit to prevent severe financial losses? (4) How can we solve this complex optimization problem in an efficient manner? In the following sections, we will address these challenges in details.

\section{Charging Demand Estimation}
In practice, EV drivers will adjust their charging demands and charging schedules in response to charging prices. The charging demand function $d_{kj}$ thus should characterize customers' response to price fluctuations. In our work, an online linear regression model \cite{rchristensen, adobson} is employed to predict the charging demand $d_{kj}$. For each charging station, the predicted charging demand is defined as

\begin{equation}\label{lrm}
\begin{cases}
d_{k1}=\gamma_{0,1}-\gamma_{1,1}p_{k1}+\gamma_{2,1}p_{k2}+\cdots+\gamma_{L,1}p_{kL}+\epsilon_{k1},\\
d_{k2}=\gamma_{0,2}+\gamma_{1,2}p_{k1}-\gamma_{2,2}p_{k2}+\cdots+\gamma_{L,2}p_{kL}+\epsilon_{k2},\\
\vdots\\
d_{kL}=\gamma_{0,L}+\gamma_{1,L}p_{k1}+\gamma_{2,2}p_{k2}+\cdots-\gamma_{L,L}p_{kL}+\epsilon_{kL},\\
\end{cases}
\end{equation}
where $\gamma_{0,j}(j=1,2,\cdots, L)$ is the intercept of the $j$-th linear regression equation, $\gamma_{i,j}=\gamma_{j,i}(i\neq j)$ are the cross-price elasticity coefficients, reflecting how the change of the charging price at station $j$ can influence the charging demand at station $i$, and $\gamma_{i,i}$ is the self-price elasticity coefficient, reflecting how the change of the charging price of station $i$ can influence its own charging demand. Finally $\epsilon_{kj}(j=1,2,\cdots,L)$ is assumed to be an independent Gaussian random variable with mean 0 and variance $\sigma_{kj}^2$. The variable $\epsilon_{kj}$ captures the unknown random charging demand which cannot be characterized by the linear terms.

Recursive least square (RLS) algorithm is a common method applied to estimate the coefficients in Eq. (\ref{lrm}) using historical data \cite{jproakis, dbertsimas}. Let $\mathbf{Y}_j=[\gamma_{0,j},\gamma_{1,j},\cdots,\gamma_{L,j}]^{\textrm{T}}$ denote the vector of price elasticity coefficients related to the $j$-th charging station. Applying RLS, we have the following update equations,

\begin{equation}\label{update1}
\begin{aligned}
\begin{cases}
&e_{kj}=d_{kj}-\mathbf{P}_k^{\textrm{T}}\mathbf{Y}_j,\\
&g_{kj}=\frac{\mathbf{H}_{(k-1)j}\mathbf{P}_k}{\nu+\mathbf{P}_{k}^{\textrm{T}}\mathbf{H}_{(k-1)j}\mathbf{P}_k},\\
&\mathbf{H}_{kj}=\nu^{-1}\mathbf{H}_{(k-1)j}-g_{kj}\mathbf{P}_k^{\textrm{T}}\nu^{-1}\mathbf{P}_k,\\
&\mathbf{Y}_j\leftarrow \mathbf{Y}_j+e_{kj}g_{kj},\\
\end{cases}
\end{aligned}
\end{equation}
where $\nu$ is the forgetting factor. Besides, $H_{0j}$ is initialized to be an identity matrix and $P_0$ is initialized to be an all-zero vector. In addition, the estimate for variance $\sigma_{kj}^2$ is given by

\begin{equation}\label{update2}
\begin{aligned}
\begin{cases}
&m_{kj}=\nu m_{(k-1)j}+\epsilon_{kj},\\
&n_{kj}=\nu n_{(k-1)j}+ 1,\\
&\bar{\epsilon}_{kj}=m_{kj}/n_{kj},\\
&u_{kj}=(\frac{m_{kj}-1}{m_{kj}})^2+(\frac{1}{m_{kj}})^2,\\
&v_{kj}=m_{kj}(1-u_{kj}),\\
&\sigma_{j}^2\leftarrow\frac{1}{v_{kj}}[(\nu v_{kj})\sigma_{j}^2+\frac{m_{kj}-1}{m_{kj}}(\bar{\epsilon}_{kj}-\epsilon_{kj})^2],
\end{cases}
\end{aligned}
\end{equation}
where $m_{kj}$ and $n_{kj}$ are initialized to be 0.

Eq. (\ref{lrm}) can characterize the spatial-temporal variation of charging demand. Different locations may have different charging demands. Thus, we use different linear regression equations to model these geographically separated charging stations. Furthermore, the price elasticity coefficients are updated continually using RLS algorithm defined in Eq. (\ref{update1}) and Eq. (\ref{update2}). The forgetting factor $\nu$ enables us to capture the most recent trend in charging demand and forget the outdated information. Thus, the RLS updating mechanism is able to track charging demand fluctuation over time.

\section{Impact on Power Grid from EV Charging}
For power flow analysis, we assume that an $N$-bus power network has 1 slack bus, $M$ load buses (PQ buses), and $N-M-1$ voltage-controlled buses (PV buses) \cite{sfrank}. Three phase balance operation and per-unit (p.u.) system are basic assumptions here. Charging stations are deployed across different PQ buses. Solving the power flow requires determining $N-1$ voltage phases (corresponding to PQ buses and PV buses) and $M$ voltage magnitude (corresponding to PQ buses). This is done by solving $N+M-1$ nonlinear power flow equations ($N-1$ active power equations and $M$ reactive power equations). The active and reactive power flow equations for each bus are given as follows,

\begin{equation}\label{networkflow1}
P_i=v_i\sum_{k=1}^Nv_k(G_{ik}\cos(\delta_{i}-\delta_{k})+B_{ik}\sin(\delta_{i}-\delta_{k})),
\end{equation}
\begin{equation}\label{networkflow2}
Q_i=v_i\sum_{k=1}^Nv_k(G_{ik}\sin(\delta_{i}-\delta_{k})-B_{ik}\cos(\delta_{i}-\delta_{k})),
\end{equation}
where $v_i$ and $\delta_i$ are, respectively, voltage magnitude and phase at the $i$-th bus; $P_i$ and $Q_i$ are real power and reactive power injections at the $i$-th bus; $G_{ik}$ and $B_{ik}$ are, respectively, conductance and susceptance of the $ik$-th element of the bus admittance matrix.

An increasing EV charging demand at PQ buses will lead to network-wide voltage variation (magnitude and phase) if the network does not provide sufficient active power and reactive power. We will use voltage variation as a metric to assess the impact of EV charging on power grid. Applying Newton's method, we can calculate the linear approximation of voltage variation in the following way,

\begin{equation}
\begin{aligned}
\left[
    \begin{array}{c}
    \mathbf{\Delta V}\\
    \mathbf{\Delta \Phi}\\
    \end{array}
\right]&=\left[
    \begin{array}{cc}
    \mathbf{\frac{\partial P}{\partial V}} & \mathbf{\frac{\partial P}{\partial \Phi}}\\
    \mathbf{\frac{\partial Q}{\partial V}} & \mathbf{\frac{\partial Q}{\partial \Phi}}\\
    \end{array}
\right]^{-1}
\left[
    \begin{array}{c}
    \mathbf{\Delta P}\\
    \mathbf{\Delta Q}\\
    \end{array}
\right]\\
&=\mathbf{J}^{-1}
\left[
    \begin{array}{c}
    \mathbf{\Delta P}\\
    \mathbf{\Delta Q}\\
    \end{array}
\right],
\end{aligned}
\end{equation}
where $\mathbf{\Delta V}$ and $\mathbf{\Delta \Phi}$ are, respectively, vectors of magnitude variation and phase variation; $\mathbf{\Delta P}$ and $\mathbf{\Delta Q}$ are, respectively, vectors of increased active power and reactive power due to EV charging. In addition, $\mathbf{\frac{\partial P}{\partial V}}$ and $\mathbf{\frac{\partial P}{\partial \Phi}}$ are partial derivatives of active power with respect to voltage magnitudes and phases, and $\mathbf{\frac{\partial Q}{\partial V}}$, $\mathbf{\frac{\partial Q}{\partial \Phi}}$ are partial derivatives of reactive power with respect to voltage magnitudes and phases. In addition, $\mathbf{J}^{-1}$ is the inverse of Jacobian matrix from power flow equations, which is given by

\begin{equation}
\mathbf{J}^{-1}=
\left[
    \begin{array}{cccc}
    b_{1,1} & b_{1,2} & \cdots & b_{1,N+M-1}\\
    b_{2,1} & b_{2,2} & \cdots & b_{2,N+M-1}\\
    \vdots & & & \vdots\\
    b_{N+M-1,1} & b_{N+M-1,2} & \cdots & b_{N+M-1,N+M-1}\\
    \end{array}
\right],
\end{equation}

Let the sequence $[a_1, a_2,\cdots,a_L]$ denote the bus indexes of all charging stations in the power network. For instance, $a_i(i=1,2\cdots,L)$ means that the $i$-th charging station is fed by the $a_i$-th bus in the power network.

Finally, we use the 2-norm voltage variation (magnitude and phase) to assess the impact of EV charging on power grid,

\begin{equation}
F_k=\left|\left|
    \mathbf{J}^{-1}
\left[
    \begin{array}{c}
    \mathbf{\Delta P}\\
    \mathbf{\Delta Q}\\
    \end{array}
\right]\right|\right|^2.
\end{equation}

Moreover, we denote $S_i^{\textrm{Ac}}$ and $S_i^{\textrm{Re}}$ as the active power sensitivity and reactive power sensitivity of the $i$-th PQ bus. And $S_i^{\textrm{Ac}}$ is defined as follows,

\begin{equation}\label{powersensitivity}
S_i^{\textrm{Ac}}=\left|\left|
    \mathbf{J}^{-1}
\left[
    \begin{array}{c}
    0\\
    \vdots\\
    0\\
    1\\
    0\\
    \vdots\\
    0
    \end{array}
\right]\right|\right|^2,
\end{equation}
where $S_i^{\textrm{Ac}}$ is the 2-norm voltage variation when the active power injection of the $i$-th PQ bus is increased by 1 W. Thus, 1 W is the $i$-th entry in the column vector in Eq. (\ref{powersensitivity}). Similarly, $S_i^{\textrm{Re}}$ is defined as the 2-norm voltage variation when the reactive power injection of the $i$-th PQ bus is increased by 1 var. A larger value of $S_i^{\textrm{Ac}}$ or $S_i^{\textrm{Re}}$ indicates that the PQ bus has a lower tolerance to load variation and more likely to disturb the whole network.

\section{Stochastic Dynamic Programming for Pricing and Electricity Procurement}
At first, this section introduces a safeguard of profit --- a minimum profit warning mechanism. In addition, major modules in SDP like renewable energy, real time wholesale electricity price, and system dynamics are discussed. Finally, we introduce the procedure to use SDP to derive pricing and electricity procurement policy.

\subsection{A Safeguard of Profit}
In practice, service providers make decisions on pricing and electricity procurement based on the estimated charging demands. Although Eq. (\ref{lrm}) provides a viable way to estimate the charging demand, uncertainties still exist in actual charging demands. This subsection aims to develop a safeguard of profit to remind that the charging service provider should make a certain amount of profit under severe circumstance of uncertainties. We incorporate the safeguard as a constraint in the optimization framework. Wherever the optimal solution touches this constraint (i.e. this constraint becomes active), a warning will be raised for the service provider. The constraint is given as follows,

\begin{equation}\label{risk}
\textrm{Prob}\left(W_k<W_{\textrm{min}}\right)<\zeta,
\end{equation}
where $W_k$ is the profit made in the $k$-th horizon, $W_{\textrm{min}}$ is a profit threshold, and $\zeta$ is a small positive number in the range of $(0,1)$. Eq. (\ref{risk}) specifies that the probability that the actual profit is less than the profit threshold should be less than $\zeta$.

Expanding $W_k$ and rearranging terms in Eq. (\ref{risk}) yields the following,

\begin{equation}
\textrm{Prob}\left(\mathbf{X}_k^{\textrm{T}}\mathbf{A}\mathbf{X}_k+\mathbf{B}^{\textrm{T}}\mathbf{X}_k+\mathbf{E}_k^{\textrm{T}}\mathbf{Z}_k+t_k<W_{\textrm{min}}\right)<\zeta,
\end{equation}
where $\mathbf{X}_k=[p_{k1},p_{k1},\cdots,p_{kL},o_k]^{\textrm{T}}$. Matrix A is given by
\begin{equation}
\mathbf{A}=
\left[
    \begin{array}{ccccc}
    -\gamma_{1,1} & \gamma_{1,2} & \cdots & \gamma_{1,L} & 0 \\
    \gamma_{2,1}  & -\gamma_{2,2} & \cdots & \gamma_{2,L} & 0 \\
    \vdots & \vdots & & \vdots &\vdots\\
    \gamma_{L,1} & \gamma_{L,2} & \cdots & -\gamma_{L,L} & 0 \\
    0 & 0 & \cdots& 0 & 0\\
    \end{array}
\right],
\end{equation}
and vector $\mathbf{B}$ is
\begin{equation}
\mathbf{B}=\left[\gamma_{0,1} + \frac{\eta_s}{\eta_d}\Gamma_1, \cdots, \gamma_{0,L} + \frac{\eta_s}{\eta_d}\Gamma_L, -c_k-\eta_s\eta_c\right]^{\textrm{T}},
\end{equation}
where $\Gamma_j$ is
\begin{equation}
\Gamma_j = -\gamma_{j,j} + \sum_{i = 1,i\neq j}^L\gamma_{j,i},
\end{equation}
and vector $E$ is
\begin{equation}
\mathbf{E}_k=\left[p_{k1} + \frac{\eta_s}{\eta_d}, \cdots,  p_{kL} + \frac{\eta_s}{\eta_d}, -\eta_s\right]^{\textrm{T}},
\end{equation}
and $\mathbf{Z}_k=[\epsilon_{k1},\epsilon_{k2},\cdots,\epsilon_{kL},w_k]^{\textrm{T}}$, and $t_k=\eta_s(\Gamma_0/\eta_d-I_k-\eta_cu_k)$, where $\Gamma_0$ is

\begin{equation}
\Gamma_0=\sum_{i=1}^L\gamma_{0,i}.
\end{equation}

Besides, we assume that $[\epsilon_{k1},\epsilon_{k2},\cdots,\epsilon_{kL},w_k]^{\textrm{T}}$ are independent Gaussian random variables. Thus, $\mathbf{E}_k^{\textrm{T}}\mathbf{Z}_k$ is also a Gaussian random variable with mean 0 and variance $\sum_{j=1}^L(p_{kj}+\eta_s/\eta_d)^2\sigma_{kj}^2+\eta_s^2\sigma_w^2$.

Finally, Eq. (\ref{risk}) can be rewritten as follows,
\begin{equation}\label{insurance}
\begin{aligned}
&\textrm{Prob}\left(\mathbf{E}_k^{\textrm{T}}\mathbf{Z}_k<W_{\textrm{min}}-\mathbf{X}_k^{\textrm{T}}\mathbf{A}\mathbf{X}_k-
\mathbf{B}^{\textrm{T}}\mathbf{X}_k-t_k\right)\\
&=\Phi\left(\frac{W_{\textrm{min}}-\mathbf{X}_k^{\textrm{T}}\mathbf{A}\mathbf{X}_k-
\mathbf{B}^{\textrm{T}}\mathbf{X}_k-t_k}{\sqrt{\sum_{j=1}^L(p_{kj}+\eta_s/\eta_d)^2\sigma_{kj}^2+\eta_s^2\sigma_w^2}}\right)
<\zeta,\\
\end{aligned}
\end{equation}
where $\Phi(.)$ is the cumulative distribution function (CDF) of a standard Gaussian random variable.

\subsection{Renewable Energy and Real Time Wholesale Price}
Literature abounds on various approaches to forecasting renewable energy, e.g., physical approach \cite{llandberg, mlange}, statistical approach \cite{yzli, eizgia}, and hybrid approach \cite{ggiebel}. In this paper, we use a Markov chain model \cite{jnorris}-\cite{pbremaud} which is a statistical approach, to demonstrate how renewable energy prediction is incorporated into our optimization model. In fact, other forecasting approaches can also be used in our model.

Markov chain characterizes the transition from the current renewable energy $u_k$ to the next $u_{k+1}$. We discretize renewable energy into $D$ levels, and the transition matrix at the $k$-th horizon is given by,

\begin{equation}
\mathbf{T}_k=
\left[
    \begin{array}{cccc}
    t_{k,1,1} & t_{k,1,2} & \cdots & t_{k,1,D}\\
    t_{k,2,1} & t_{k,2,2} & \cdots & t_{k,2,D}\\
    \vdots & & & \vdots\\
    t_{k,D,1} & t_{k,D,2} & \cdots & t_{k,D,D}\\
    \end{array}
\right],
\end{equation}
where $t_{k,i,j}$ is the transition probability of renewable energy from level $i$ to level $j$ in the $k$-th horizon, and $\sum_{j=1}^Dt_{k,i,j}=1$. All transition probabilities can be estimated from historical data.

Similar to renewable energy, real time wholesale price forecasting has also been extensively studied through time series analysis, machine learning, big data, or hybrid approach in \cite{yji, rweron}. Real time price forecasting is a topic beyond the technical scope of our paper. Thus, we do not study specific real time price forecasting approaches in this paper.

\subsection{Stochastic Dynamic Programming}
Eq. (\ref{eq1}) is a complex multi-variable optimization problem involving $K(L+1)$ variables. It may be mathematically cumbersome and difficult to solve in a brute-force manner. We observe that the original problem exhibits the properties of overlapping subproblems and optimal substructure, which can be solved efficiently using SDP. SDP solves a large-scale complex problem by partitioning it into a set of smaller and simpler subproblems \cite{gnemhauser, dbertsekas}. The solution to the original problem is constructed by solving and combining the solutions of subproblems in a forward or backward manner. In contrast to a brute-force algorithm, SDP can greatly reduce computation and save storage.

In a wholesale real time electricity market, electricity is sold on an hourly basis. So our problem should have a finite number of planning horizons with $K=24$. System dynamics are governed by the evolution of system states, under the influence of decision variables and random variables. In our case, system dynamics are expressed by the following equations

\begin{equation}
\begin{aligned}
&I_{k+1}=I_{k}+\eta_cu_k+\eta_co_k-\frac{1}{\eta_d}\phi_k+w_k,\\
&u_{k+1}=h(u_k,\upsilon_k),\\
\end{aligned}
\end{equation}
where $I_k$ represents electricity storage at the beginning of the $k$-th horizon, $u_k$ is renewable energy, $o_k$ is the electricity procurement, $\eta_c$ is the charging efficiency, $\eta_d$ is the discharging efficiency, and $\phi_k$ is the total charging demand. Besides, $w_k$ and $\upsilon_k$ are independent process noises for the energy storage system and the renewable energy generation.

The aggregated expected utility from the first horizon to the $K$-th horizon is given by

\begin{equation}
\mathbb{E}\left\{\Pi_{K+1}(I_{K+1},u_{K+1})+\sum_{k=1}^K\Pi_k(I_k,u_k)\right\},
\end{equation}
where $\Pi_{K+1}(I_{K+1},u_{K+1})$ is a terminal utility occurred at the end of this process, and the expectation is taken over  $\epsilon_{kj}(j=1,\cdots,L)$ defined in Eq. (\ref{lrm}), $w_k$, and $\upsilon_k$. Therefore, the maximum aggregated expected utility $J(I_1,u_1)$ is given by

\begin{equation}\label{eq2}
\begin{aligned}
&J_1(I_1,u_1) =\max_{\mathbf{X}_1,
\cdots,\mathbf{X}_K}\mathbb{E}\left\{\Pi_{K+1}+\sum_{k=1}^K\Pi_k\right\},\\
&s.t.\\
&\begin{cases}
\textrm{Prob}(W_k<W_{\textrm{min}})<\zeta\\
0\leq o_k\leq o_{\textrm{max}};k=1,2,\cdots,N\\
p_{kj}\geq0;j=1,2,\cdots,L\\
 I_k + u_k + o_k-\sum_{j=1}^Ld_{kj}\geq0\\
I_k + u_k+ o_k-\sum_{j=1}^Ld_{kj}\leq E\\
d_{kj}\geq0;j=1,2,\cdots,L.
\end{cases}
\end{aligned}
\end{equation}
Applying SDP we can partition the problem into multiple small subproblems, which can be calculated recursively as follows,

\begin{equation}
\begin{aligned}
J_k(I_k,u_k) &=\max_{\mathbf{X}_k\in U_k(\mathbf{X}_k)}\mathbb{E}\left\{\Pi_k+J_{k+1}(I_{k+1},u_{k+1})\right\}\\
&=\max_{\mathbf{X}_k\in U_k(\mathbf{X}_k)}\left\{\mathbb{E}\{\Pi_k(I_k,u_k)\}+\right.\\
&\;\;\;\;\;\;\;\;\;\;\;\;\;\;\left.\mathbb{E}\{J_{k+1}(I_{k+1},u_{k+1})\}\right\}.
\end{aligned}
\end{equation}

Furthermore, we can rewrite each subproblem into a nice quadratic form by combining like terms as follows,
\begin{equation}
\begin{aligned}
J_k(I_k,u_k)=&\max_{\mathbf{X}_k\in U_k(\mathbf{X}_k)}\Big\{\mathbb{E}\{\frac{1}{2}\mathbf{X}_k^{\textrm{T}}\mathbf{Q}\mathbf{X}_k\\
&+\mathbf{B}^{\textrm{T}}_k\mathbf{X}_k\}+ \mathbb{E}\{r_k\}\Big\},
\end{aligned}
\end{equation}
where $\mathbf{Q}$, $\mathbf{B}_k$, and $r_k$ are given by

\begin{equation}
\scalebox{0.8}{$
\mathbf{Q}=
\left[
    \begin{array}{cccc}
    -2\gamma_{1,1}\lambda_1 - \alpha\lambda_2\Gamma_1^2-2\lambda_3\sum_{j=1}^{N+M-1}\Theta_{1,j}^2 & \cdots & 2\gamma_{1,L}\lambda_1 - \alpha\lambda_2\Gamma_1\Gamma_L-2\lambda_3\sum_{j=1}^{N+M-1}\Theta_{1,j}\Theta_{L,j}&0\\
    2\gamma_{2,1}\lambda_1-\alpha\lambda_2\Gamma_2\Gamma_1 - 2\lambda_3\sum_{j=1}^{N+M-1}\Theta_{2,j}\Theta_{1,j}& \cdots & 2\gamma_{2, L}\lambda_1 -\alpha\lambda_2\Gamma_2\Gamma_L-2\lambda_3\sum_{j = 1} ^ {N+M-1}\Theta_{2,j}\Theta_{L,j}& 0 \\
    \vdots & & & \vdots\\
    2\gamma_{L,1}\lambda_1 - \alpha\lambda_2\Gamma_L\Gamma_1-2\lambda_3\sum_{j=1}^{N+M-1}\Theta_{L,j}\Theta_{1,j}& \cdots & -2\gamma_{L,L}\lambda_1 - \alpha\lambda_2\Gamma_L^2-2\lambda_3\sum_{j=1}^{N+M-1}\Theta_{L,j}^2& 0\\
    0 & \cdots & 0 & 0\\
    \end{array}
\right],$}
\end{equation}

\begin{equation}
\mathbf{B}_k =
\left[
    \begin{array}{c}
    \lambda_1(\gamma_{0,1}+\frac{\eta_s}{\eta_d}\Gamma_1) + \lambda_2\left(\omega\sum_{j=1}^L\gamma_{1,j}-\alpha\Gamma_0\Gamma_1\right)-
    2\lambda_3\sum_{j=1}^{N+M-1}\Theta_{0,j}\Theta_{1,j}\\
    \vdots\\
    \lambda_1(\gamma_{0,L}+\frac{\eta_s}{\eta_d}\Gamma_L)+ \lambda_2\left(\omega\sum_{j=1}^L\gamma_{L,j}-\alpha\Gamma_0\Gamma_L\right)-
    2\lambda_3\sum_{j=1}^{N+M-1}\Theta_{0,j}\Theta_{L,j}\\
    -(c_k+\eta_s\eta_c)\lambda_1\\
    \end{array}
\right],
\end{equation}

\begin{equation}
\begin{aligned}
r_k=&\lambda_1\eta_s(\Gamma_0/\eta_d-I_k-\eta_cu_k)+\lambda_2\left(\omega\Gamma_0-\frac{\alpha}{2}\left(\Gamma_0^2+\sum_{j=1}^L\sigma_{k,j}^2\right)\right)-\\
&\lambda_3\left(\sum_{j=1}^{N+M-1}\Theta_{0,j}^2+\lambda_3\sum_{j=1}^{N+M-1}\sum_{i=1}^Lb_{j,a_i}^2\sigma_{k,j}^2\right)
+\mathbb{E}_{u_{k+1}}\{J_{k+1}(I_{k+1},u_{k+1})\},
\end{aligned}
\end{equation}

\noindent where $\Theta_{n,j}$ is

\begin{equation}
\Theta_{n,j}=-b_{j,a_n}\gamma_{n,n}+\sum_{i=1,i\neq n}^Lb_{j,a_i}\gamma_{i,n},
\end{equation}
and $a_n$ is the bus index of the power network for the $n$-th charging station, and

\begin{equation}
\Theta_{0,j}=\sum_{i=1}^Lb_{j,a_i}\gamma_{0,i}.
\end{equation}

\begin{figure}[htbp]
\centering
\includegraphics[width=5.5in]{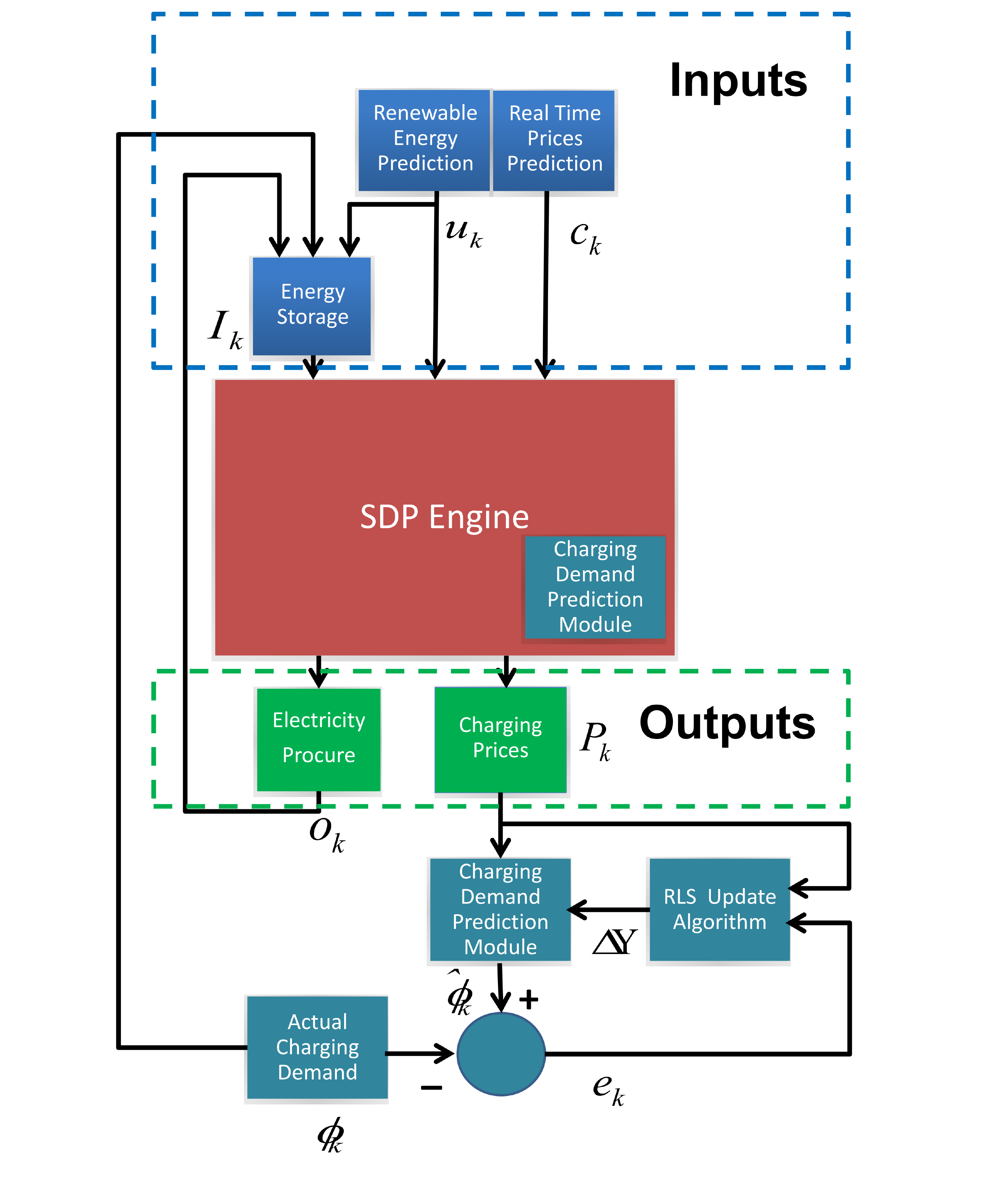}
\caption{Dynamic Pricing and Energy Management Algorithm}
\label{dpemalgo}
\end{figure}

Finally, $J_{k+1}(I_{k+1},u_{k+1})$ is the total aggregated utility starting from the $(k+1)$-th horizon to the $K$-th horizon. Fig. \ref{dpemalgo} illustrates the schematic of the entire optimization framework. The charging service provider should run the SDP engine at the beginning of every planning horizon.

\section{Simulations and Discussions}
\begin{figure}[b]
  \centering
    \includegraphics[width=4.5in]{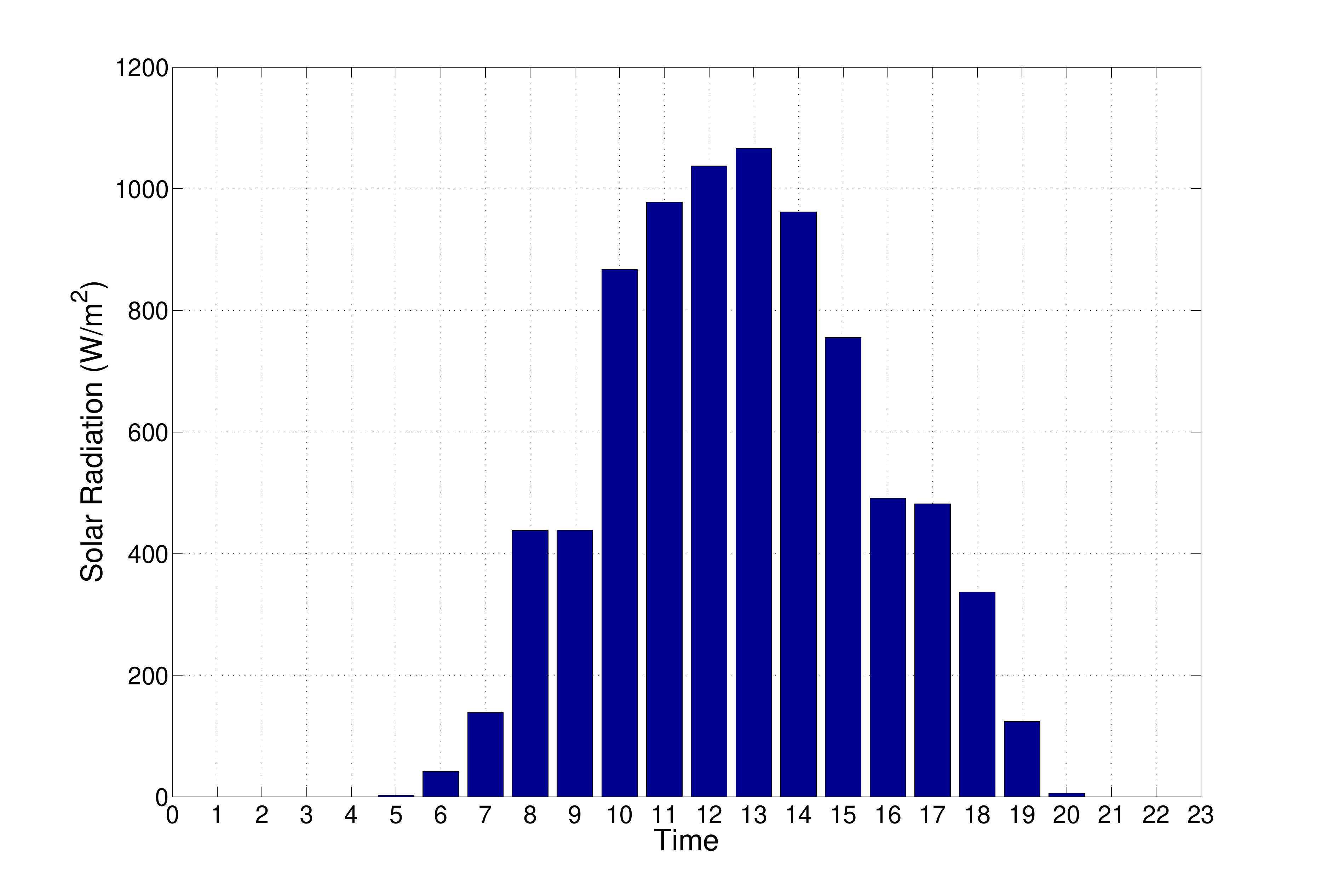}
    \caption{Typical Daily Solar Radiation}
    \label{solar}
  \end{figure}

The simulation coefficients are given in Table \ref{simulationpara}. Tesla's home rechargeable Lithium-ion battery system --- Powerwall has a 92.5\% round-trip DC efficiency with 100\% depth of discharge \cite{powerwall}. Eos Energy Storage has a battery-based energy storage with a round-trip efficiency of 75\% and a 100\% depth of discharge \cite{esoenergy}. In our simulation, we assume the charging efficiency $\eta_c$ and discharging efficiency $\eta_d$ are both 0.9. For simplicity, we use the day-ahead wholesale electricity price data from PJM \cite{pjmcom} to represent the real time wholesale price forecasting in the simulations, but other forecasting approaches can be used. In addition, we assume that the charging service provider procures electricity at a single locational marginal price (LMP). We use solar power to represent the renewable energy source. The solar radiation data is from National Renewable Energy Laboratory (NREL) \cite{emckenna}, and the typical daily solar radiation is depicted in Fig. \ref{solar}. Note that solar radiation begins at 6:00 am and ends at 8:00 pm. Additionally, we assume that solar cell efficiency is 20\%. We use IEEE 57 Bus Test case for the power network in our simulations \cite{testcase}.

\begin{table}[!htbp]\label{simulationpara}
\caption{SIMULATION PARAMETERS}
\begin{center}
\begin{tabular}{llll}
\hline
Coefficient & Description & Unit & Value\\

$N$ & Horizon number & - &24\\

$E$ & Energy storage capacity & MWh & 200\\

$\lambda_1$ & Profit weight & - & 0 to 1\\

$\lambda_2$ & Customer weight & - & 0 to 1\\

$\lambda_3$ & Impact weight & - & 0 to 1\\

$\zeta$ & Revenue safeguard prob. & - & 0.2\\

$\alpha$ & Shape parameter &-&5e-5\\

$\omega$ & Shape parameter &-&0.01\\

$\eta_s$ & Unit storage cost  & \$/MWh &0 to 4\\

$\eta_c$ & Charging efficiency & - & 0.9\\

$\eta_d$ & Discharging efficiency & - & 0.9\\

$\rho_0$ & Knee point threshold & - & 1\\
\hline
\end{tabular}
\end{center}
\end{table}

    \begin{figure}[b]
    \centering
    \includegraphics[width=4.5in]{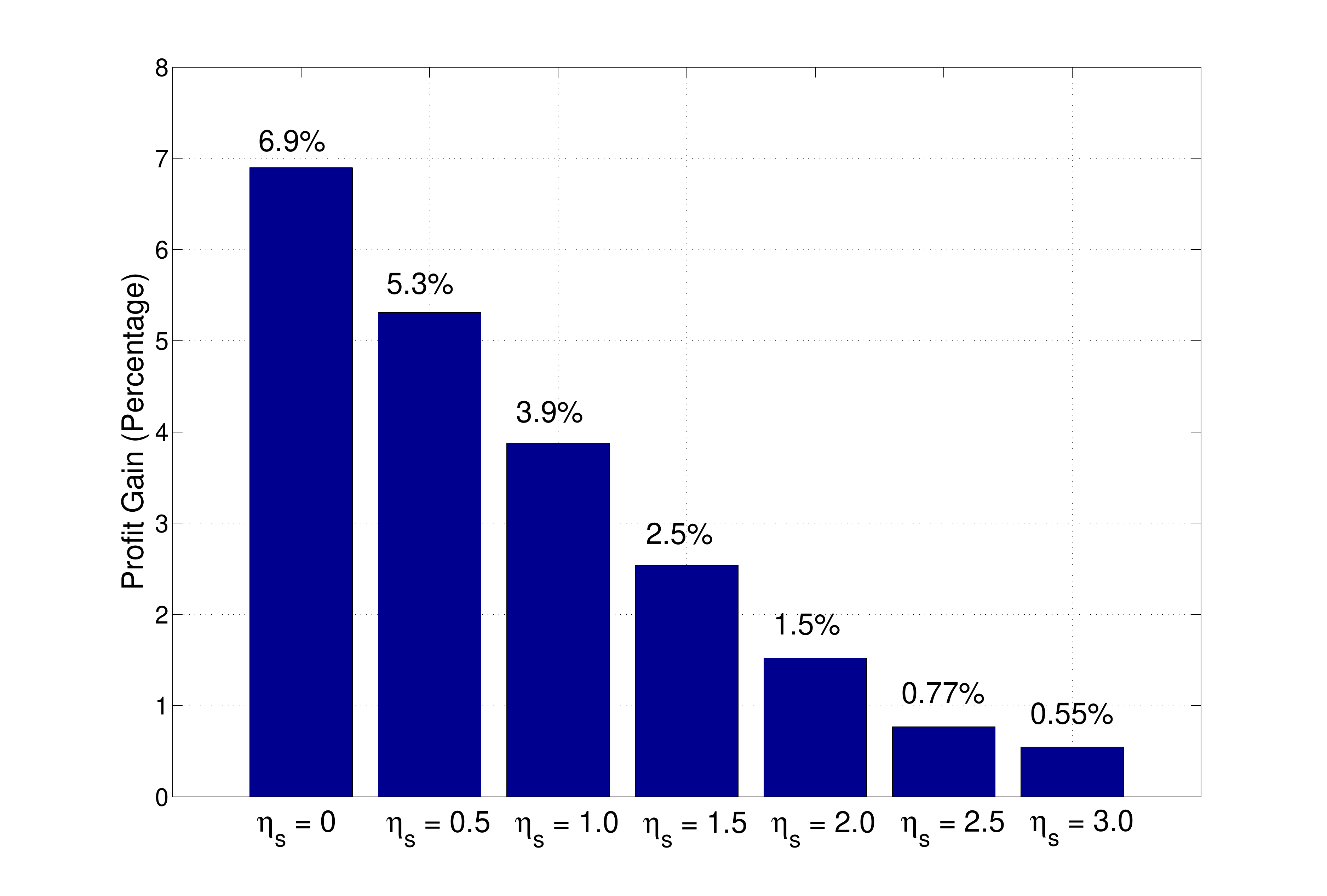}
    \caption{SDP Profit Increase Percentage}
    \label{dp_vs_greedy}
    \end{figure}

  \begin{figure}[b]
  \centering
    \includegraphics[width=4.5in]{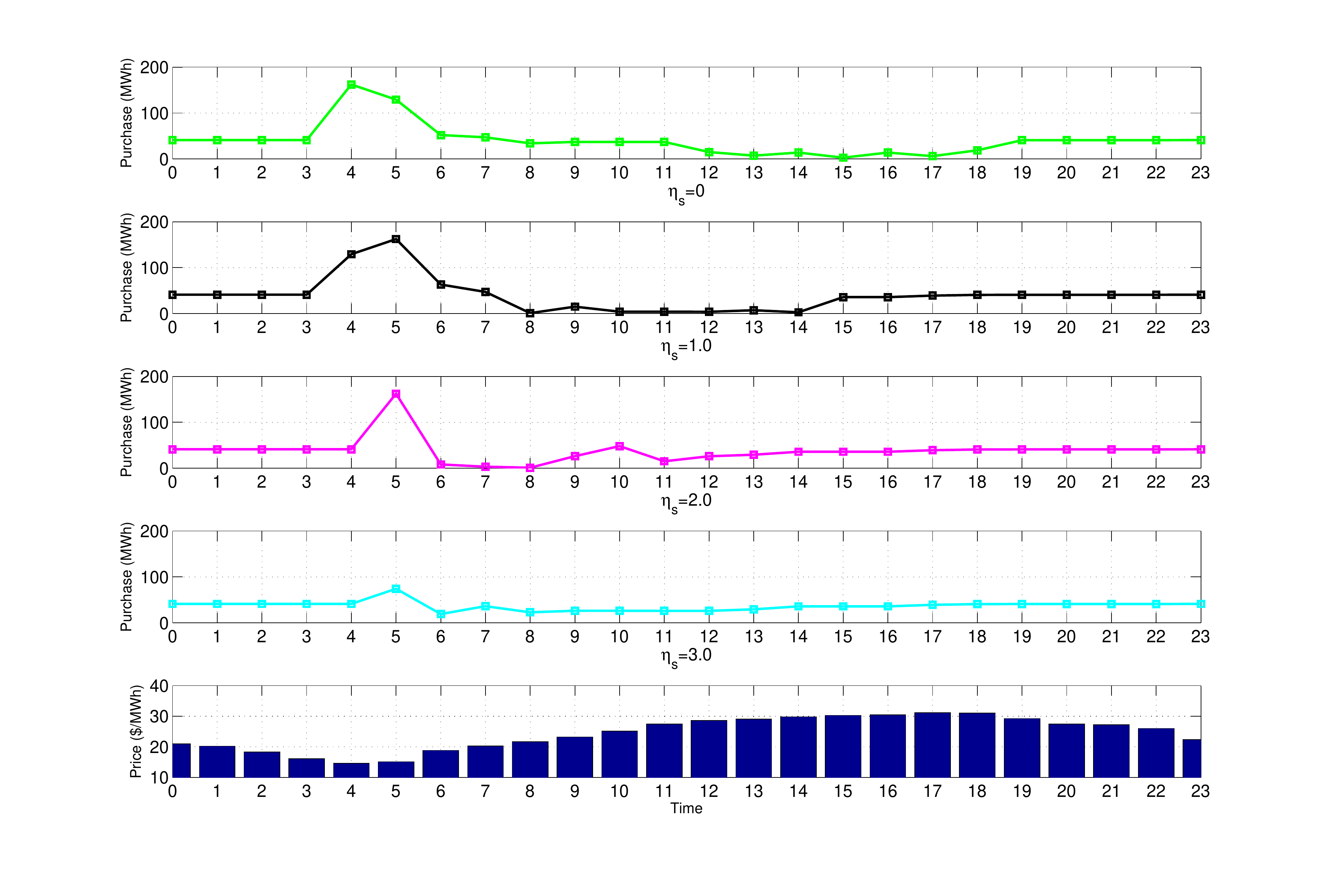}
    \caption{Electricity Procurement with Different Storage Cost}
    \label{purchase_vs_storage}
    \end{figure}

    \begin{figure}[b]
    \centering
    \includegraphics[width=4.5in]{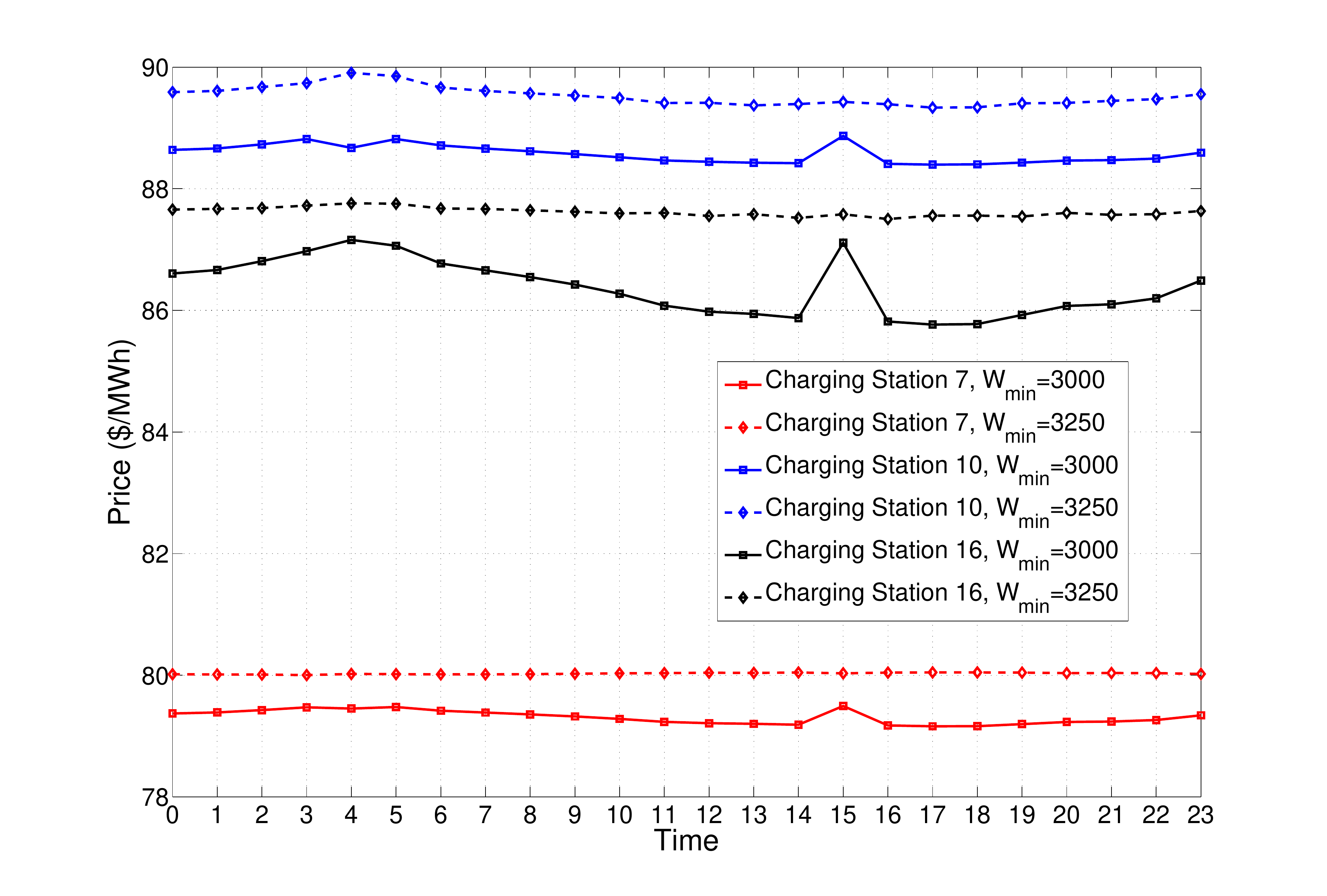}
    \caption{Charging Prices with Different $W_{\textrm{min}}$}
    \label{price_vs_rmin}
  \end{figure}

  \begin{figure}[b]
  \centering
    \includegraphics[width=4.5in]{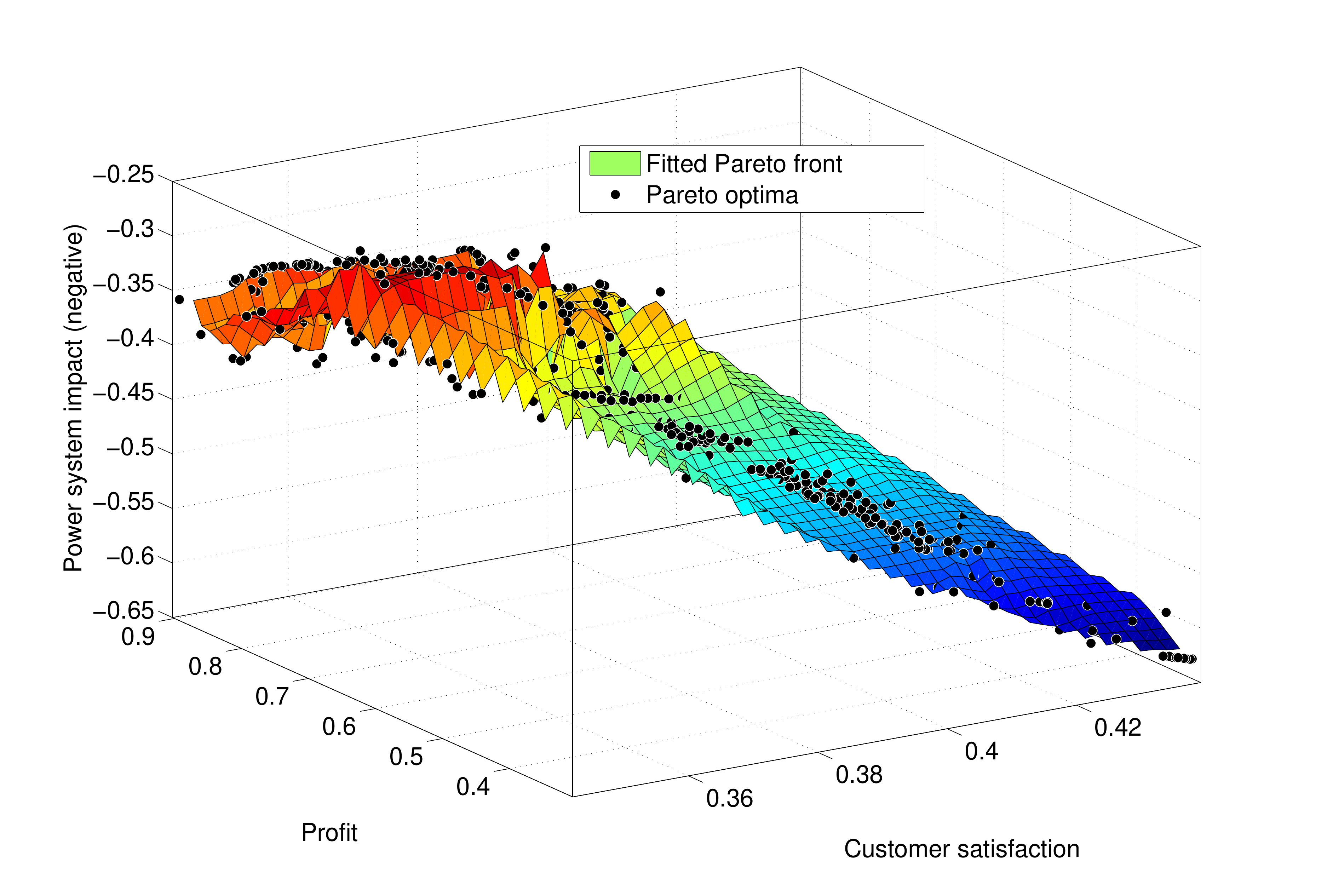}
    \caption{Pareto Front}
    \label{pareto_front}
  \end{figure}

    \begin{figure}[b]
    \centering
    \includegraphics[width=4.5in]{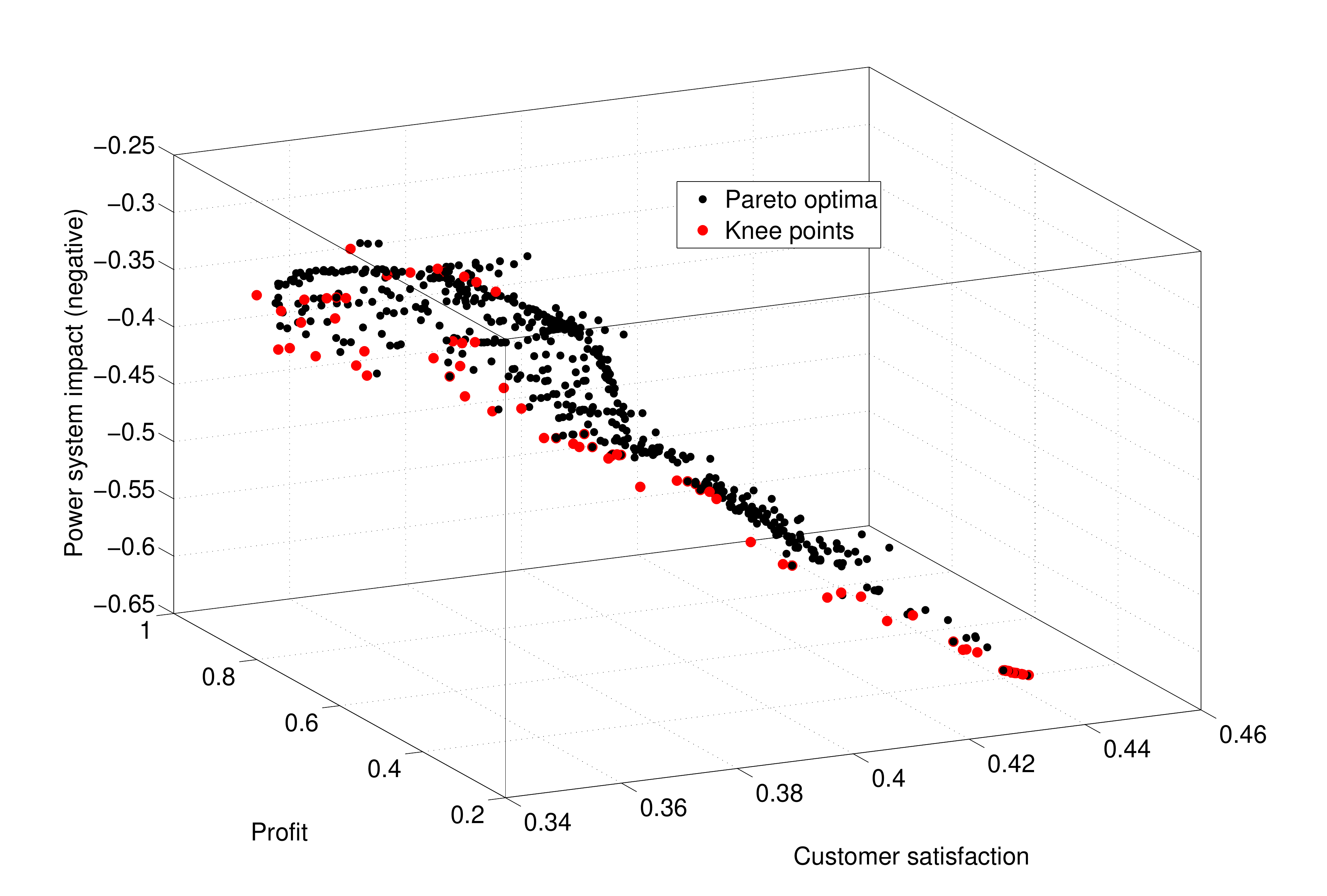}
    \caption{Pareto Front with Knee Points}
    \label{pareto_front_knee_points}
  \end{figure}

  \begin{figure}[b]
  \centering
    \includegraphics[width=4.5in]{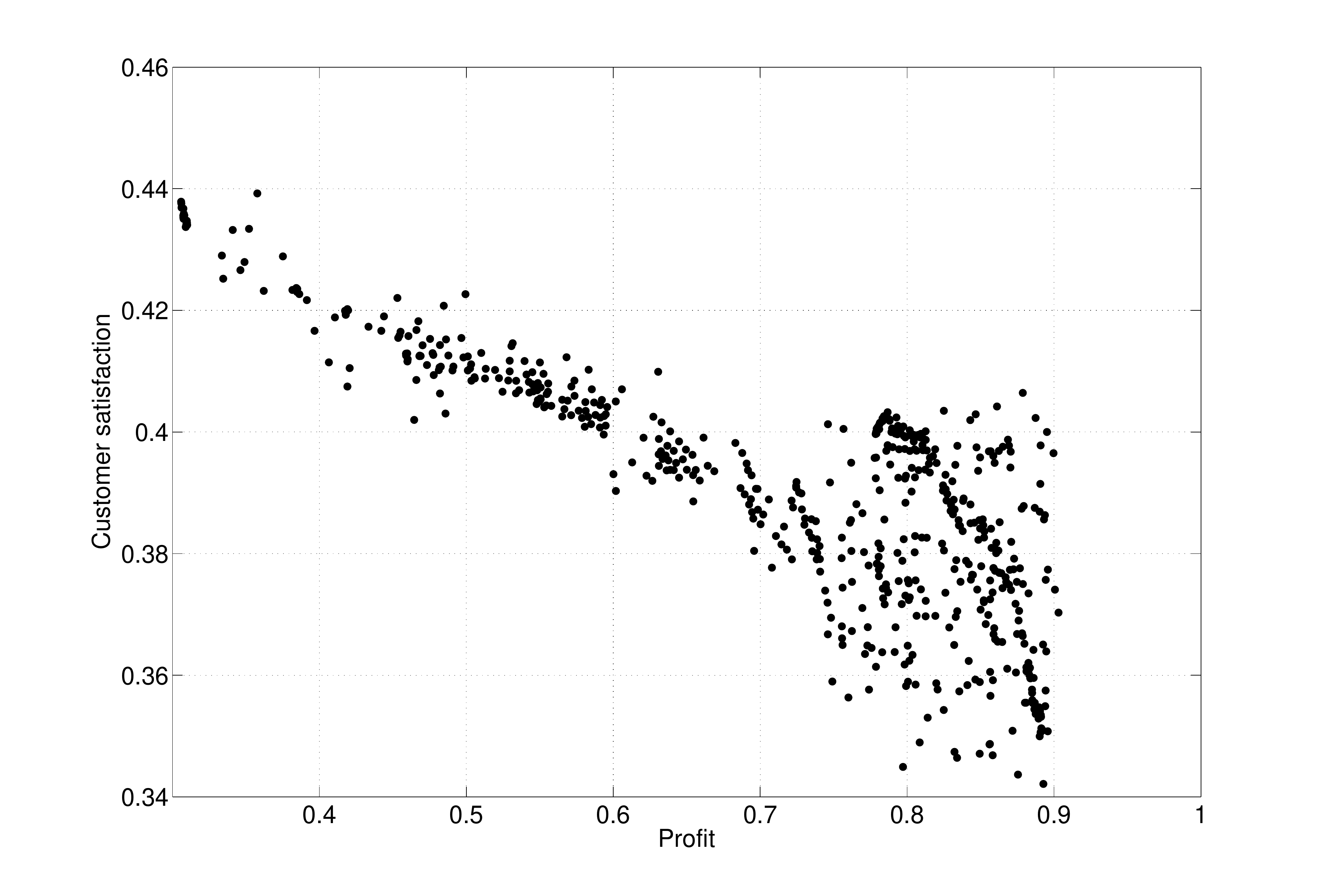}
    \caption{Profit vs Customer Satisfaction}
    \label{profit_vs_customer}
  \end{figure}

  \begin{figure}[b]
  \centering
    \includegraphics[width=4.5in]{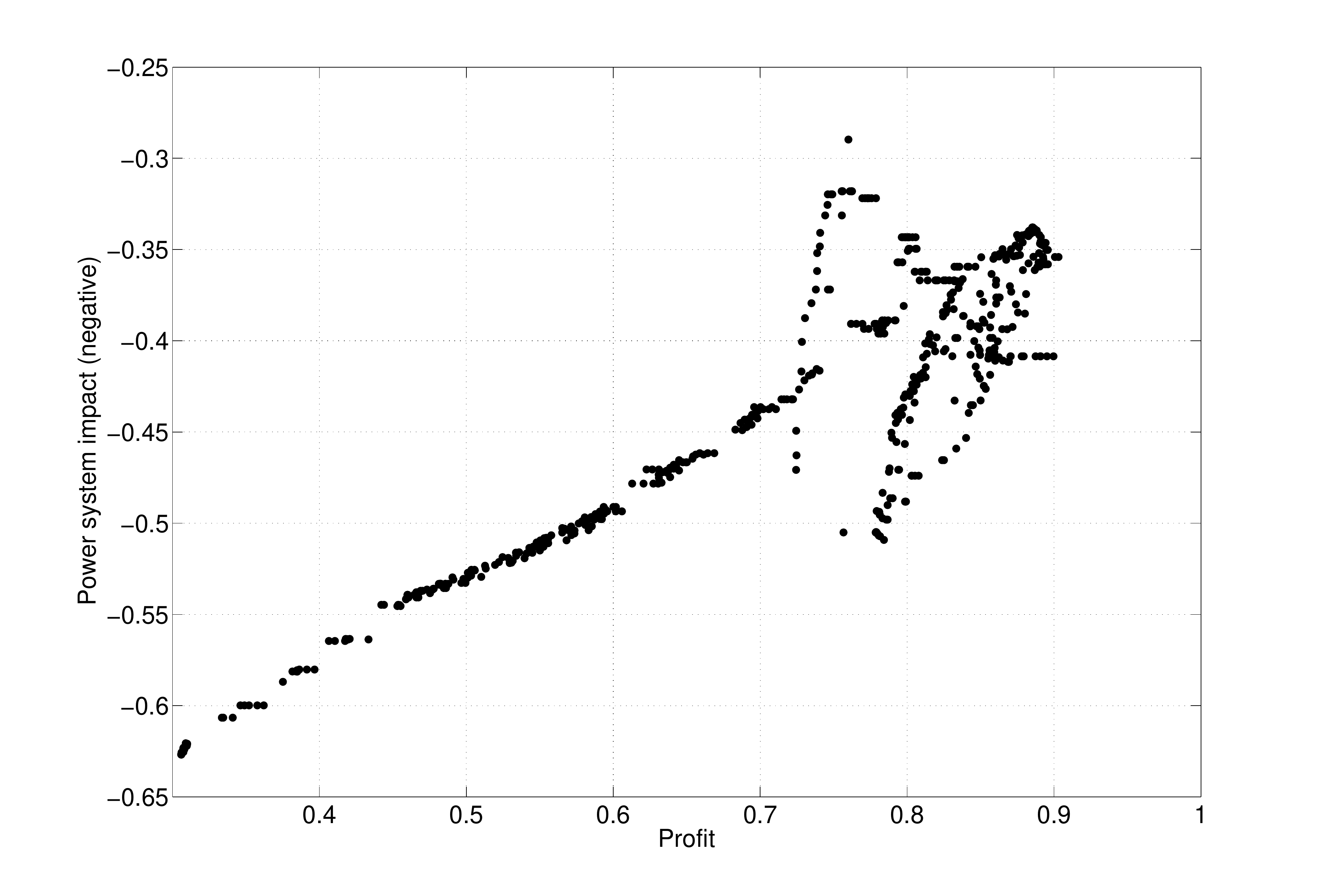}
    \caption{Profit vs Impact on Power Grid}
    \label{profit_vs_impact}
  \end{figure}

  \begin{figure}[b]
  \centering
    \includegraphics[width=4.5in]{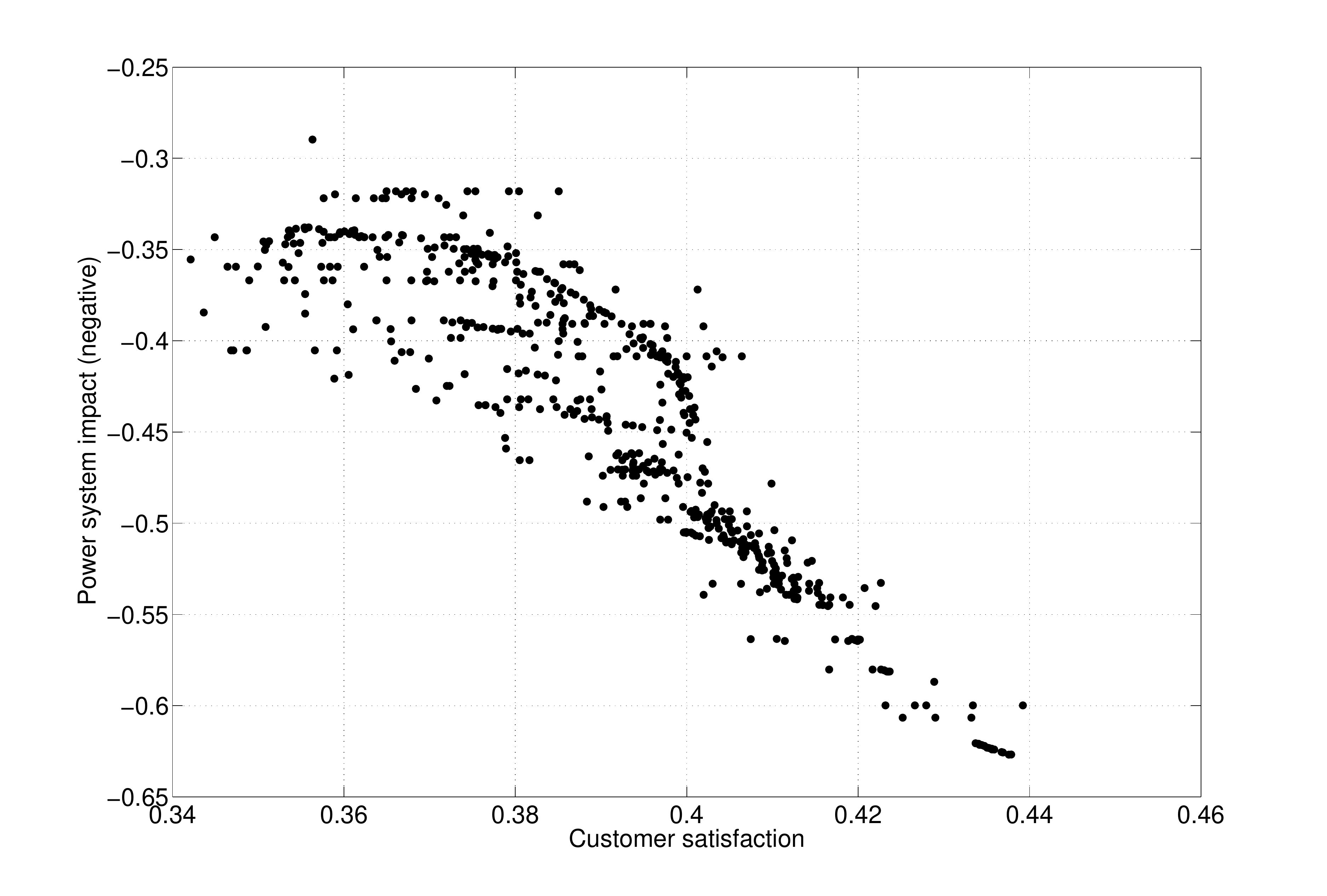}
    \caption{Customer Satisfaction vs Impact on Power Grid}
    \label{customer_vs_impact}
  \end{figure}

  \begin{figure}[b]
  \centering
    \includegraphics[width=4.5in]{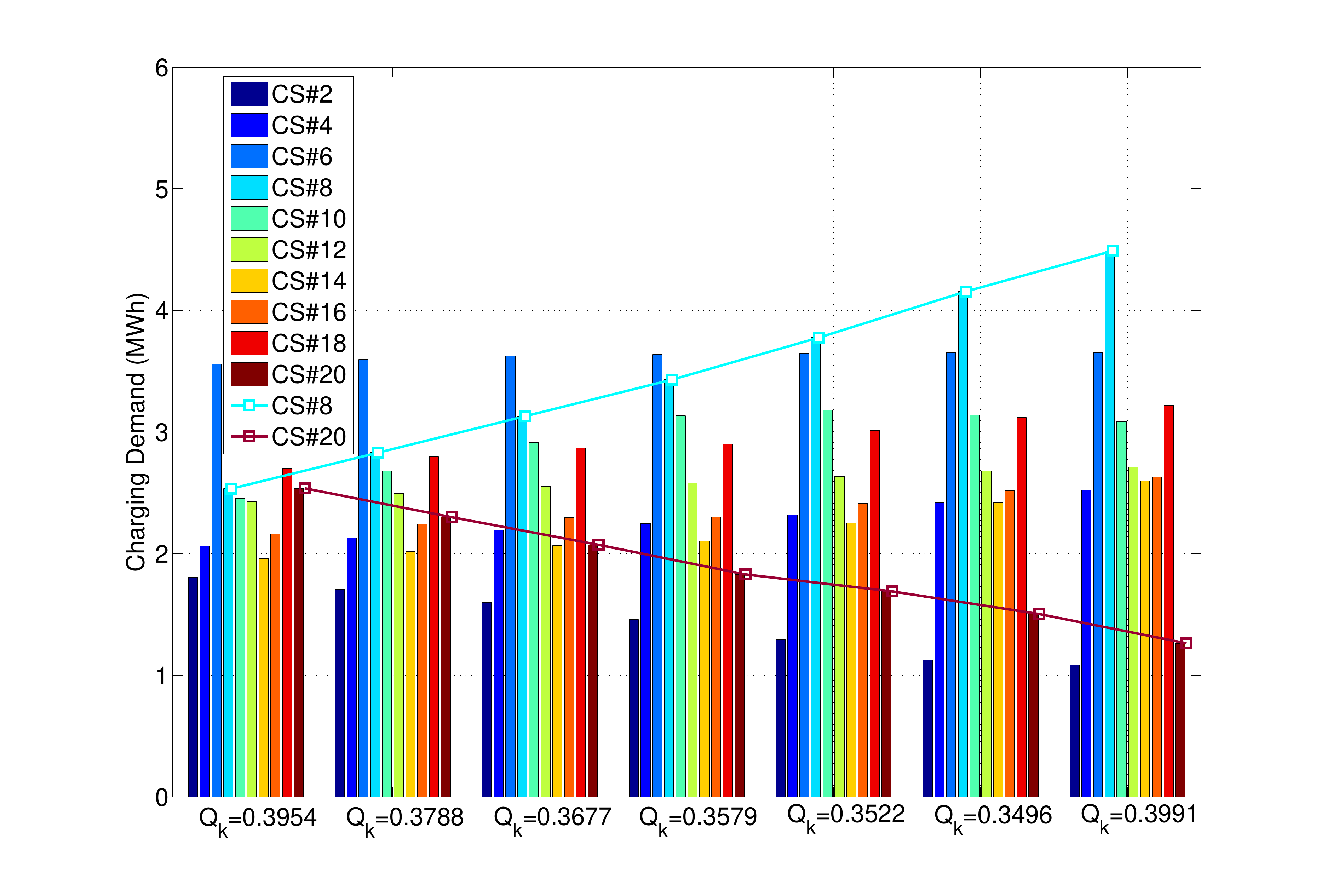}
    \caption{Charging Demand Redistribution with Different Impacts}
    \label{charging_vs_parameter}
  \end{figure}

\subsection{SDP Algorithm versus Greedy Algorithm}
The greedy algorithm aims to optimize the current planning horizon without considering the future. We use the greedy algorithm as a benchmark, to which we compare the SDP algorithm in terms of profitability. The profit percentage gain of SDP algorithm compared to greedy algorithm is shown in Fig. \ref{dp_vs_greedy}. The simulation reveals that the SDP algorithm can achieve up to 7\% profit gain compared to the greedy algorithm. The reason why SDP is able to obtain a higher profit is that it fully exploits the information of day-ahead wholesale electricity prices and renewable energy prediction, and makes decisions to optimize the aggregated utility over multiple horizons. However, the greedy algorithm lacks a forward-looking vision, which solely maximizes the utility of the current horizon. As far as the computational complexity is concerned, greedy algorithm has a linear time complexity with $O(K)$, and SDP has a quadratic time complexity with $O(K^2)$, where $K$ is the number of planning horizons. This is because the greedy algorithm only involves one loop from horizon 0 to horizon 23. However, the SDP algorithm has two loops with the outer loop starting from horizon 0 to horizon 23 and the inner loop for backward recursive SDP calculation. In essence, the SDP algorithm trades complexity for a higher profit.

\subsection{Aggressive or Conservative Electricity Procurement Strategy}
An electricity storage enables the charging service provider to store the intermittent renewable energy or excessive electricity when the wholesale price is low, and sell it to EVs when the wholesale price is high. In this subsection, we analyse how this ``buy low and sell high" strategy will change when the unit storage cost ($\eta_s=0$ to 4) changes. In Fig. \ref{purchase_vs_storage}, electricity procurement strategies with different unit storage costs are depicted in the first four subplots, and the last subplot shows the real time wholesale electricity prices. We can make three observations: (1) From 8:00 to 16:00, the service provider tends to procure less electricity from the wholesale market because of renewable energy generation at this period of time, and (2) The service provider tends to procure more electricity during the low wholesale price period (from 3:00 to 6:00) and procure less electricity during the high wholesale price period (from 11:00 to 17:00), and (3) When $\eta_s$ is small, the service provider becomes aggressive in electricity procurement during low price period, and when $\eta_s$ is large, it becomes more conservative.

\subsection{Charging Price with Safeguard of Profit}
In the simulation, we investigate the interplay between charging prices and the safeguard of profit. From Fig. \ref{price_vs_rmin}, we note that the charging prices increase as the profit threshold $W_{\textrm{min}}$ increases. According to Eq. (\ref{insurance}), we must ensure the probability $\zeta$ does not change even if $W_{\textrm{min}}$ increases. In other words, $(W_{\textrm{min}}-\mathbf{X}_k^{\textrm{T}}\mathbf{A}\mathbf{X}_k-
\mathbf{B}^{\textrm{T}}\mathbf{X}_k-t_k)/(\sqrt{\sum_{j=1}^L(p_{kj}+\eta_s/\eta_d)^2\sigma_{kj}^2+\eta_s^2\sigma_w^2})$ should not change as $W_{\textrm{min}}$ increases. The simulation results show that the charging service provider ends up raising charging prices to ensure the probability $\zeta$.

\subsection{Pareto Optima and Knee Points}
We need to simultaneously maximize multiple objectives --- profit, customer satisfaction, and the negative of impact on power grid. Each point in Fig. \ref{pareto_front} is a Pareto optimum in which it is impossible to increase any one individual objective without decreasing at least one of the other objectives \cite{clhwang}. The Pareto front is obtained by using the linear interpolation fitting method \cite{pdavis}.

Knee points in the Pareto front provide the best tradeoff among multiple objectives, which yield largest improvement per unit degradation. Following the metric discussed in \cite{lrachmawati, xzhang}, we define $\rho(Y_i,S)$ to represent the least improvement per unit degradation by replacing any other Pareto optima in $S$ with $Y_i$. The entries in $Y_i=[y_{1i},y_{2i},y_{3i}]^{\textrm{T}}$ represent the profit, the customer satisfaction, and the impact on power grid, respectively.

\begin{equation}
\rho(Y_i,S)=\min_{Y_j\in S, j\neq i}\frac{\sum_{k=1}^3\max(0,y_{ki}-y_{kj})}{\sum_{k=1}^3\max(0,y_{kj}-y_{ki})}.
\end{equation}
Then we set a threshold $\rho_0$ to select the knee points as follows,

\begin{equation}
S_{\textrm{knee}}^{\rho_0}=\{Y_i|\rho(Y_i,S)>\rho_0;Y_i\in S\}.
\end{equation}
We use $\rho=1$ in the simulations. The knee points are marked in red in Fig. \ref{pareto_front_knee_points}. We notice that there are several knee regions among the Pareto optima, which reflect different preference over the three objectives --- profit, customer satisfaction, and the impact on power grid.

\subsection{Interplays between Profit, Customer Satisfaction, and Impact on Power Grid}
The projection of Pareto optima on the Profit-Customer plane is plotted in Fig. \ref{profit_vs_customer}. We observe that customer satisfaction decreases when profit increases. This is because the charging service provider raises charging prices to decrease the total charging demand. The decreased total charging demand leads to a decreased customer satisfaction. However, the net effect of raising charging prices is that the service provider achieves a higher profit. Therefore, the service provider should strike a balance between the two competing objectives of profit and customer satisfaction.

The projection of Pareto optima on the Profit-Impact plane is plotted in Fig. \ref{profit_vs_impact}. It turns out that the impact and the profit are not competing objectives since the impact on power grid decreases when profit increases. The increased charging prices cause a decrease in total charging demand, relieving the stress on power grid. However, the profit is improved even though the total charging demand decreases.

Fig. \ref{customer_vs_impact} shows the projection of Pareto optima on the Customer-Impact plane. Note that customer satisfaction and the impact are competing objectives since the impact on power grid increases as customer satisfaction increases. It is obvious that customer satisfaction and impact on power grid are both related to the total charging demand. According to Eq. (\ref{satisfactioneq}), customer satisfaction increases when the total charging demand increases. However, the increased charging demand will inevitably pose a heavier stress on the power grid.

\subsection{Spatial Charging Demand versus Impact on Power Grid}
The relationship between spatial charging demand and impact on power grid is shown in Fig. \ref{charging_vs_parameter}. Due to limited space, we only plotted the charging stations with even indices. We observe that as the impact on power grid ($Q_k$) decreases, the charging demands of Charging Station \#2 (CS\#2) and Charging Station \#20 (CS\#20) decrease while the charging demands of other charging stations increase. This is because the PQ buses feeding CS\#2 and CS\#20 have larger active power sensitivity metric $S_i^{\textrm{Ac}}$ than the others. The active power sensitivity for charging stations with even indices are $[0.80, 0.61, 0.33, 0.17, 0.31, 0.29, 0.22, 0.60, 0.29, 1.33]$. Note that CS\#8 has the smallest active power sensitivity 0.17, its charging demand increases very fast as the impact decreases. While CS\#20 has the largest active power sensitivity 1.33, its charging demand decreases fast. Thus, the service provider has to shift the charging demands from the PQ buses with large $S_i^{\textrm{Ac}}$ to those with small $S_i^{\textrm{Ac}}$ to reduce the impact on power grid.

\section{Conclusion}
This chapter proposes a multi-objective optimization framework for EV charging service provider to determine retail charging prices and appropriate amount of electricity to purchase from the real time wholesale market. A linear regression model is employed to estimate EV charging demands. To cope with multiple uncertainties, SDP algorithm is applied to simplify the optimization problem. Compared to greedy algorithm (benchmark), SDP algorithm can make a higher profit at the cost of increased algorithm complexity. A lost-cost electricity storage is beneficial for the service provider to harvest the intermittent renewable energy and exert the ``buy low and sell high" strategy to improve profits. In addition, the service provider can shift charging demands from high-sensitive buses to low-sensitive buses to alleviate the impact on power grid by changing charging prices.

%
%

%
%
%
%
%
%
%
%
%
%

%
%

\chapter{THE DESIGN OF A LIBERAL ELECTRICITY TRADING MARKET AND DISTRIBUTION NETWORK --- A SHARING ECONOMY PERSPECTIVE}

\section{Overview}
 In this chapter, we explore ways to apply the prevalent sharing economy model to the electricity market to stimulate different entities to exchange and monetize their underutilized electricity. The design of a liberal electricity trading market and distribution network is presented, which enhances liberal electricity trading, fosters renewable energy generation, facilitates distributed storage and integration of electric vehicles (EVs), and manages network congestion. In our study we treat electricity as a heterogenous commodity which has multiple characteristics such as active power, reactive power, power factor, power quality (voltage, frequency), energy sources (i.e. coal, wind, solar), delivery rate, delivery time, etc. An online advertisement-based peer-to-peer (P2P) electricity trading mechanism is proposed for producers and consumers to publish their supply offers and demand bids with different requirements and preferences. Accordingly, a fitness-score (FS) based matching algorithm is developed to select the best supply-demand pairs by taking into account consumer surplus, network congestion, and economic power dispatch. We compare the FS matching algorithm with a first-come-first-serve (FCFS) based matching algorithm (the benchmark algorithm). The simulation reveals that the FS matching algorithm outperforms the FCFS algorithm in terms of network congestion management, electricity delivery delay probability, energy efficiency, and consumer surplus.

\section{Motivation and Related Work}
The process of electricity market deregulation has provided a fair and competitive wholesale market for giant power plants and load serving entities (LSEs), which facilitates free electricity trading, enhances power grid stability, and ensures economic power dispatch \cite{tjamasb, fsioshansi, caiso, pjmiso}. Although a deregulated electricity market exhibits some advantages over a regulated market, it still has inherent disadvantages arising from the inner market and infrastructure. First, it remains a challenge for the existing electricity delivery infrastructure to naturally integrate volatile and intermittent renewable energy sources (RESs). Second, the nature of bulk electricity transactions in a wholesale market impedes small/micro power producers to participate in the market. Third, the strict membership requirements of a wholesale market participant excludes end customers and small business from directly accessing the market, who can actually play a significant role in an electricity network like a microgrid.

\subsection{Related Work}
There is increasing research aimed at addressing the problem of electricity trading in the context of a smart grid. In \cite{dilic}, a stock exchange based local electricity market was proposed, where electricity is traded during fixed trading intervals and delivered after a
predefined time interval like in a stock exchange market. In \cite{slamparter, pvytelingum}, agent-based market designs were studied and a double auction model was employed to determine the clearing price. In \cite{jkang}, the authors proposed a local P2P electricity trading of plug-in electric vehicles (PHEV) based on a consortium blockchain model. They also used a double auction to derive optimal electricity price. A game theory based approach was applied to analyze the interplay between electricity producers and consumers in \cite{wlee, tdai}. In \cite{wlee}, the authors applied a coalitional game to model the direct electricity trading between small-scale electricity suppliers and end users, where the Shapley value was used to derive the price in the market. In \cite{tdai}, a model of stochastic programming and game theory was studied to generate optimal bidding strategy for wind and conventional power producers. In \cite{mcrosby}, the author first discussed the possibility of applying a sharing economy model to the smart grid, which sets up a P2P platform for producers and consumers to trade their underutilized electricity. Furthermore, \cite{dkalathil} explored the applicability of sharing economy for a smart grid and formulated the investment decisions as a non-convex non-cooperative game. In addition, a security analysis of a local electricity trading market was conducted in \cite{mmustafa}.

Note that the existing literature mostly focused on electricity market design in terms of market operation, economics analysis, and security issues. Other important factors such as network congestion forecast and avoidance, delivery cost and distribution strategy have been mostly neglected. Besides, electricity is usually treated as a homogeneous commodity, but it can actually have different characteristics in terms of active power, reactive power, power factor, energy sources (i.e. wind, solar, coal, etc.), power quality (voltage, frequency), delivery starting/ending time, and delivery rate, etc. In this chapter we will propose an online advertisement based electricity market design which takes into account network congestion, delivery cost, and consumer surplus. In addition, we will present a new electricity distribution strategy which can naturally support variable electricity delivery and discontinuous renewable energy integration.

\subsection{Main Contributions}
The motivation of our work is to design a liberal electricity trading market and distribution network which provides a flexible, reliable and economical platform to support seamless RES integration, grid to vehicle (G2V), vehicle to grid (V2G), vehicle to vehicle (V2V), and distributed generation (DG) and storage. Recently, a sharing economy model is prevailing in a few industries which have fundamentally changed our life, e.g. Uber/Lyft in the transportation industry \cite{uberintro, lyftintro}, Airbnb in the accommodation industry \cite{airbnb}, and LendingClub in the online loan and lending industry \cite{lendingclub}. Basically, a sharing economy model is a P2P platform for individuals to trade and monetize their underutilized services or resources \cite{tpuschmann, rbotsman}. Similarly, in this study we apply a sharing economy model in designing electricity and distribution network, where electricity is regarded as a type of commodity and everyone can sell excessive electricity via this platform. In contrast to the traditional electricity network, there is not a clear delimitation between the roles of a consumer and a producer in our design. Instead, every individual can switch from a consumer to a producer, or vise versa freely in the network. For instance, an EV is a consumer when it is being charged, whereas it becomes a producer when it discharges. A household with a solar panel and a local energy storage installed can be a producer during sunny days and a consumer during rainy days. To build up this electricity market network, we need to develop new algorithms and mechanisms for both market operation and electricity delivery.

The main contributions of our work are summarized as follows,

\begin{itemize}
\item Propose an online advertisement-based mechanism for producers and consumers to submit their electricity supply offers and demand bids, which treats electricity as a heterogeneous commodity with multiple characteristics.
\item Develop a fitness score-based supply and demand matching algorithm by considering consumer surplus, network congestion, and renewable energy integration.
\item Design a novel electricity distribution policy which naturally supports intermittent renewable energy generation and variable electricity delivery rate.
\item Propose a demand and surge pricing prediction algorithm to alleviate network congestion.
\end{itemize}

\section{Problem Formulation}
\begin{figure}[htbp]
\centering
\includegraphics[width=1\textwidth]{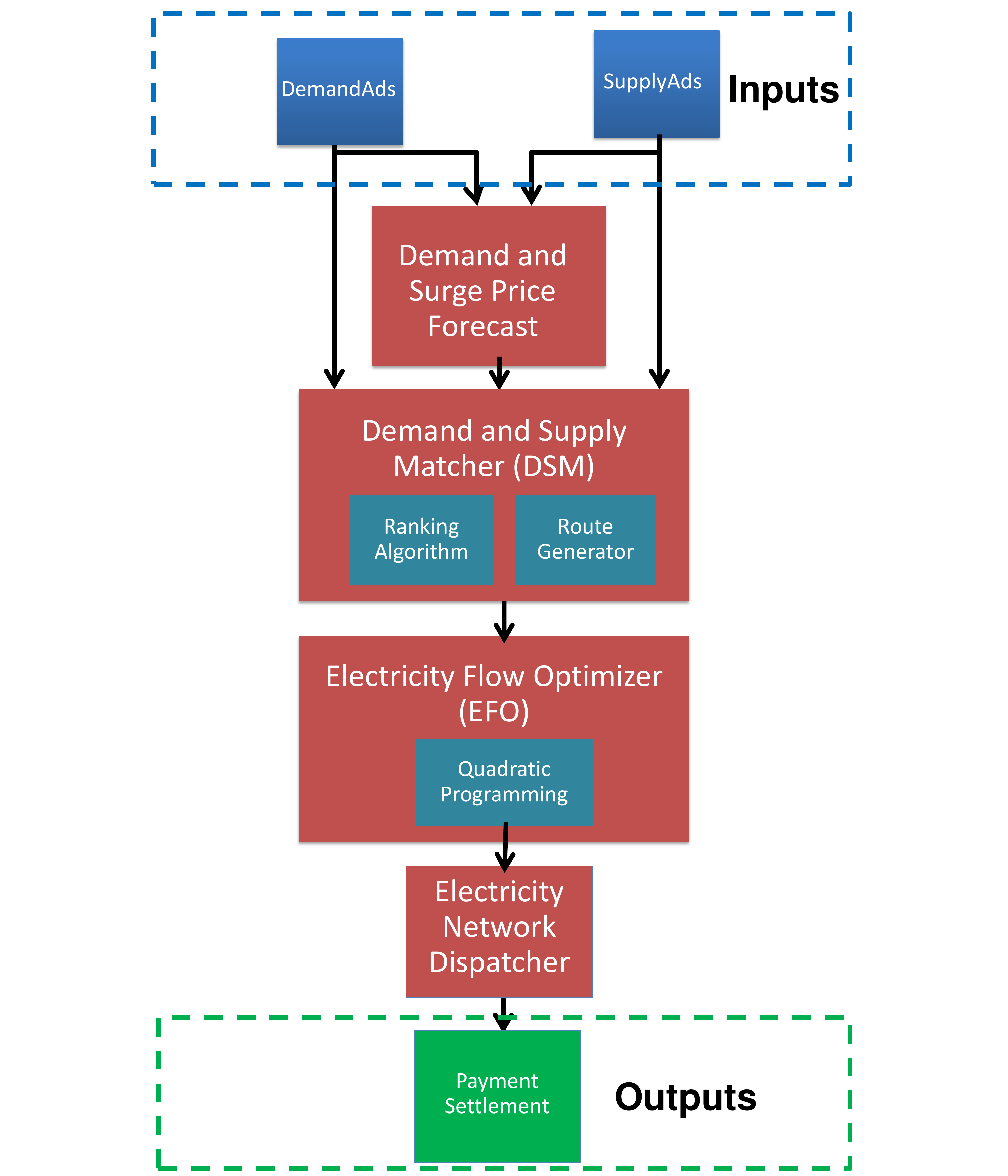}
\caption{Market and Network Architecture}
\label{marketarchitecture}
\end{figure}

The proposed electricity trading market and distribution network consists of four modules: (1) demand and surge price prediction module, (2) supply and demand matching module, (3) electricity flow optimization module, and (4) electricity network scheduling module. Fig. \ref{marketarchitecture} shows the modules of the entire system.

\subsection{Supply and Demand Matcher}
Inspired by the ClassAd and Matchmaking mechanism from HTCondor in the distributed computing community \cite{dthain}, we will use a similar mechanism to match the demands and supplies in the electricity trading market. In HTCondor, ClassAd offers a flexible representation of characteristics and constraints of both computation jobs and host machines. Jobs can specify their requirements (i.e. CPU, GPU, memory, I/O, etc.) and job preferences. Likewise, host machines can state the requirements and preferences they would like to admit to run. Basically, the ClassAd and Matchmaking mechanism behaves like an e-commerce where buyers and sellers post their advertisements and search for potential peers. The supply and demand matcher (SDM) module in our design collects the electricity demand advertisement --- DemandAd from consumers, and supply advertisement --- SupplyAd from producers, and employs a ranking algorithm to identify the best pairs. Examples of  DemandAd and SupplyAd are shown in table \ref{demandad} and table \ref{supplyad}, respectively.

\begin{table}
\caption{DEMANDAD EXAMPLE}
\label{demandad}
\begin{tabular}{ |p{3cm}|p{3cm}|p{3cm}|  }
\hline
\multicolumn{3}{|c|}{DemandAd} \\
\hline
Entry & Unit & Value \\
\hline
DID & - & 123 \\
\hline
BID & -   & 10 \\
\hline
ZID & - & 1\\
\hline
$Q$ & MWh & 15 \\
\hline
$\theta$ & - & 0.7\\
\hline
$T^{\textrm{min}}$    & YYYY-MM-DD HH:MI:Sec & 2017-6-11 17:00:00  \\
\hline
$T^{\textrm{max}}$  & YYYY-MM-DD HH:MI:Sec & 2017-6-11 18:00:00 \\
\hline
$p^{\textrm{bid}}$ & \$/MWh & 10   \\
\hline
$R$ & MW & 2 \\
\hline
DM & Boolean & False\\
\hline
\end{tabular}
\end{table}

\begin{table}
\caption{SUPPLYAD EXAMPLE}
\label{supplyad}
\begin{tabular}{ |p{3cm}|p{3cm}|p{3cm}|  }
\hline
\multicolumn{3}{|c|}{SupplyAd} \\
\hline
Entry & Unit & Value \\
\hline
SID & - & 100 \\
\hline
BID & -   & 9 \\
\hline
ZID & - & 2\\
\hline
$Q$ & MWh & 100 \\
\hline
$\theta^{\textrm{min}}$ & - & 0.4 \\
\hline
$\theta^{\textrm{max}}$ & - & 0.9\\
\hline
$T^{\textrm{min}}$   & YYYY-MM-DD HH:MI:Sec & 2017-6-12 17:00:00  \\
\hline
$T^{\textrm{max}}$ & YYYY-MM-DD HH:MI:Sec & 2017-6-12 18:00:00 \\
\hline
$p^{\textrm{offer}}$ & \$/MWh & 8   \\
\hline
$R^{\textrm{min}}$ & MW & 1 \\
\hline
$R^{\textrm{max}}$ & MW & 5\\
\hline
DM & Boolean & False\\
\hline
ET&Boolean & False\\
\hline
\end{tabular}
\end{table}

\noindent The DemandAd consists of 9 elements:
\begin{itemize}
\item Demand ID (DID) --- The unique ID assigned for every demand advertisement which is submitted to the SDM.
\item Bus ID (BID) --- The Bus ID from which the demand advertisement is submitted.
\item Zone ID (ZID) --- The zone ID where the bus belongs to.
\item Quantity (Q) --- The required electricity.
\item Power factor ($\theta$) --- The power factor of this electricity delivery task which is the ratio of active power to apparent power.
\item Minimum start time ($T^{\textrm{min}}$) --- The earliest time when the bus expects to receive electricity.
\item Maximum start time ($T^{\textrm{max}}$) --- The latest time when the bus expects to receive electricity.
\item Bid price ($p^{\textrm{bid}}$) --- The highest price the bus is willing to pay for a supplier.
\item Delivery rate ($R$) --- The delivery rate at which the bus expects to receive electricity from a supplier.
\item Delivery method (DM) --- This element specifies that if the bus can accept discontinuous electricity delivery. If DM = True, the bus can accept discontinuous electricity delivery, which DM = False indicates that it cannot accept discontinuous electricity delivery.
\end{itemize}

\noindent The SupplyAd consists of 10 elements:
\begin{itemize}
\item Supply ID (SID) --- The unique ID for every supply advertisement which is submitted to SDM.
\item Bus ID (BID) --- The Bus ID from which the supply advertisement is submitted.
\item Zone ID (ZID) --- The zone ID where the bus belongs to.
\item Quantity (Q) --- The maximum electricity that the bus can supply.
\item Minimum power factor ($\theta^{\textrm{min}}$) --- The minimum power factor.
\item Maximum power factor ($\theta^{\textrm{max}}$) --- The maximum power factor.
\item Minimum start time ($T^{\textrm{min}}$) --- The earliest time when the bus expects to supply electricity.
\item Maximum start time ($T^{\textrm{max}}$) --- The latest time when the bus expects to supply electricity.
\item Offer price ($p^{\textrm{offer}}$) --- The lowest price at which the bus is willing to supply electricity.
\item Minimum delivery rate ($R^{\textrm{min}}$) --- The minimum delivery rate at which the bus expects to supply electricity.
\item Maximum delivery rate ($R^{\textrm{max}}$) --- The maximum delivery rate at which the bus expects to supply electricity.
\item Delivery method (DM) --- This element specifies that if the bus will supply discontinuous electricity. If DM = True, the bus supplies discontinuous electricity, which DM = False indicates that it supplies continuous electricity.
\item Energy type (ET) --- This element specifies that if the electricity is generated from renewable sources (i.e. solar, wind). ET = True indicates it is renewable energy, which ET = False indicates it is not renewable energy.
\end{itemize}

For every DemandAd, SDM generates a set of SupplyAds which satisfy the requirement specified by this DemandAd. Then, SDM applies a matching algorithm to those SupplyAds to find the optimal one for this DemandAd. If such a SupplyAd exists, SDM claims to have a supply-demand pair. Basically, the matching algorithm uses a formula to calculate a fitness score for every SupplyAd and ranks those SupplyAds by scores. The formula is given as follows,

\begin{equation}\label{ranking}
S(i,j) = \alpha_1\frac{p^{\textrm{bid}}_i-p^{\textrm{offer}}_j}{p^{\textrm{bid}}_{\textrm{max}}-p^{\textrm{offer}}_{\textrm{min}}}-
\alpha_2\frac{D_{i,j}}{D_{\textrm{max}}}+\alpha_3Z(i,j)+\alpha_4E(j),
\end{equation}
where $p^{\textrm{bid}}_i$ and $p^{\textrm{offer}}_j$ are the bid electricity price of the $i$-th DemandAd and the offer electricity price of the $j$-th SupplyAd, respectively. And $p^{\textrm{bid}}_{\textrm{max}}$ and $p^{\textrm{offer}}_{\textrm{min}}$ are the maximum bid price among all DemandAds and the minimum offer price among all SupplyAds, respectively. Hence $(p^{\textrm{bid}}_i-p^{\textrm{offer}}_j)$ represents the cost saving that the $i$-th DemandAd can obtain from the $j$-th SupplyAd, which is normalized by $(p^{\textrm{bid}}_{\textrm{max}}-p^{\textrm{offer}}_{\textrm{min}})$, while $\alpha_1$ is the corresponding weight parameter. In addition, $D_{i,j}$ is the predicted electricity delivery price from the $j$-th SupplyAd to the $i$-th DemandAd, which will be paid to the electricity network operator, and $D_{\textrm{max}}$ is the maximum electricity delivery price with $\alpha_2$ being the corresponding weight parameter. In the following subsection, we will discuss in more details on how to predict electricity delivery price. Furthermore, $Z(i,j)$ is an indicator function, which is equal to 1 if buses of the DemandAd and SupplyAd belong to the same zone in the electricity network. The idea here is that SDM prefers to match supply-demand pairs that are geographically close to each other so that delivery cost can be reduced and electricity distribution lines can be wisely utilized. And $\alpha_3$ is the weight parameter for $Z(i,j)$. Besides, $E(j)$ is an indicator function, which is equal to 1 if the $j$-th SupplyAd bus supplies renewable energy. And $\alpha_4$ is the weight parameter for $E(j)$. After normalization, the four terms all have the range from 0 to 1, and the corresponding weight parameters are non-negative and have the constraint of $\alpha_1+\alpha_2+\alpha_3+\alpha_4=1$.

Once the demand-supply pairs are matched, SDM generates a predefined number of routes for each demand-supply pair. SDM employs the breath-first-search (BFS) \cite{tcormen} algorithm to find the routes from a SupplyAd bus to a DemandAd bus in the electricity network. SDM sends the route information to the next module --- electricity flow optimizer.

\subsection{Electricity Flow Optimizer}
The objective of the electricity flow optimizer (EFO) is to minimize power loss in the electricity network by choosing the optimal routes and determine the optimal electricity flow along the distribution lines. First, we assume that electricity delivery is scheduled at an interval of $T^{\textrm{int}}$. We use $\mathcal{K}^h$ to represent the set of active supply-demand pairs during the $h$-th interval. We denote $L^h$ as the number of distribution lines (branches) to be used in the $h$-th interval. Note that the branch indices used in the following analysis may not necessarily be the same as the indices in the electricity network for simulations. For the electricity network, it has internal branch indices, so we need to convert the branch indices in theoretical analysis to internal branch indices in simulations. We use $\mathcal{J}^h_k$ to represent the set of routes for the $k$-th demand-supply pairs during the $h$-th interval. We denote $\beta_{k,j,l}^h$ as the power flow (in MW) of the $l$-th branch of the $j$-th route of the $k$-th demand-supply pair. Let $Q_k^h$ represent the scheduled power delivery (in MW) of the $k$-th supply-demand pair. Let $r_l$ denote the resistance (in $\Omega$) of the $l$-th branch. To minimize energy loss, EFO seeks the most economic way to deliver electricity over the network. For ease of notation, we drop the interval superscript in the following analysis. Mathematically, EFO aims to solve an optimization problem formulated as follows,

\begin{subequations}\label{efo}
\begin{align}
&\argmin_{\beta_{k,j,l}} \sum\limits_{l = 1}^{L}\left \{r_l\left(\frac{\sum \limits_{k \in \mathcal{K}_l}
\sum \limits_{j \in \mathcal{J}_{k,l}}\beta_{k,j,l}}{U_l}\right)^2 \right \},\label{objfun}\\
\intertext{$\textrm{\;\;\;\;\;\;\;\;\;\;\;\;\;\;s.t.}$}
&\;\;\;\;\;\;\sum \limits_{k \in \mathcal{K}_l}\sum \limits_{j \in \mathcal{J}_{k,l}}\beta_{k,j,l}\leq C_l; \;\;\;l=1,2,\cdots,L,\label{capacity}\\
&\;\;\;\;\;\; \beta_{k,j,l_1}=\beta_{k,j,l_2}=\beta_{k,j,l_3}=\cdots=\beta_{k,j,l_{k,j}}; \;\;\;\textrm{for any } k \in \mathcal{K}, j \in \mathcal{J}_k,\label{route}\\
&\;\;\;\;\;\; \sum \limits_{j \in \mathcal{J}_k}\beta_{k,j,l_1}=Q_k; \;\;\;\textrm{for any } k \in \mathcal{K},\label{pair}
\end{align}
\end{subequations}
where $\mathcal{K}_l$ is the set of supply-demand pairs whose routes contain the $l$-th branch, $\mathcal{J}_{k,l}$ is a set of routes of the $k$-th demand-supply pair in $\mathcal{K}_l$ whose route path contains the $l$-th branch, and $U_l$ is the voltage of the $l$-th branch. Hence, the objective function defined by Eq. (\ref{objfun}) is the sum of power loss of all involved branches.

Additionally, Eq. (\ref{capacity}) assures that the power flow along each branch does not exceed its capacity. In Eq. (\ref{route}), a sequence of branches $(l_1,l_2,l_3,\cdots,l_{k,j})$ form the the $j$-th route of the $k$-th demand-supply pair. Obviously, Eq. (\ref{route}) requires that the power flow of branches of the same route should be the same. Besides, Eq. (\ref{pair}) stipulates that the sum of all routes' power flow be equal to the scheduled power delivery for every demand-supply pair.

Note that the objective function in Eq. (\ref{objfun}) only depends on the quadratic terms of the variables, so we can formulate it as a quadratic programming problem by arranging and grouping like terms.

\begin{subequations}\label{quadeq}
\begin{align}
&\argmin_{X}\frac{1}{2}X^{\textrm{T}}\mathbf{P}X,\\
\intertext{$\textrm{\;\;\;\;\;\;\;\;\;\;\;\;\;\;\;\;\;\;\;\;\;\;\;\;\;\;\;\;\;\;\;\;\;\;\;\;\;\;\;\;\;\;\;\;\;\;\;\;\;\;\;\;\;\;\;\;\;\;s.t.}$}
&\;\;\;\;\;\;\mathbf{A}X=b,\\
&\;\;\;\;\;\;\mathbf{G}X\leq h,
\end{align}
\end{subequations}
where $X=[X_1^{\textrm{T}},X_2^{\textrm{T}},\cdots,X_L^{\textrm{T}}]^{\textrm{T}}$ is the variable vector, and $X_l (l=1,2,\cdots,L)$ is a sub-vector consisting of power flows $\beta_{k,j,l}$ on the $l$-th branch of different supply-demand pairs. Specifically,

\begin{equation}
X_l=\left[
\begin{array}{c}
\beta_{1,l}\\
\beta_{2,l}\\
\vdots\\
\beta_{n_l,l}
\end{array}
\right],
\end{equation}
where $\beta_{n,j}$ denotes the $n$-th power flow on the $l$-th branch.

Matrix $\mathbf{P}$ has the following form,
\begin{equation}
\mathbf{P}=
\left[
\begin{array}{ccccc}
P_1&&&&\\
&P_2&&\text{\fontsize{40}{50}\selectfont 0}&\\
&&\ddots&&\\
&\text{\fontsize{40}{50}\selectfont 0}&&P_{L-1}&\\
&&&&P_L\\
\end{array}
\right]
\end{equation}
where the submatrix $P_l$ is given by

\begin{equation}
P_l=\frac{r_l}{U_l^2}\mathbf{J}_{n_l},
\end{equation}
and $\mathbf{J}_{n_l}$ is an $n_l\times n_l$ all-one square matrix, whose entries are all 1's.

We notice that $\mathbf{P}$ is a symmetric block diagonal matrix. All diagonal submatrices are all-one square matrices multiplied by a scalar, which are positive semidefinite. So matrix $\mathbf{P}$ is also positive semidefinite \cite{rhorn}. Therefore, the quadratic programming problem defined in Eq. (\ref{quadeq}) is convex and we can obtain a global optimum using common optimization packages in Matlab, Cplex, and Joptimizer.

\subsection{Demand and Surge Price Prediction Module}
In this study, the payment of a DemandAd consists of two components --- electricity procurement cost and electricity delivery cost. The electricity procurement cost is the money that a DemandAd pays to the SupplyAd for electricity procurement. The electricity delivery cost is the money that a DemandAd pays to the electricity network operator for electricity delivery. However, a surge price multiplier may be added to the electricity delivery cost if the supply-demand pair uses certain congested branches in the delivery of electricity. The surge price mechanism has been used in a few ride-hailing companies like Uber (i.e. Surge Price) or Lyft (i.e. Prime Time) \cite{uber, lyft}. For Uber Surge Price or Lyft Prime Time, a customer in a high demand region is likely to be charged at a higher price (surge price). Likewise, if a supply-demand pairs uses some congested branches, the DemandAd must pay a higher electricity delivery price to the network operator.

The surge price mechanism follows the basic supply and demand principles. For instance, we consider a potential supple-demand pairs $(i,j)$ --- the $i$-th DemandAd and the $j$-th SupplyAd, whose delivery starting time is $t_{i,j}$ and scheduled electricity delivery is $Q_{i,j}$. First SDM uses the BFS algorithm to compute the routes $\mathcal{J}_{i,j}$ for this pair. To predict the delivery cost, we first find the set of supply-demand pairs $\mathcal{S}_{i,j}$ which will be active during the $t_{i,j}$-th interval. All supply-demand pairs in $S_{i,j}$ are from previous matched pairs which have an earlier delivery starting time than $t_{i,j}$. Then SDM sends the information of $S_{i,j}$ to EFO. EFO uses $S_{i,j}$ to compute the power flow $F_l^{\textrm{base}}$ on each branch. For each branch $l$ in $\mathcal{J}_{i,j}$, we add $Q_{i,j}$ to the corresponding branch flow $F_l^{\textrm{base}}$.

\begin{equation}
F_l^{\textrm{now}}=Q_{i,j}+F_l^{\textrm{base}}.
\end{equation}

We define threshold brackets for the aggregated branch flow. If the aggregated branch flow $F_l^{\textrm{now}}$ falls into a certain threshold bracket, a surge price multiplier will be applied accordingly. Furthermore, if the aggregated branch flow exceeds the largest threshold, this supply-demand pair will be rejected by SDM. If the supply-demand pair is not rejected, the delivery price can be calculated using the following formula,

\begin{equation}\label{deliverycost}
D(i,j)=\gamma_{i,j} vR_{i,j}T^{\textrm{int}},
\end{equation}
where $R_{i,j}$ is the delivery rate (in MW) of this supply-demand pair, $T^{\textrm{int}}$ (in hour) is the interval duration, and $v$ is the flat delivery price (in \$/MWh). Additionally, $\gamma_{i,j}$ is the surge price multiplier which is defined as follows,

\begin{equation}\label{surge}
\gamma_{i,j}=
\begin{cases}
\gamma_1\;\;\;F_l<0.5C_l\\
\gamma_2 \;\;\; 0.5C_l\leq F_l <0.75C_l\\
\gamma_3 \;\;\; 0.75\leq F_l < 0.9C_l\\
\textrm{rejected} \;\;\; F_l \geq 0.9C_l
\end{cases}
\end{equation}

Note that the aggregated branch flow is computed and compared to its corresponding branch capacity for each branch in $\mathcal{J}_{i,j}$, where we use the largest $\gamma_{i,j}$ value of all used branches in Eq. (\ref{deliverycost}). Besides, the predicted delivery cost may not be exactly equal to the real delivery cost used in the payment settlement module. This is because new supply-demand pairs can be added to the network after the delivery cost is predicted. The new pairs may eventually increase the aggregated branch flows in real electricity scheduling, which makes previous predictions inaccurate. Although there may exist prediction error in delivery cost, it can still reflect the electricity flow congestion in the network at a particular time. Hence the delivery cost prediction is used as a factor in the ranking algorithm defined in Eq. (\ref{ranking}).

\subsection{Payment Settlement Module}
Electricity delivery is financially settled at every time interval. The final payment can be calculated using the following formula,

\begin{equation}
M(i,j)=p^{\textrm{offer}}_jR_{i,j}T^{\textrm{int}}+\gamma_{i,j} vR_{i,j}T^{\textrm{int}},
\end{equation}
where $R_{i,j}$ is the delivery rate (in MW) of the $j$-th SupplyAd and the $i$-th DemandAd, $T^{\textrm{int}}$ (in hour) is the interval duration, and $p^{\textrm{offer}}_j$ is the electricity procurement price (in \$/MWh). So the first term represents the electricity procurement cost. The second term is the electricity delivery cost. The payment settlement module will determine the final $\gamma$ based on true power flow from the electricity network scheduler.

\section{Simulations and Discussions}
We compare the proposed FS matching algorithm defined by Eq. (\ref{ranking}) with a FCFS matching algorithm, which we take as a benchmark algorithm in the simulations. For the benchmark matching algorithm, we first sort the DemandAds and SupplyAds by their submission time in ascending order. The benchmark matching algorithm involves two loops. In the outer loop, it iterates every DemandAd over the DemandAd list. In the inner loop, for each DemandAd, it iterates over the SupplyAd list until we find a match for this DemandAd or we reach the end of the SupplyAd list. We use IEEE 118 Bus Test Case as a proxy for the electricity network in simulations \cite{rchristie}. The simulation parameters are given in Table \ref{simulationpara2}. Specifically, $p_{\textrm{active}}$ is an important parameter to simulate network traffic. In the simulations, each idle bus has a probability of $p_{\textrm{active}}/2$ to submit a DemandAd, and a probability of $p_{\textrm{active}}/2$ to submit a SupplyAd, and a probability of $1-p_{\textrm{active}}$ to remain idle. Besides, the electricity quantity for either DemandAd or SupplyAd is in the range of $[20, 45]$ MWh and the bid or offer price is in the range of $[5, 10]$ \$/MWh. The comparison is done based on the following metrics: network congestion probability, electricity delivery delay probability, network throughput, power loss, and consumer surplus.

\begin{table}[!htbp]\label{simulationpara2}
\caption{SIMULATION PARAMETERS}
\begin{center}
\begin{tabular}{llll}
\hline
Coefficient & Description & Unit & Value\\

$p_{\textrm{active}}$ & Bus active probability & - & 0.1 to 1.0\\

$T^{\textrm{int}}$ & Interval duration & minutes & 15\\

$\gamma_1$ & Surge price multiplier & - &1.0\\

$\gamma_2$ & Surge price multiplier & - & 2.0\\

$\gamma_3$ & Surge price multiplier & - & 4.0\\

$\gamma_4$ & Surge price multiplier & - & 8.0\\

$\alpha_1$ & Para. for price gap & - & 0.2\\

$\alpha_2$ & Para. for delivery cost & - & 0.4\\

$\alpha_3$ & Para. for bus zone &-&0.2\\

$\alpha_4$ & Para. for renewable  &-& 0.2\\

$Q^{\textrm{max}}$ & Max supply/demand & MWh &45\\

$Q^{\textrm{min}}$ & Min supply/demand & MWh & 20\\

$N_r$ & Route number & - & 1, 2, 3, 4, 5\\

$p^{\textrm{bid}}_{max}$ & Max bid price & \$/MWh & 100\\

$p^{\textrm{bid}}_{min}$ & Min bid price & \$/MWh & 50\\

$p^{\textrm{bid}}_{max}$ & Max offer price & \$/MWh & 100\\

$p^{\textrm{offer}}_{min}$ & Min offer price & \$/MWh & 50\\

$N_{\textrm{sim}}$ & Simulation interval number & - & 100000\\

$C_l$ & Branch capacity & MW & 200, 500\\
\hline
\end{tabular}
\end{center}
\end{table}

\subsection{Network Congestion Probability}
\begin{figure}[b]
  \centering
    \includegraphics[width=4.5in]{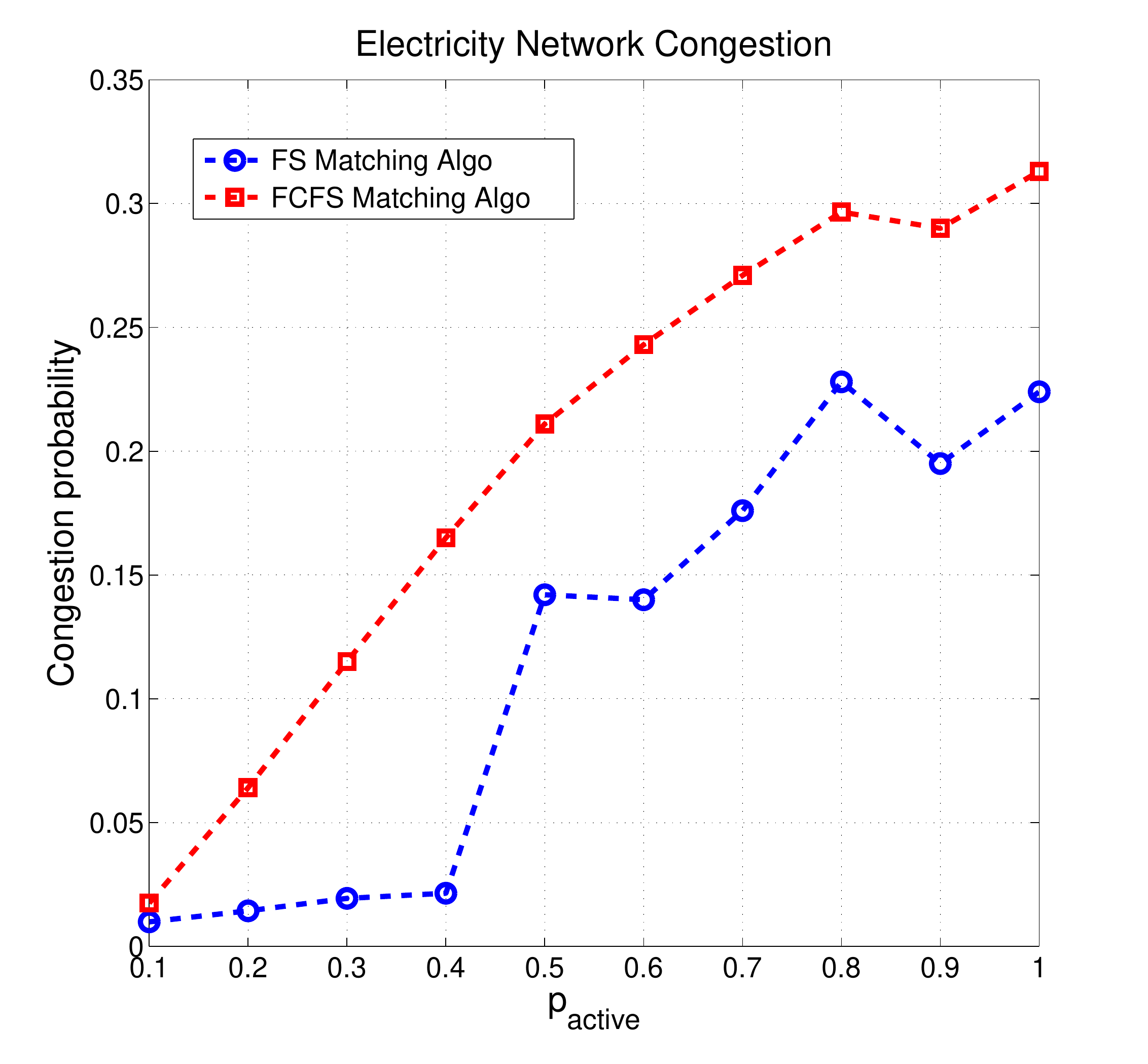}
    \caption{Electricity Network Congestion Comparison}
    \label{networkcongestion}
\end{figure}

In this simulation, we compare the FS matching algorithm and the FCFS matching algorithm in terms of the network congestion performance under different network traffic scenarios. A network congestion occurs when the EFO module cannot find a feasible power flow schedule plan for a given set of supply-demand pairs, which means one or more branch flow must violate branch capacities. The simulation reveals that as the electricity network traffic changes from idle to busy ($p_{\textrm{active}}$ increases from 0.1 to 1.0), the network congestion probability increases for both the FS and the FCFS matching algorithms. However, the FS matching algorithm can always achieve 6\% to 12\% lower congestion probability than the FCFS algorithm for the same network traffic, see Fig. \ref{networkcongestion}. This is because the FS matching algorithm leverages the information from demand and surge price prediction module to select the appropriate supply-demand pairs. These supply-demand pairs avoid using the congestion branches, therefore it is less likely to cause network congestions.

\subsection{Electricity Delivery Delay Probability}
\begin{figure}[b]
  \centering
    \includegraphics[width=4.5in]{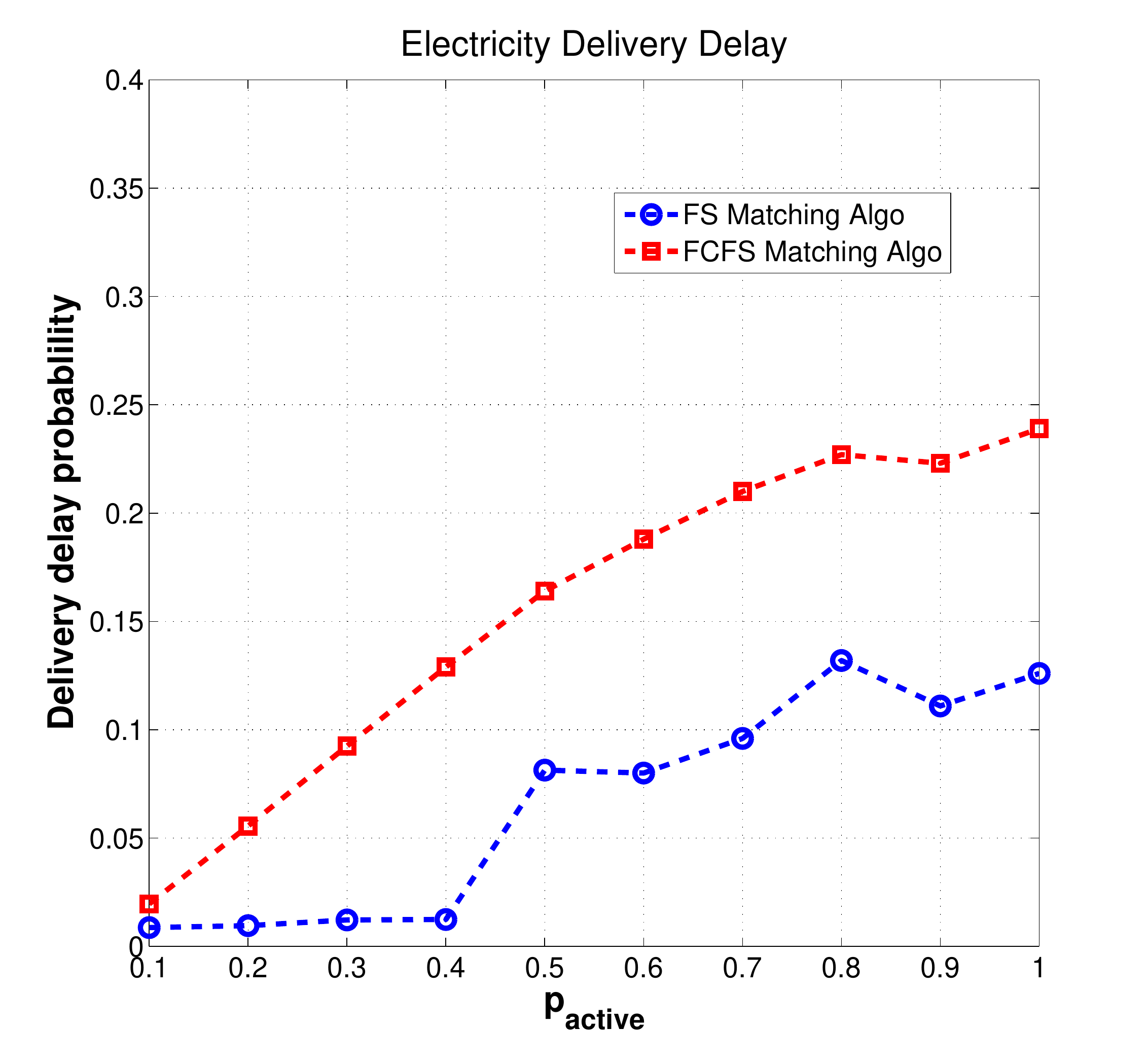}
    \caption{Electricity Delivery Delay Comparison}
    \label{deliverydelay}
\end{figure}

Once the SDM matches a SupplyAd with a DemandAd, the delivery starting time and ending time are also determined. However, sometimes a few supply-demand pairs may not complete the electricity delivery as scheduled due to network congestion or network traffic control, so their original delivery plans have to be delayed. Obviously, customer satisfaction decreases if the electricity market accepts the supply-demand pairs but fails to schedule and finish electricity delivery on time. Here, we want to compare the FS matching algorithm and the FCFS algorithm in their electricity delivery delay performance. From Fig. \ref{deliverydelay}, we note that the electricity delivery delay performance has a similar trend to the network congestion performance. As the electricity network traffic load increases ($p_{\textrm{active}}$ increases from 0.1 to 1.0), the delay probability of the FCFS algorithm increases from 2.0\% to 23.9\%, while the delay probability of the FS algorithm increases from 0.9\% to 12.6\%. It is clear that the FS matching algorithm outperforms the FCFS algorithm for all network traffic scenarios. In particular, the FS matching algorithm performs very well under the low traffic network scenarios ($p_{\textrm{active}}$ from 0.1 to 0.4), since the delay probability almost does not change.

\subsection{Network Throughput}
Another metric for network performance analysis is network throughput. We define the electricity network throughput as the average power (in MW) the network can successfully deliver. The simulation results show that both algorithms have an increase in network throughput as $p_{\textrm{active}}$ increases from 0.1 to 1.0. The FCFS algorithm achieves a larger network throughput than the FS algorithm. This is because the FCFS algorithm does not use any traffic control mechanism. Instead, it will accept any supply-demand pair as long as a SupplyAd matches the requirements of a DemandAd. However, the FS algorithm compares the aggregated branch flow and rejects some supply-demand pairs if their delivery rates are very likely to overload the branches. Thus, the FS algorithm may accept fewer supply-demand pairs than the FCFS algorithm. So the FS algorithm has a smaller throughput than the FCFS algorithm. This is a tradeoff between throughput and network congestion. While the FCFS algorithm has a larger throughput, it suffers severe electricity delivery delay and network congestion.

\begin{figure}[b]
  \centering
    \includegraphics[width=4.5in]{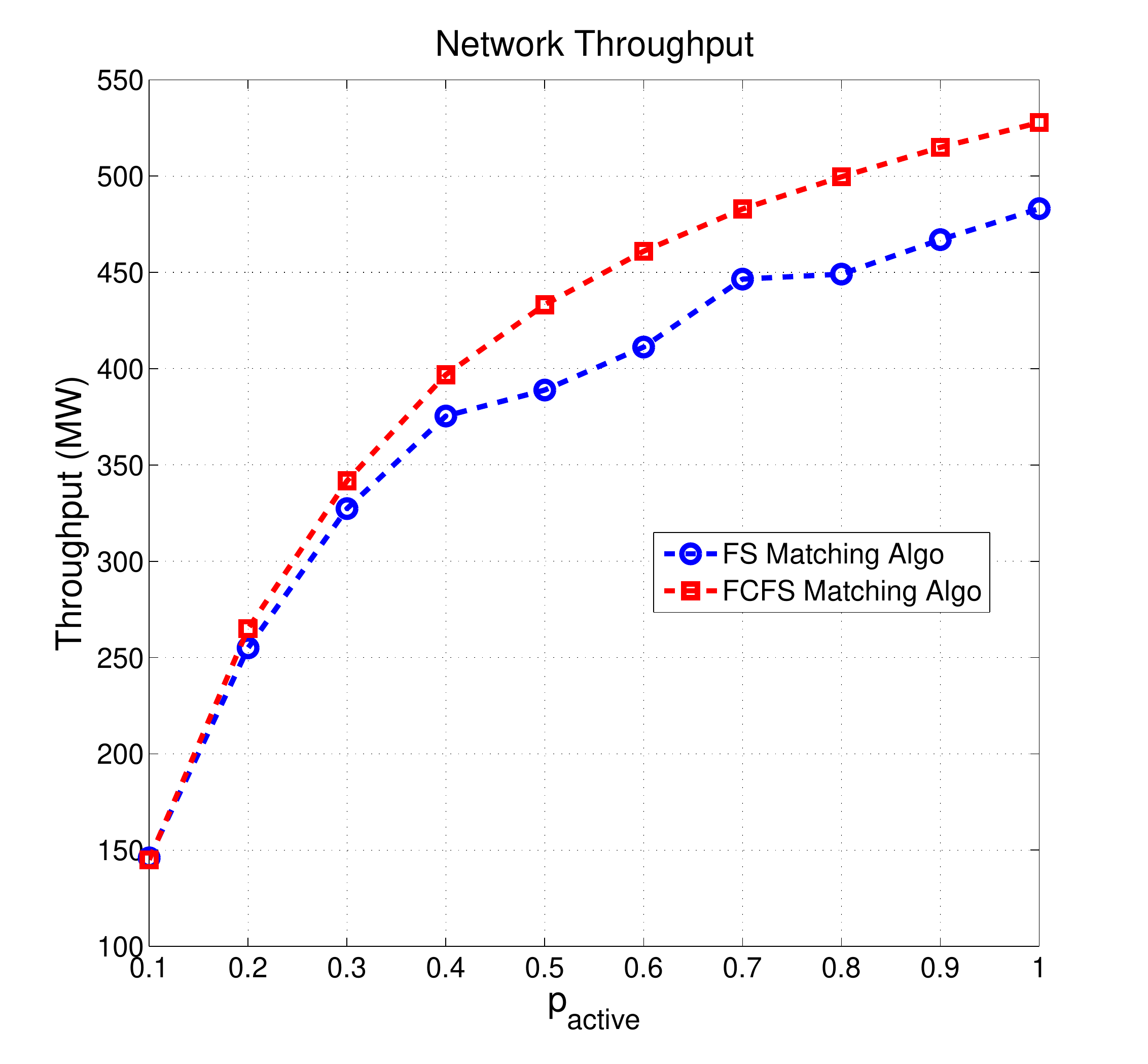}
    \caption{Electricity Network Throughput Comparison}
    \label{throughput}
\end{figure}

\subsection{Power Loss}
Thermal power loss occurs if electricity is transferred over a resistive distribution line. In this simulation, the power loss of the electricity network with FS and FCFS algorithms are compared. We define the power loss as the ratio of the thermal power loss to the network throughput,

\begin{equation}
PL=\frac{P_{\textrm{loss}}}{\Pi},
\end{equation}
where $P_{\textrm{loss}}$ is the thermal power loss (in MW) and $\Pi$ is the network throughput (in MW).

\begin{figure}[b]
  \centering
    \includegraphics[width=4.5in]{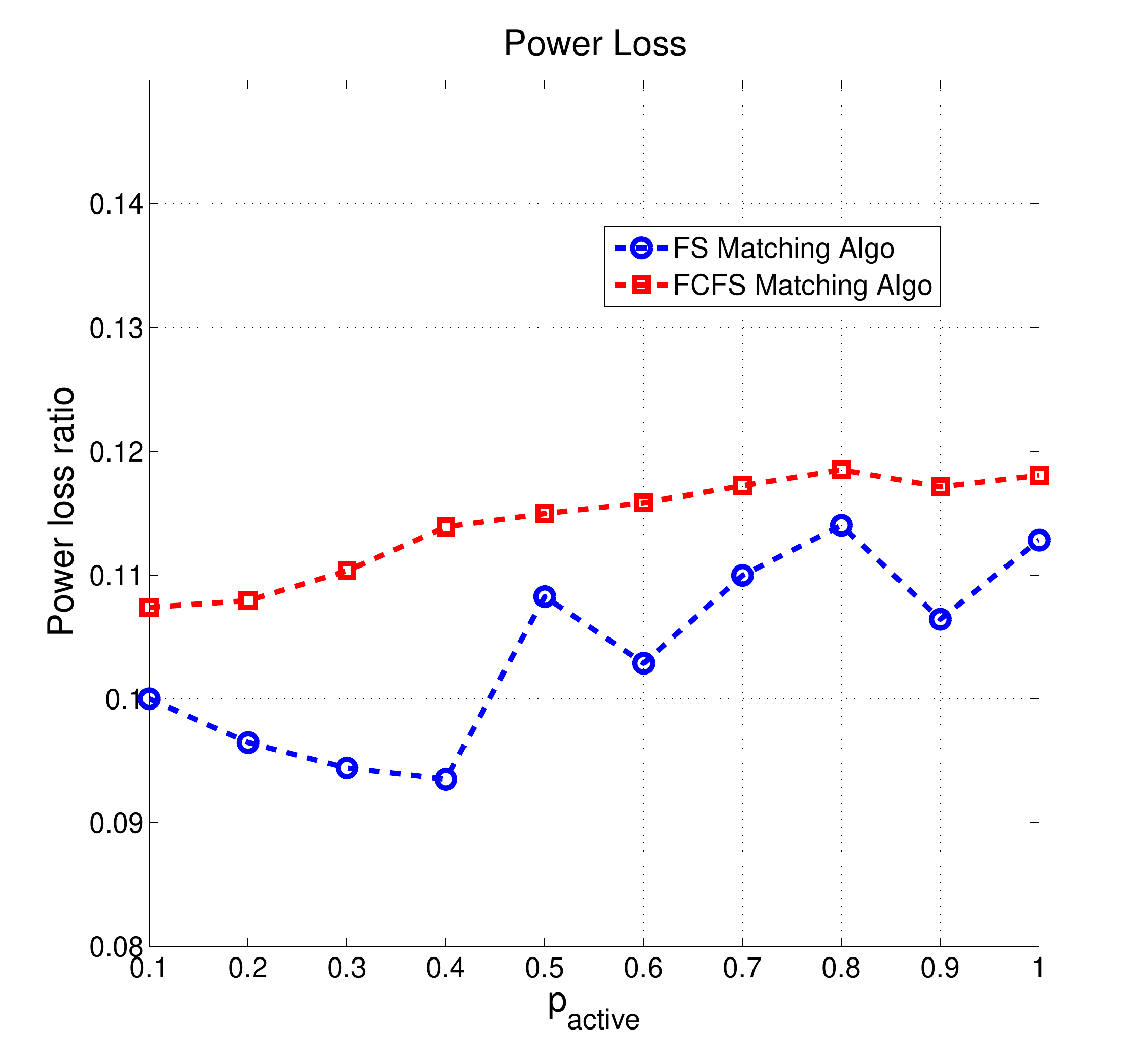}
    \caption{Power Loss Comparison}
    \label{powerloss}
\end{figure}

Fig. \ref{powerloss} shows that the FS matching algorithm have 0.5\% to 2\% lower power loss ratio than the FCFS matching algorithm. This is because the FS matching algorithm matches a SupplyAd for a DemandAd by looking at the electricity procurement cost, the electricity delivery cost, the relative geographical locations and the electricity generation method of this SupplyAd. The supply-demand pair within the same zone and lower delivery cost is favored by the FS algorithm. If a supply-demand pair belongs to the same zone, it does not use inter-zone distribution lines, which can reduce power loss due to long-distance transmission. On the other hand, the surge price multiplier defined in Eq. (\ref{surge}) imposes a heavy penalty to those supply-demand pairs using congestion branches, which can reduce the branch currents, thereby reducing the overall power loss because the power loss is directly proportional to the square of current.

\subsection{Consumer Surplus}
\begin{figure}[b]
  \centering
    \includegraphics[width=4.5in]{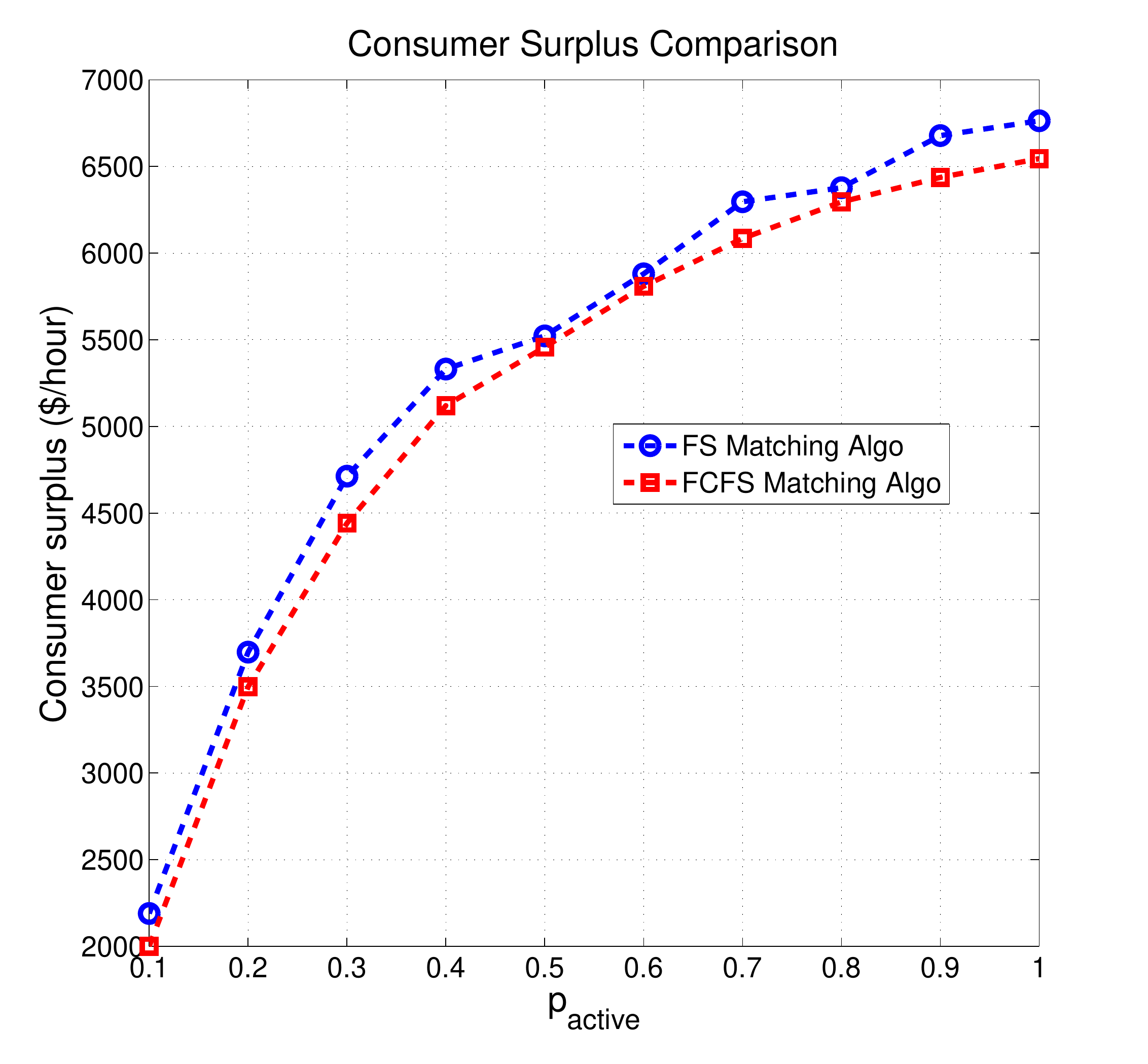}
    \caption{Consumer Surplus Comparison}
    \label{surplus}
\end{figure}

Consumer surplus is defined as the difference between the total amount that consumers are willing to pay for a product or service and the total amount that they actually pay \cite{rbraeutigam}. In our study, the bid price in a DemandAd is the highest price the consumer is willing to pay for electricity, and the offer price in a SupplyAd is the lowest price the consumer actually needs to pay. In this simulation, we use the average consumer surplus to compare the two algorithms, which is defined as follows,

\begin{equation}
CS = \Pi * (p^{\textrm{bid}}_{\textrm{avg}}-p^{\textrm{offer}}_{\textrm{avg}}),
\end{equation}
where $\Pi$ in MW is the network throughput, $p^{\textrm{bid}}_{\textrm{avg}}$ and $p^{\textrm{offer}}_{\textrm{avg}}$ are the average bid price and average offer price, respectively. Thus, the average consumer surplus $CS$ has a unit of $\$/hour$. Fig. \ref{surplus} shows that the FS matching algorithm achieves a larger consumer surplus than the FCFS algorithm. The FS algorithm has a lower network throughput than the FCFS algorithm, but it has a much larger price saving ($p^{\textrm{bid}}_{\textrm{avg}}-p^{\textrm{offer}}_{\textrm{avg}}$). Thus it turns out that the FS algorithm still outperforms the FCFS algorithm in consumer surplus.

\section{Conclusion}
This chapter studies a novel electricity trading market and distribution network for electricity producers and consumers to exchange their underutilized electricity in a flexible, efficient, and economic manner. In particular, we treat electricity as a heterogenous commodity which has different characteristics such as active power, reactive power, power factor, power quality, delivery rate, etc. We propose an online advertisement mechanism for producers and consumers to publicize their demand bids and supply offers, with which they can specify their delivery requirements and constraints. A demand and surge price prediction algorithm is developed to alleviate network congestion. In addition, we propose a FS matching algorithm to match supply-demand pairs. Compared to the FCFS matching algorithm (the benchmark algorithm), the FS matching algorithm achieves better performance in terms of network congestion management, electricity delivery delay probability, energy efficiency, and consumer surplus.

%
%

%
%
%
%
%
%
%
%
%
%

%
%

\chapter{MAIN CONTRIBUTIONS AND FUTURE WORK}

\section{Main Contributions}
This dissertation aims to solve three critical issues arising from the EV proliferation: (1) optimal EV charging station placement, (2) dynamic pricing and electricity management policy of EV charging station, and (3) an electricity trading market and distribution network redesign for large-scale EV and renewable energy integration. Our main contributions are summarized in the following subsections.

\subsection{Optimal EV Charging Station Placement}
To solve the problem of EV charging station placement, we developed a multi-stage placement strategy considering incremental EV market penetration rates and the 3-way interactions among EVs, the traffic network, and the electricity network. A nested logit model was first employed to characterize the EV driver's satisfaction and charging preference, which was used to predict overall charging demand. An oligopoly market form was then applied to model the EV charging market. We analyzed how the EV charging service providers will interact with each other. In addition, we developed a simulation toolkit called ``The EV Virtual City" based on Repast to simulate the real-time EV traffic and how EVs would interact with the changing surroundings. Finally, we developed an Android application called ``Charging Butler" which serves as an assistant for EV drivers to search for the optimal charging station and provides turn-by-turn navigation.

\subsection{Dynamic Pricing and Electricity Management Policy of EV Charging Station}
In Chapter 4, we proposed a multi-objective optimization framework to jointly optimize multiple objectives of improving profitability, enhancing customer satisfaction, and reducing the charging impact on the power system. In addition, we developed a new metric to evaluate how EV charging will impact the power system, which is fast-computing and frees us from solving nonlinear power flow equations. Additionally, we derived the active power and reactive power sensitivity formula for the load buses in a power system which can serve as a guideline for EV charging station placement to alleviate the charging stress on the power grid. In terms of market risk, we introduced a safeguard of profit for EV charging service providers, which raises a warning when the profit is likely to reach a dangerous threshold. This mechanism is beneficial for the charging service provider to manage its capital and avoid severe financial losses.

\subsection{Electricity Trading Market and Distribution Network Redesign}
To foster large-scale EV integration and renewable energy generation, we propose to design a liberal electricity trading market model and a distribution network from a perspective of sharing economy. We designed a peer-to-peer (P2P) electricity trading market based on an online advertisement mechanism. Accordingly, a fitness score (FS)-based matching algorithm was studied to pair up the best supply ads and demand ads by considering electricity network congestion control, renewable energy integration, and consumer surplus. A new electricity distribution mechanism was also investigated to naturally support multi-rate electricity delivery and discontinuous renewable energy generation.

\section{Future Work}
There are many other interesting problems on EV charging and integration which are worth studying in the future. Some future research topics include: (1) apply machine learning and data mining to help EVs choose optimal EV charging stations and charging scheduling, (2) energy aware routing for EVs, and (3) decentralized P2P electricity trading market and distribution network design.

%
%

%
%

\appendix

%
%
%
%
%
%
%
%

%
%

\chapter{DISCRETE CHOICE MODEL FORMULA}

\section{Derivation of Eq. (3.10)}
We assume the unobservable utility $\delta$ has extreme value distribution with a probability density function given by
\begin{equation}
f(\delta)=e^{-\delta}e^{-e^{-\delta}}
\end{equation}
and the cumulative distribution function is
\begin{equation}\label{cdf2}
F(\delta)=e^{-e^{-\delta}}
\end{equation}
The probability that a consumer chooses alternative $k$ is given by
\begin{equation}
\begin{aligned}
P_k&=\textrm{Prob}(\overline{M}_k+\delta_k>\overline{M}_l+\delta_l,\;\;\;\forall l\neq k)\\
&=\textrm{Prob}(\delta_l<\delta_k+\overline{M}_k-\overline{M}_l,\;\;\;\forall l\neq k)
\end{aligned}
\end{equation}
If $\delta_k$ is assumed to be known, the expression above is the cumulative distribution function for each $\delta_l$ evaluated at $\delta_k+\overline{M}_k-\overline{M}_l$. According to Eq. (\ref{cdf2}), this is equal to $\exp(-\exp(-(\delta_k+\overline{M}_k-\overline{M}_l)))$. Since the unobservable utility is i.i.d., the probability for all $l \neq k$ is the product of each cumulative distribution
\begin{equation}
P_{k|\delta_k}=\prod_{l \neq k}e^{-e^{-(\delta_k+\overline{M}_k-\overline{M}_l)}}
\end{equation}
Then, the choice probability is the integral of $P_{k|\delta_k}$ over the distribution of $\delta_k$

\begin{equation}
P_k=\int_{-\infty}^{\infty}\left(\prod_{l \neq k}e^{-e^{-(\delta_k+\overline{M}_k-\overline{M}_l)}}\right)
e^{-\delta_k}e^{-e^{-\delta_k^n}}d\delta_k
\end{equation}
Define $u=\delta_k$, and we shall have
\begin{equation}
P_k=\int_{-\infty}^{\infty}\left(\prod_{l \neq k}e^{-e^{-(u+\overline{M}_k-\overline{M}_l)}}\right)
e^{-u}e^{-e^{-u}}du
\end{equation}
Note that $\overline{M}_k-\overline{M}_k=0$ and we can collect terms in the following way
\begin{equation}
\begin{aligned}
P_k&=\int_{-\infty}^{\infty}\left(\prod_le^{-e^{-(u+\overline{M}_k-\overline{M}_l)}}\right)
e^{-u}du\\
&=\int_{-\infty}^{\infty}\exp\left(-\sum_le^{-(u+\overline{M}_k-\overline{M}_l)}\right)e^{-u}du\\
&=\int_{-\infty}^{\infty}\exp\left(-e^{-u}\sum_le^{-(\overline{M}_k-\overline{M}_l)}\right)e^{-u}du
\end{aligned}
\end{equation}
Then we define $s=e^{-u}$,  so $ds=-e^{-u}du$
\begin{equation}
\begin{aligned}
P_k&=\int_0^\infty\exp\left(-s\sum_le^{-(\overline{M}_k-\overline{M}_l)}\right)ds\\
&=\left.\frac{e^{-s(\overline{M}_k-\overline{M}_l)}}{-\sum_le^{-(\overline{M}_k-\overline{M}_l)}}\right|_0^\infty\\
&=\frac{1}{\sum_le^{-(\overline{M}_k-\overline{M}_l)}}=\frac{e^{\overline{M}_k}}{\sum_le^{\overline{M}_l}}
\end{aligned}
\end{equation}

\begin{equation}
\begin{aligned}
\Phi_{j,k}^n&=\frac{e^{\overline{U}_{j,k}^n/\sigma_k}\left(\sum_{l=1}^Le^{\overline{U}_{l,k}^n/\sigma_k}\right)^{\sigma_k-1}}
{\sum_{t=1}^3\left(\sum_{l=1}^Le^{\overline{U}_{l,t}^n/\sigma_t}\right)^{\sigma_t}}\\
&=\frac{e^{\overline{U}_{j,k}^n/\sigma_k}}{\sum_{l=1}^Le^{\overline{U}_{l,k}^n/\sigma_k}}\frac{\left(
\sum_{l=1}^Le^{\overline{U}_{l,k}^n/\sigma_k}\right)^{\sigma_k}}{\sum_{t=1}^3\left(\sum_{l=1}^Le^{\overline{U}_{l,t}^n/\sigma_t}\right)^{\sigma_t}}\\
&=\frac{e^{\left(\overline{W}_k^n+\overline{V}_{j,k}^n\right)/\sigma_k}}{\sum_{l=1}^Le^{\left(\overline{W}_k^n+\overline{V}_{l,k}^n\right)/\sigma_k}}
\frac{\left(\sum_{l=1}^Le^{\left(\overline{W}_k^n+\overline{V}_{l,k}^n\right)/\sigma_k}\right)^{\sigma_k}}{\sum_{t=1}^3\left(\sum_{l=1}^Le^{\left(\overline{W}_t^n+\overline{V}_{l,t}^n\right)/\sigma_t}\right)^{\sigma_t}}\\
&=\frac{e^{\overline{W}_k^n/\sigma_k}e^{\overline{V}_{j,k}^n/\sigma_k}}{e^{\overline{W}_k^n/\sigma_k}\sum_{l=1}^Le^{\overline{V}_{l,k}^n/\sigma_k}}
\frac{e^{\overline{W}_k^n}\left(\sum_{l=1}^Le^{\overline{V}_{l,k}^n/\sigma_k}\right)^{\sigma_k}}{\sum_{t=1}^3e^{\overline{W}_t^n}\left(\sum_{l=1}^Le^{\overline{V}_{l,t}^n/\sigma_t}\right)^{\sigma_t}}\\
&=\frac{e^{\overline{V}_{j,k}^n/\sigma_k}}{\sum_{l=1}^Le^{\overline{V}_{l,k}^n/\sigma_k}}
\frac{e^{\overline{W}_k^n+\sigma_kI_k^n}}{\sum_{t=1}^3e^{\overline{W}_t^n+\sigma_tI_t^n}}\\
&=P_{j|k}^nP_k^n
\end{aligned}
\end{equation}
where the second last equality holds because $I_t^n=\log\left(\sum_{l=1}^Le^{\overline{V}_{l,t}^n/\sigma_k}\right)$.

%

%
%
%
%
%
%
%
%

%
%

\chapter{MOBILE APPLICATION DEVELOPMENT --- EV CHARGING BUTLER}
We develop a location-aware EV charging station recommender android application. The main functionality of this application is to assist EV driver search for optimal EV charging station and provide turn-by-turn navigation to get there.

\begin{figure}[htbp]
\centerline{\includegraphics[width=4in]{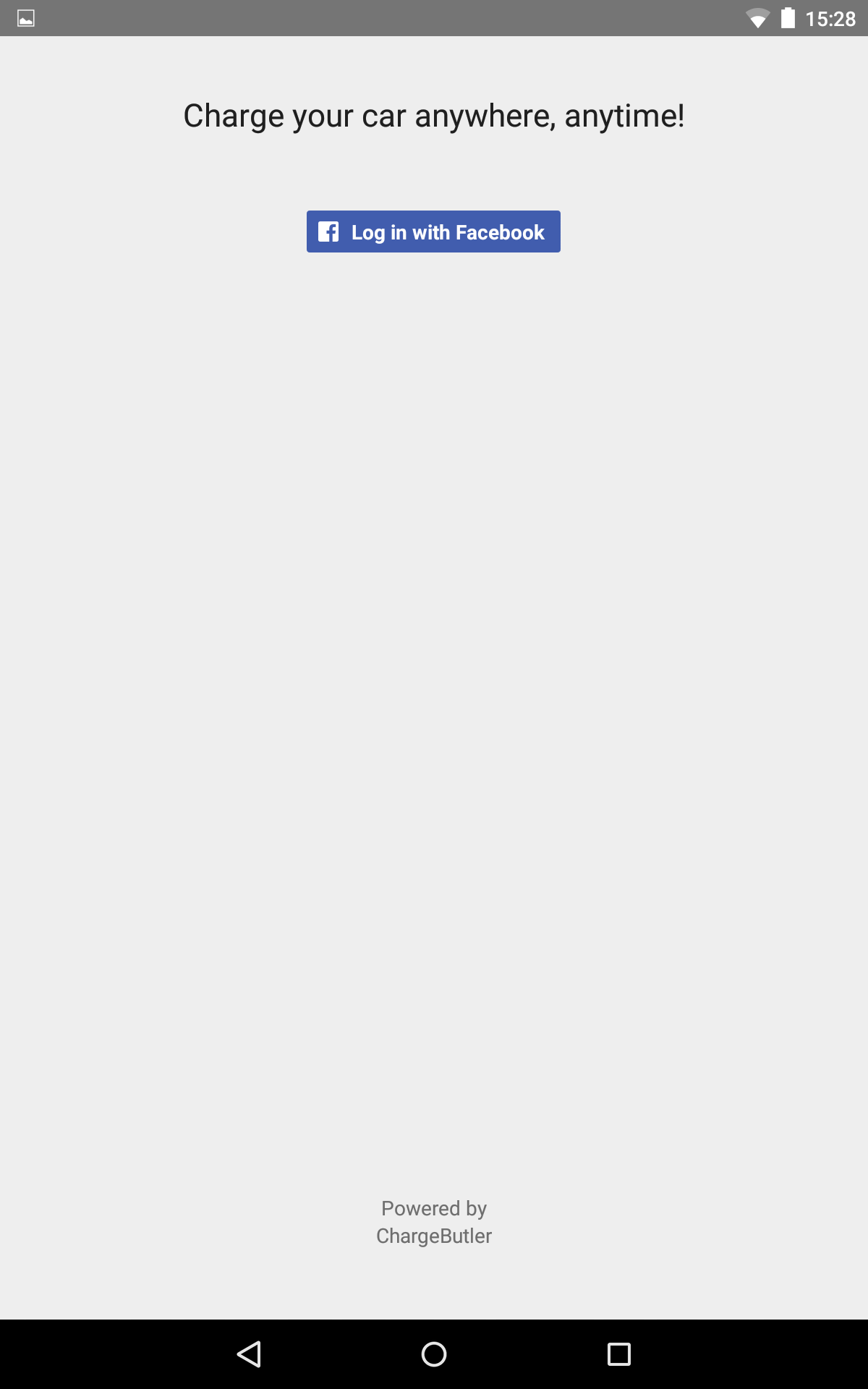}}
\center
\caption{Facebook Login Activity}
\label{login}
\end{figure}

\begin{figure}[htbp]
\centerline{\includegraphics[width=4in]{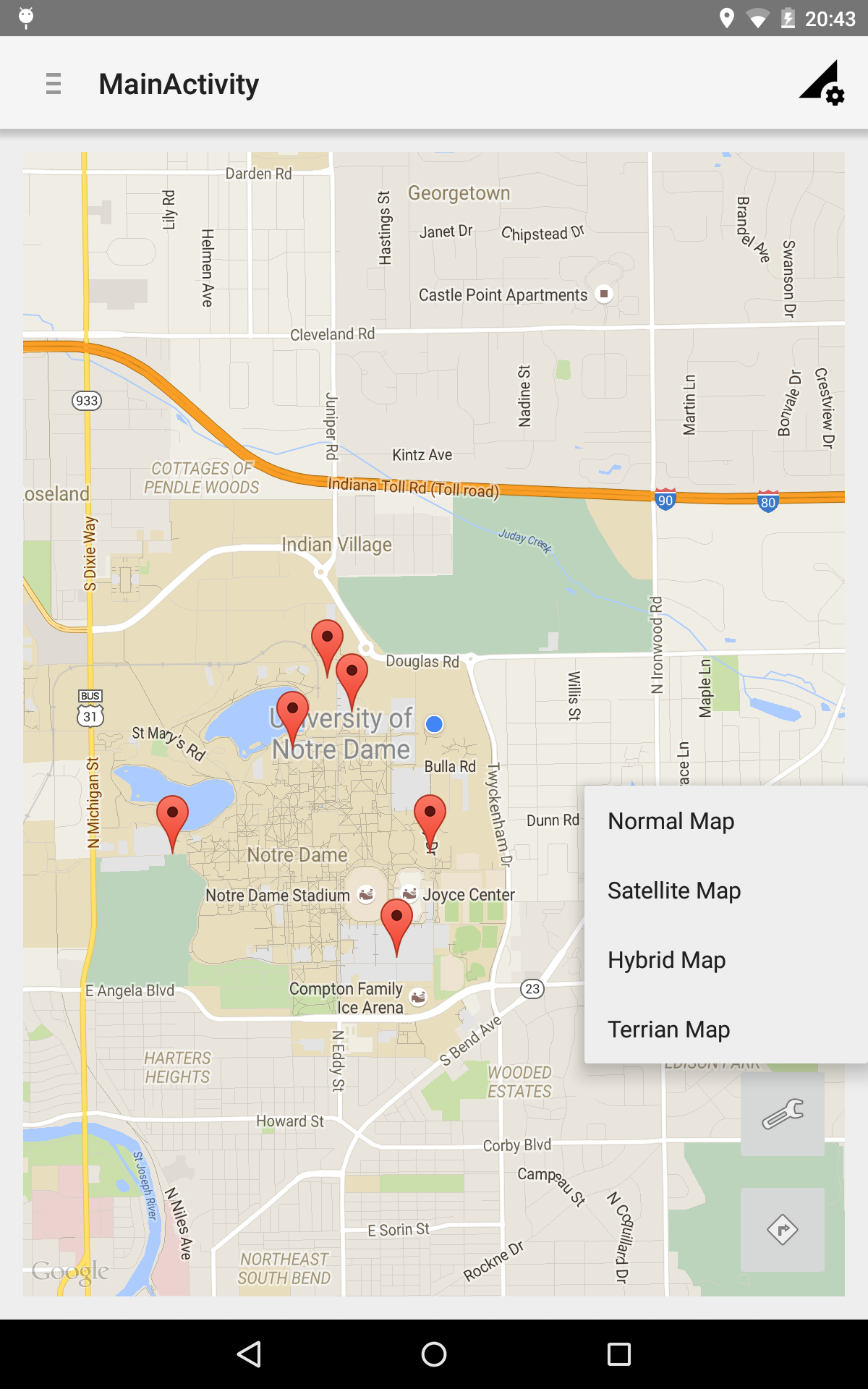}}
\center
\caption{Main Activity}
\label{mainactivity}
\end{figure}

First, an EV driver needs to log into the application using his/her Facebook account, see Fig. \ref{login}. Then, the user sees the main activity, where a Google Map is displayed. The user can choose map types (i.e. normal, satellite, hybrid, and terrain), see Fig. \ref{mainactivity}. In addition, an EV charging station map is integrated into the application. We use the application programming interfaces (APIs) to query and fetch the charging station information from the charging station map which is provided by the Open Charge Map Organization. The red markers in Fig. \ref{mainactivity} represent the nearby EV charging stations.

After selecting a charging station, the Google Maps will provide turn-by-turn navigation to the user, see Fig. \ref{navigation}. The application has a main menu, where the user can see his/her profile (i.e. picture, name, email) and edit the preference and settings, and perform logout action, see Fig. \ref{menu}.

\begin{figure}[htbp]
\centerline{\includegraphics[width=4in]{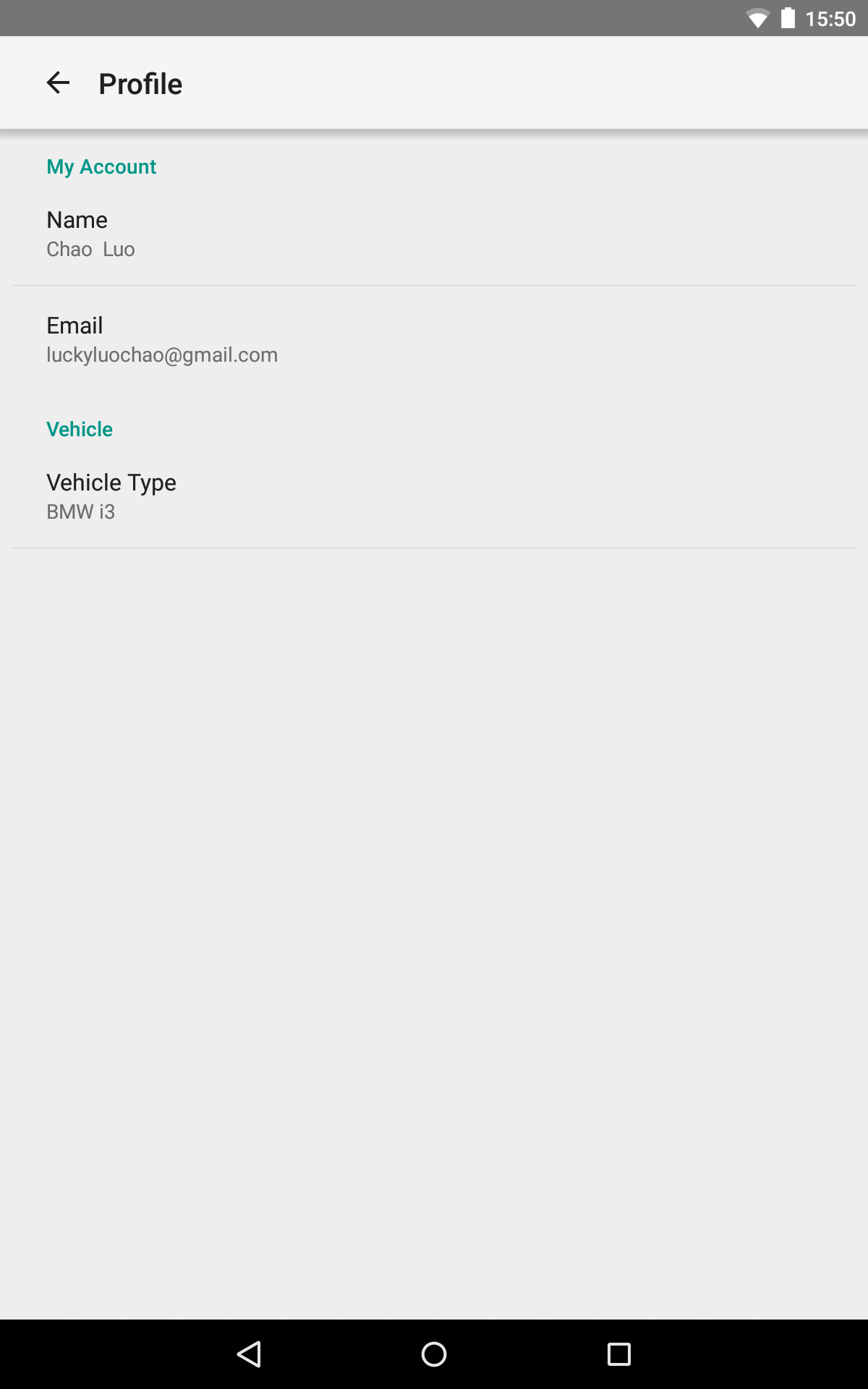}}
\center
\caption{Profile Settings}
\label{profile}
\end{figure}

\begin{figure}[htbp]
\centerline{\includegraphics[width=4in]{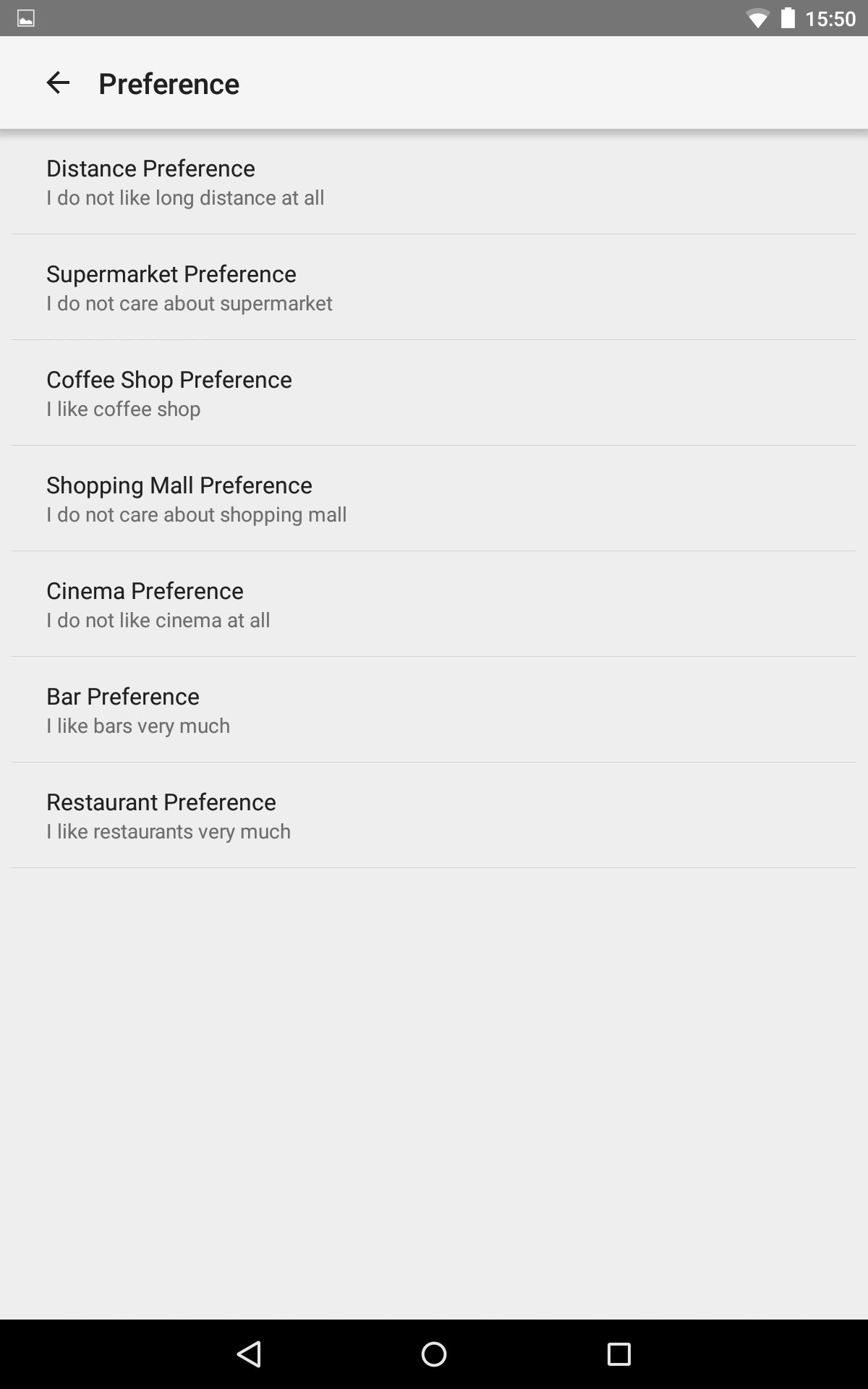}}
\center
\caption{Charging Preference Settings}
\label{preference}
\end{figure}

\begin{figure}[htbp]
\centerline{\includegraphics[width=4in]{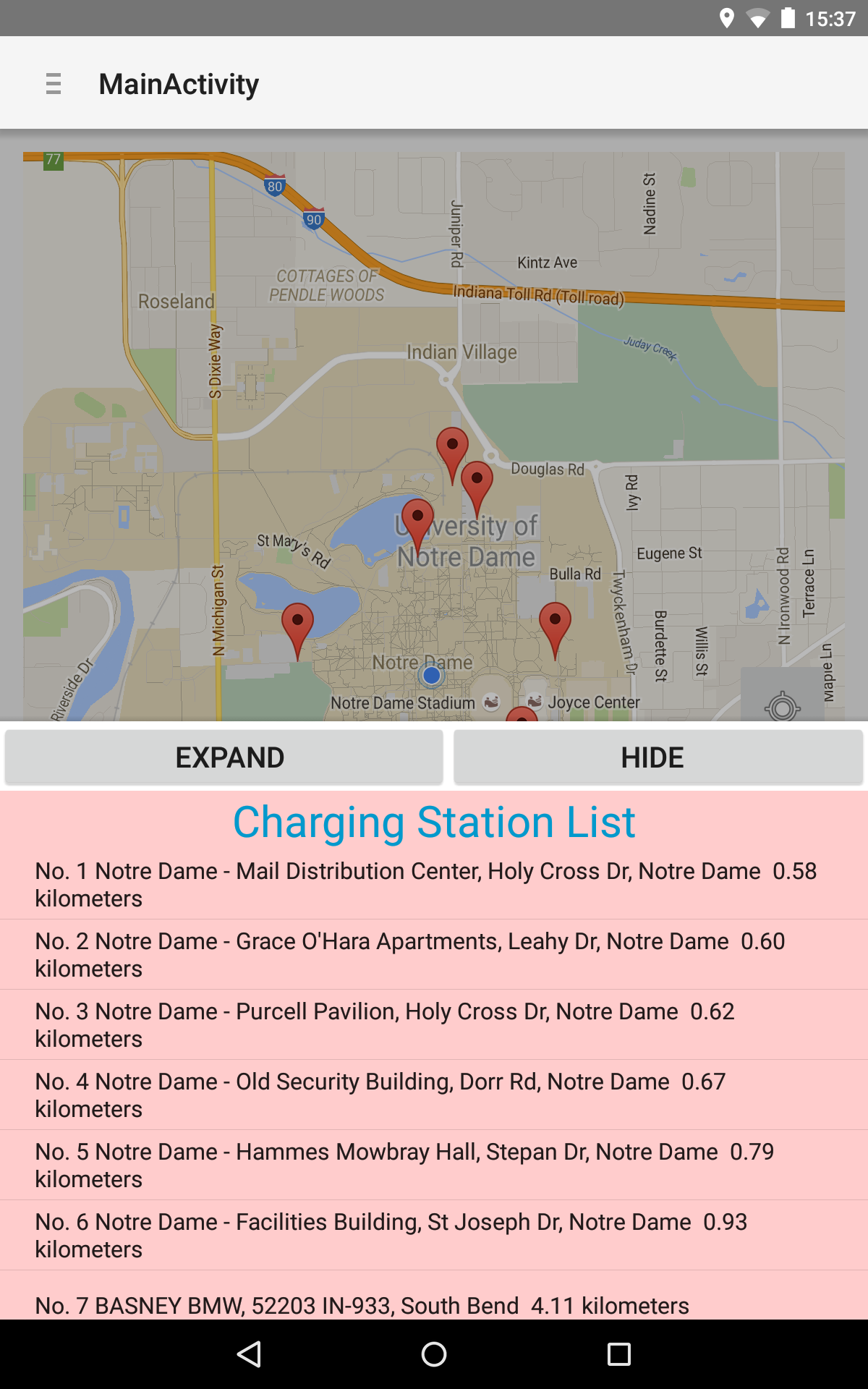}}
\center
\caption{Recommendation List}
\label{list}
\end{figure}

\begin{figure}[htbp]
\centerline{\includegraphics[width=4in]{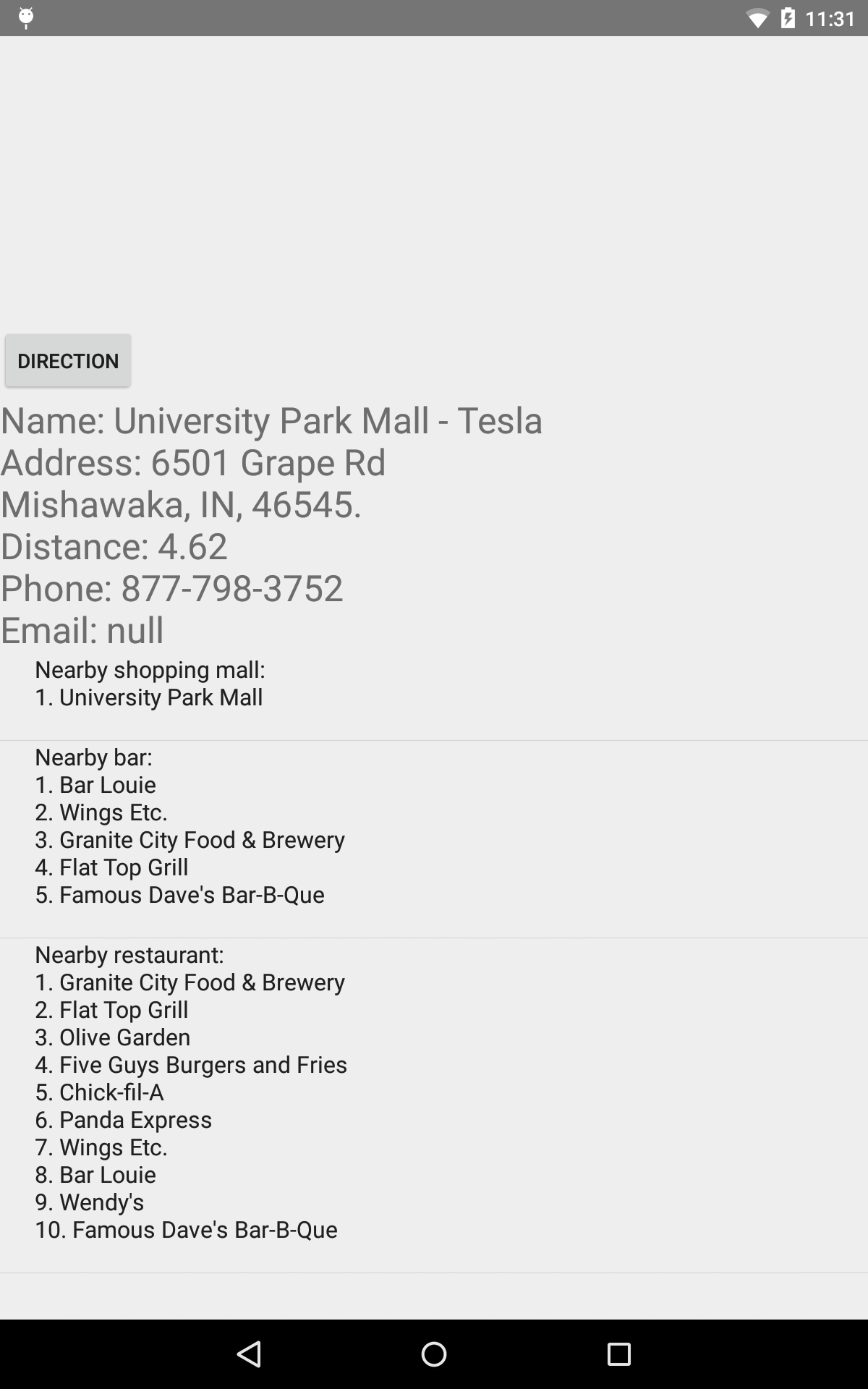}}
\center
\caption{Charging Station Information}
\label{panel}
\end{figure}

I implemented the profile setting activity and charging preference activity, see Fig. \ref{profile} and Fig. \ref{preference}. In profile setting activity, the user can choose the vehicle type. In the charging preference activity, the user can edit his/her preferences about distance, cafe, cinema, supermarket, etc. I also implemented the recommendation engine algorithm and the recommendation list panel. See Fig. \ref{list} and Fig. \ref{panel}. Fig. \ref{list} displays the list of recommended charging stations which are ordered according the scores from recommendation engine. Fig. \ref{panel} displays the detailed information of a charging station if the user clicks one from the recommendation list.

%

%
%


\backmatter
\begingroup
\raggedright
 \bibliographystyle{IEEEtran} 
 \bibliography{example}       
\endgroup

\end{document}